# Machine learning to tame divergent density functional approximations: a new path to consensus materials design principles

Chenru Duan[1,2], Shuxin Chen[1,2], Michael G. Taylor[1], Fang Liu[1], and Heather J. Kulik[1,*]

[1]*Department of Chemical Engineering, Massachusetts Institute of Technology, Cambridge, MA 02139*

[2]*Department of Chemistry, Massachusetts Institute of Technology, Cambridge, MA 02139*

ABSTRACT: Computational virtual high-throughput screening (VHTS) with density functional theory (DFT) and machine-learning (ML)-acceleration is essential in rapid materials discovery. By necessity, efficient DFT-based workflows are carried out with a single density functional approximation (DFA). Nevertheless, properties evaluated with different DFAs can be expected to disagree for the cases with challenging electronic structure (e.g., open shell transition metal complexes, TMCs) for which rapid screening is most needed and accurate benchmarks are often unavailable. To quantify the effect of DFA bias, we introduce an approach to rapidly obtain property predictions from 23 representative DFAs spanning multiple families and "rungs" (e.g., semi-local to double hybrid) and basis sets on over 2,000 TMCs. Although computed properties (e.g., spin-state ordering and frontier orbital gap) naturally differ by DFA, high linear correlations persist across all DFAs. We train independent ML models for each DFA and observe convergent trends in feature importance; these features thus provide DFA-invariant, universal design rules. We devise a strategy to train ML models informed by all 23 DFAs and use them to predict properties (e.g., spin-splitting energy) of over 182k TMCs. By requiring consensus of the ANN-predicted DFA properties, we improve correspondence of these computational lead compounds with literature-mined, experimental compounds over the single-DFA approach typically employed. Both feature analysis and consensus-based ML provide efficient, alternative paths to overcome accuracy limitations of practical DFT.

## 1. Introduction

Virtual high-throughput screening (VHTS)[1-8] both with direct physics-based simulation and with machine learning (ML)[9-12] is essential in the accelerated discovery of new molecules and materials. Approximate density functional theory (DFT) has become indispensable either for property prediction in VHTS or to serve as training data for ML models. Although the favorable combination of cost and accuracy in DFT has motivated its use in screening workflows, the failures of DFT are prominent for the cases where chemical discovery efforts are most needed (e.g., open-shell radicals, transition metal containing systems, and strained bonds).[13-17] One solution to overcome these limits is to climb up a "Jacob's ladder"[18] of density functional approximations (DFAs), where approximations on higher rungs include more ingredients such as higher-order expansions of the density, Hartree-Fock exchange, and correlation from perturbation theory (i.e., MP2). Doing so has been shown to increase accuracy for organic molecules with a modest increase in computational cost, but simply climbing up to higher rungs does not always guarantee improvements in challenging systems.[19-20] Furthermore, choosing the "right" rung *a priori* in computational materials discovery efforts is impractical if benchmarks are unavailable, and, instead, usually a single low-cost, heavily-tested DFA (e.g., PBE or B3LYP) is employed.

Transition metal complexes (TMCs) are exemplary of such challenges in single-DFA-based high-throughput screening. While TMCs are of interest for efficient VHTS and ML-accelerated discovery due their wide applications in catalysis[4, 21-27] and energy utilization (e.g., in redox flow batteries[28], solar cells[29], and molecular switches[30]), their electronic structure is challenging to describe accurately.[15] The variable nature of metal, oxidation state, and spin of TMCs introduces combinatorial explosion in design spaces[11, 31] that cannot be exhaustively explored by either first-principles methods or experiments, motivating ML-acceleration.[32-38] Open shell TMCs are particularly difficult to study due to their near-degenerate *d* orbitals that may introduce significant multireference (MR) character[39-43]. Furthermore, properties are highly sensitive to selected DFA. For example, semi-local-DFT-derived ML acceleration will discover TMCs targeted for specific spin state properties (i.e., spin-crossover or SCO) with weaker field ligands than hybrid-DFT-derived ML models will.[36]

Despite this challenge, a single, low-cost DFA choice is needed to automate workflows and generate large data sets. When careful studies of smaller data sets have been carried out, they have revealed DFA dependence (e.g., including fraction of Hartree-Fock, HF, exchange) on property evaluations on both organic molecules[44-46] and TMCs[47-51]. To address this issue, some have tried to optimize a DFA for specific properties with respect to the experimental or correlated wavefunction theory (WFT) reference data[52-54] or suggest DFAs in a system and property specific manner[55-56]. Recently, we have built an ML decision engine[57-58] at DFT costs that classify systems with strong MR character and thus identify regions of chemical spaces that are safe to compute with DFT[59]. One may expect single-reference WFT and DFAs with high HF exchange fraction to fail when strong MR character is present, but high errors may not be present for all DFAs even when MR character is present.[13-14, 17] Bayesian inference has been used to analyze errors from different DFAs and design new DFAs for systems that potentially contain strong static correlation (i.e., MR character).[60-63] To date, however, no one has thoroughly investigated how this systematic DFA bias in the datasets would further influence ML model training and lead compounds in chemical discovery.

Here, we carry out the first large-scale study on over 2,000 TMCs of 23 DFAs from numerous rungs of "Jacob's ladder" (i.e., from semi-local DFT to double hybrids) for three distinct chemical properties. We show that while absolute properties computed by different DFAs disagree, their good overall linear correlation is observed. We show how design rules obtained from the most important features in feature-selected ML models are invariant to DFA choice or basis set, providing a robust tool for materials screening. We introduce a new fine-tuning strategy for transfer learning to train 23 artificial neural networks (ANNs) with comparable latent spaces. We show how exploiting the consensus of all 23 ANNs to discover complexes (e.g., SCOs) ensures improved agreement with experiment over a single-DFA approach.

## 2. Results and discussion

### 2a. Statistical analysis on properties derived at different DFAs

We study a broad range of 23 density functional approximations (DFAs) that are distributed among multiple rungs of "Jacob's ladder"[18] (ESI Table S1). We employ three popular semi-local generalized gradient approximations (GGAs) that are widely used to study both molecular and solid-state systems (i.e., BLYP[64-65], BP86[66-67], and PBE[68]) and their corresponding global GGA hybrids (B3LYP[69-71], B3P86[66, 69], B3PW91[69, 72], and PBE0[73]). We include two few-parameter, meta-GGAs (TPSS[74] and SCAN[75]) and two more highly parameterized ones (M06-L[76] and MN15-L[77]) that have been empirically tuned to improve performance on a range of benchmark properties[76-77], such as bond energies, reaction barrier heights, and noncovalent interactions. We also include popular hybrid variants of these meta-GGAs (i.e., TPSSh[74], SCAN0[78], M06[79], M06-2X[79], and MN15[80]). In addition to the GGA and meta-GGA hybrids, we employ two range-separated (RS) hybrids (i.e., LRC-ωPBEh[81] and ωB97X[82]) that consist of GGA hybrids in the short range and full non-local exchange in the long-range. Lastly, we incorporate both non-empirical, double hybrids (B2GP-BLYP[83] and PBE0-DH[84]) and parameterized, spin-component-scaled double hybrids with dispersion corrections (DSD-BLYP-D3BJ[85], DSD-PBEB95-D3BJ[85], and DSD-PBEP86-D3BJ[85]). Over this set of DFAs, we cover a number of semi-local exchange or correlation functional families, an extended range of HF exchange fractions (0.100 to 0.710) in hybrids, and a large range of MP2 correction fractions (0.125 to 1.000) in double hybrids (ESI Table S1).

To evaluate the relative agreement among these DFAs, we focus on three properties that depend on multiple geometries, charges, or spin states in a large set of transition metal complexes (TMCs, see Sec. 4). These include: i) the adiabatic high-spin (HS) to low-spin (LS) splitting energy, $\Delta E_{\text{H-L}}$; ii) the vertical ionization potential, IP, of the complex; and iii) the frontier orbital gap from Δ-SCF[86] (hereafter, the Δ-SCF gap) obtained as the difference of the vertical IP and vertical electron affinity (EA) of the TMC. We selected $\Delta E_{\text{H-L}}$ because it is known[47, 87-95] to be strongly sensitive to DFA choice. We also selected the Δ-SCF[86] evaluation of the HOMO-LUMO gap and vertical IP evaluated from total energy differences that are expected to be less sensitive to the lack of piecewise linearity[96-97] in a DFA in comparison to the same properties obtained from frontier orbital energies.[13] All properties are evaluated on a single parent functional and basis set choice (B3LYP/LACVP*) we typically employ for its efficiency in high-throughput screening (see Sec. 4). By both using a consistent geometry and developing a strategy for preserving the qualitative description of the wavefunction across DFAs (see Sec. 4), we isolate the role of the DFA parameterization in altering predicted energetic properties.

Although the computed values differ with change in DFA for each of the three properties, the obtained values from different DFAs generally have high linear correlations, as quantified by Pearson's correlation coefficients (Fig. 1, ESI Fig. S1 and Table S2). We observe this strong positive correlation between all pairs of DFAs, even for DFAs from different rungs on "Jacob's ladder", and for properties (i.e., $\Delta E_{\text{H-L}}$) that can be expected to be strongly functional-dependent. Across all functionals, the vertical IP correlations are consistently high (i.e., 0.99–1.00) so we focus further analysis on the Δ-SCF gap and especially the most DFA-sensitive $\Delta E_{\text{H-L}}$ property (ESI Fig. S1).

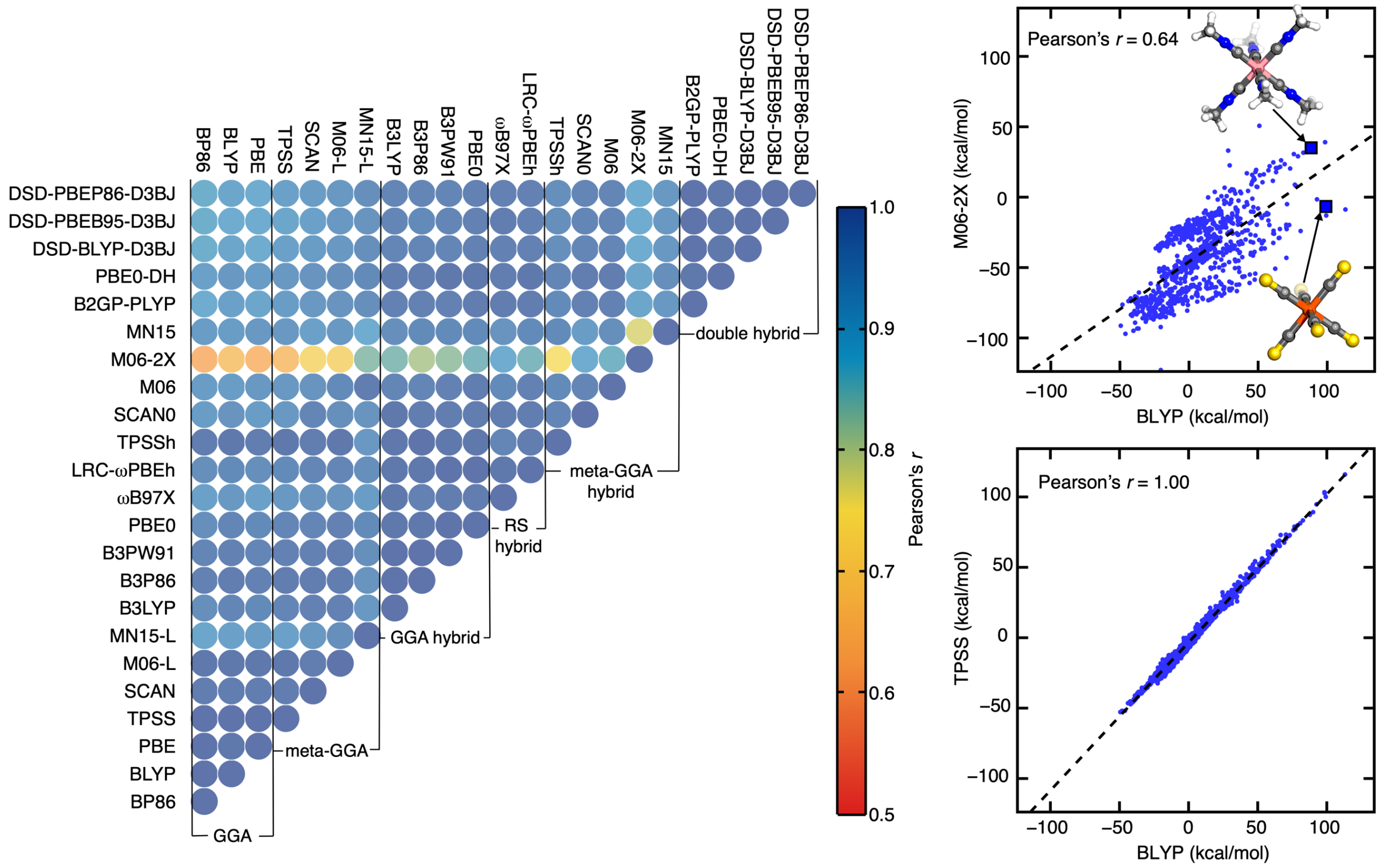

**Fig. 1** (left) An upper triangular matrix of Pearson's $r$ for pairs of $\Delta E_{\text{H-L}}$ derived from 23 DFAs with the LACVP* basis set colored according to inset colorbar (i.e., blue for 1.0 to red for 0.5). (right) Parity plot of $\Delta E_{\text{H-L}}$ for pairs of DFAs with the lowest Pearson's $r$ (0.64, BLYP and M06-2X, top) and the highest (1.00, BLYP and TPSS, bottom). In each parity plot, a black dashed linear regression line is shown. Two representative complexes are shown (top pane): Co(III)$(C_2H_4N)_6$ and Fe(II)$(CS)_6$. Atoms are colored as follows: orange for Fe, pink for Co, blue for N, yellow for S, gray for C, and white for H.

Despite strong DFA sensitivity, three GGAs have near-perfect linear correlations with each other (Pearson's $r > 0.99$) for $\Delta E_{\text{H-L}}$ (Fig. 1). For this property, most of the meta-GGAs also have extremely high (Pearson's $r > 0.98$) linear correlation with GGAs, with MN15-L being the sole exception (e.g., Pearson's $r$ of 0.89 with BLYP, ESI Fig. S2). One might expect a functional like the SCAN meta-GGA that has been demonstrated to make more accurate predictions on formation enthalpy[98] and reaction energy[99] to correlate poorly with less accurate semi-local GGA functionals, but Δ-SCF gap or $\Delta E_{\text{H-L}}$ properties computed with SCAN correlate just as highly to the GGAs as other few-parameter meta-GGAs and better than the more highly parameterized MN15-L (Fig. 1 and ESI Fig. S1). Although the family of double hybrids have the lowest correlations with pure GGAs in comparison to their correlation with hybrids (i.e., GGA or meta-GGA) for both Δ-SCF gap and $\Delta E_{\text{H-L}}$, even the low correlations are still high (Pearson's $r$ = 0.8–0.9, Fig. 1 and ESI Fig. S1).

As could be expected, the HF exchange influences $\Delta E_{\text{H-L}}$ correlations significantly: within the LYP correlation family, B3LYP and B2GP-PLYP correlate better than the latter does with BLYP (ESI Fig. S2). This observation for $\Delta E_{\text{H-L}}$ extends to the more highly parameterized Minnesota (e.g., M06) functionals, e.g., Pearson's $r$ coefficients are lowest between the pure meta-GGA M06-L and the high HF exchange M06-2X (ESI Table S3). Overall, functional agreement, as quantified by Pearson's $r$ values, is surprisingly strong across our data set with the 23 distinct functionals regardless of property. Even when deviations occur (e.g., for Δ-SCF gap or $\Delta E_{\text{H-L}}$), they appear to be due most to HF exchange fraction variation rather than due to any effect that can be attributed pure DFA parameter or correlation family choice. These good correlations can also likely be attributed to our workflow that ensures qualitative correspondence and limited spin contamination of the converged electronic state with change of DFA (see Sec. 4).

To explore the limits of this observation, we consider a worst-case scenario corresponding to the pair of DFAs with the greatest disagreement on the most DFA-sensitive $\Delta E_{\text{H-L}}$ property. Among all possible DFA pairings in our set, we observe the lowest Pearson's $r$ for the pure GGA BLYP and the highly-parameterized meta-GGA hybrid M06-2X (Pearson's *r*: 0.64, Fig. 1). This pair of DFAs is also among the most poorly correlated for Δ-SCF gap (Pearson's *r* ca. 0.8, ESI Fig. S1). As an example of this poor correlation, differences between the two DFAs of over 100 kcal/mol are observed for $\Delta E_{H-L}$ (BLYP: 98 kcal/mol, M06-2X: -6 kcal/mol) for the $Fe(II)(CS)_6$ TMC. While BLYP predicts the low-spin (LS) ground state expected for the strong-field CS ligand, M06-2X predicts a likely incorrect high-spin (HS) ground state. For another TMC with similarly strong field ligands, $Co(III)(C_2H_3N)_6$, BLYP and M06-2X both predict a LS ground state, albeit with large variations in the $\Delta E_{H-L}$ (BLYP: 88 kcal/mol, M06-2X: 35 kcal/mol) predicted.

For the DFAs that demonstrate surprisingly low linear correlations for $\Delta E_{H-L}$ and the Δ-SCF gap, we note there are also large variations in the DFA-computed property distribution shapes (Fig. 2 and ESI Figs. S3–S5). In contrast to the two DFA-sensitive properties, vertical IPs computed with different DFAs tend to differ by a small rigid shift in value with a preserved distribution (Fig. 2 and ESI Fig. S4). The large difference in the $\Delta E_{H-L}$ distribution means there is more disagreement among the ranking of compounds between functionals from the same family (e.g., M06-L and M06-2X in Fig. 2). M06-L and M06-2X predict the strong-field $Fe(II)(HNO)_6$ TMC, to have the same $\Delta E_{H-L}$ of 51 kcal/mol, but they differ strongly in the $\Delta E_{H-L}$ for the mixed-ligand-field the $Mn(II)(CO)_4(I^-)_2$ TMC. In this case, a different ground state is obtained with M06-L (LS; $\Delta E_{H-L}$: 32 kcal/mol) and M06-2X (HS; $\Delta E_{H-L}$: -61 kcal/mol), and the predicted spin splitting values differ by 93 kcal/mol. Returning to the vertical IP for the same pair of DFAs, consistent distributions and ranks are instead observed, with HF exchange in M06-2X rigidly shifting the IP up by around 5% (i.e., 1-2 eV over a 30 eV range, Fig. 2). The greatest disagreement, which is observed for the LS $Co(II)(H_2O)_4(NH_3)_2$ TMC, is only twice this amount (i.e., 4 eV) when comparing M06-L (17.1 eV) to M06-2X (21.4 eV, Fig. 2).

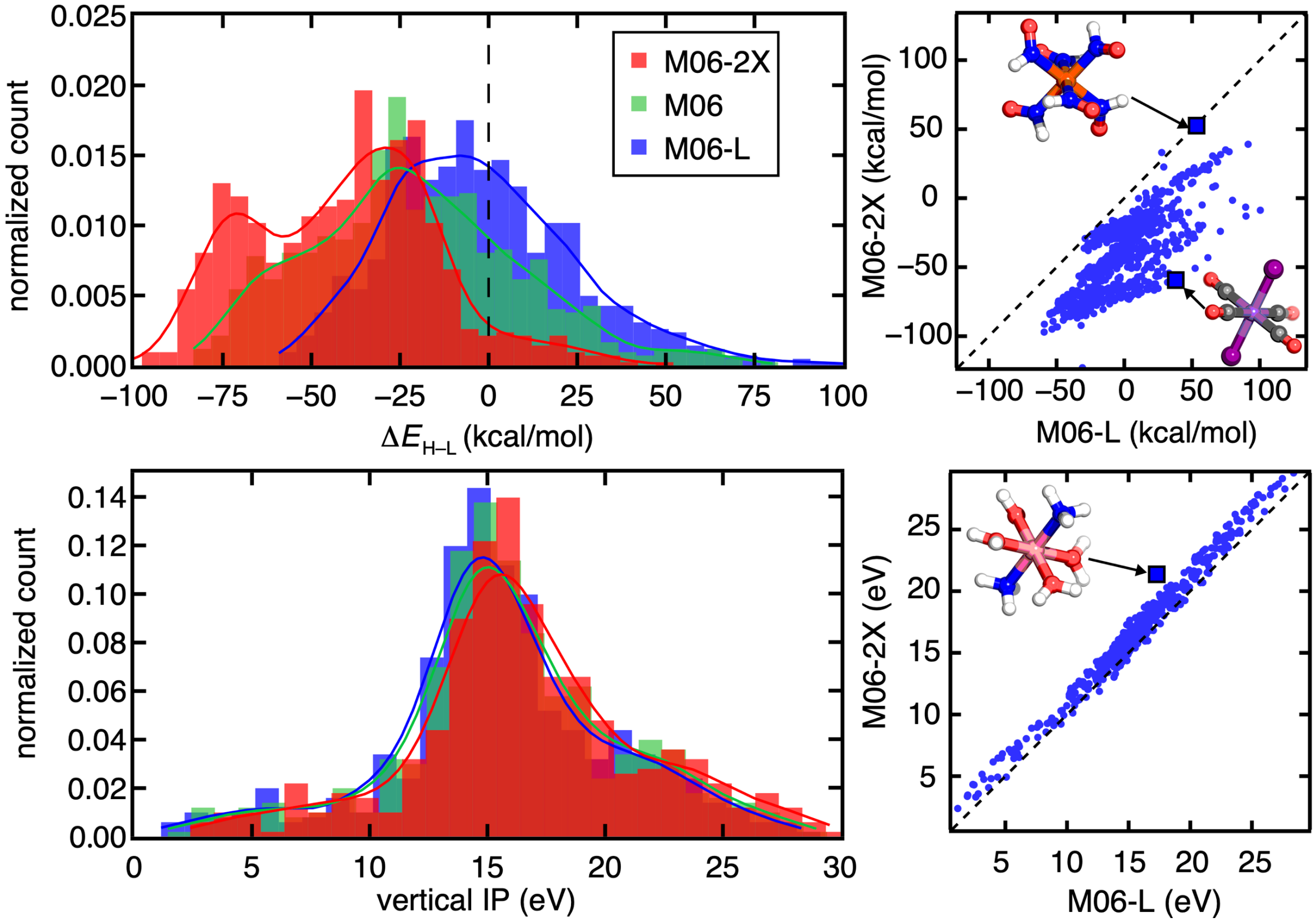

**Fig. 2** (left) The distribution of $\Delta E_{\text{H-L}}$ (top) and vertical IP (bottom) for three DFAs in the M06 family: the pure meta-GGA M06-L (blue) and the hybrid meta-GGAs M06 (27% HF, green) and M06-2X (54% HF, red). For $\Delta E_{\text{H-L}}$, a vertical dashed line is shown at 0 kcal/mol. (right) Parity plots of $\Delta E_{\text{H-L}}$ (top) and vertical IP (bottom) between M06-L and M06-2X with a black dashed parity line shown. Representative complexes are shown (top right pane): Fe(II)$(HNO)_6$ (top inset) and Mn(II)$(CO)_4(I^-)_2$ (middle inset) and (bottom right pane): LS Co(II)$(H_2O)_4(NH_3)_2$ (inset). Atoms are colored as follows: orange for Fe, purple for Mn, pink for Co, blue for N, red for O, gray for C, dark purple for I, and white for H.

As could be expected from the shapes of property distributions, for the gap, the percentile rank of $\Delta E_{\text{H-L}}$ and Δ-SCF properties for some compounds varies with DFA choice, but little variation is observed for the vertical IP percentile ranks (Fig. 3 and ESI Fig. S6). For $\Delta E_{\text{H-L}}$, the TMCs with the strongest (e.g., Co(III)$(CO)_6$) and weakest ligand fields (e.g., Mn(II)$(H_2O)_5(C_5H_5N)$) contain property values at the extremes and correspondingly have the lowest standard deviation of percentile ranks obtained by 23 DFAs (e.g., for $\Delta E_{\text{H-L}}$, ESI Table

S4). Complexes with moderate ligand field strengths instead have the largest standard deviation of percentile ranks, suggesting that the relative ordering of TMCs with moderate ligand field strengths for $\Delta E_{\text{H-L}}$ is the most strongly functional dependent. For example, Mn(II)$(CO)_4(H_2O)(C_5H_5N)$ is an intermediate-rank $\Delta E_{\text{H-L}}$ complex (average across the 23 DFAs: 49$^{\text{th}}$ percentile) but has a percentile rank ranging from the lowest quartile (e.g., 20 for M06-2X or 26 for DSD-BLYP-D3BJ) to the highest (e.g., 73 for BP86 and 65 for M06-L) depending on the DFA (ESI Table S4). Generally, the pure GGAs and pure meta-GGAs predict this compound to have the highest percentile rank while double hybrids predict it to have among the lowest. Thus, DFAs can broadly be expected to agree at extremes but have divergent behavior for these compounds that have intermediate $\Delta E_{\text{H-L}}$ properties arising from a mixture of strong-field and weak-field axial ligands (Fig. 3 and ESI Table S4).

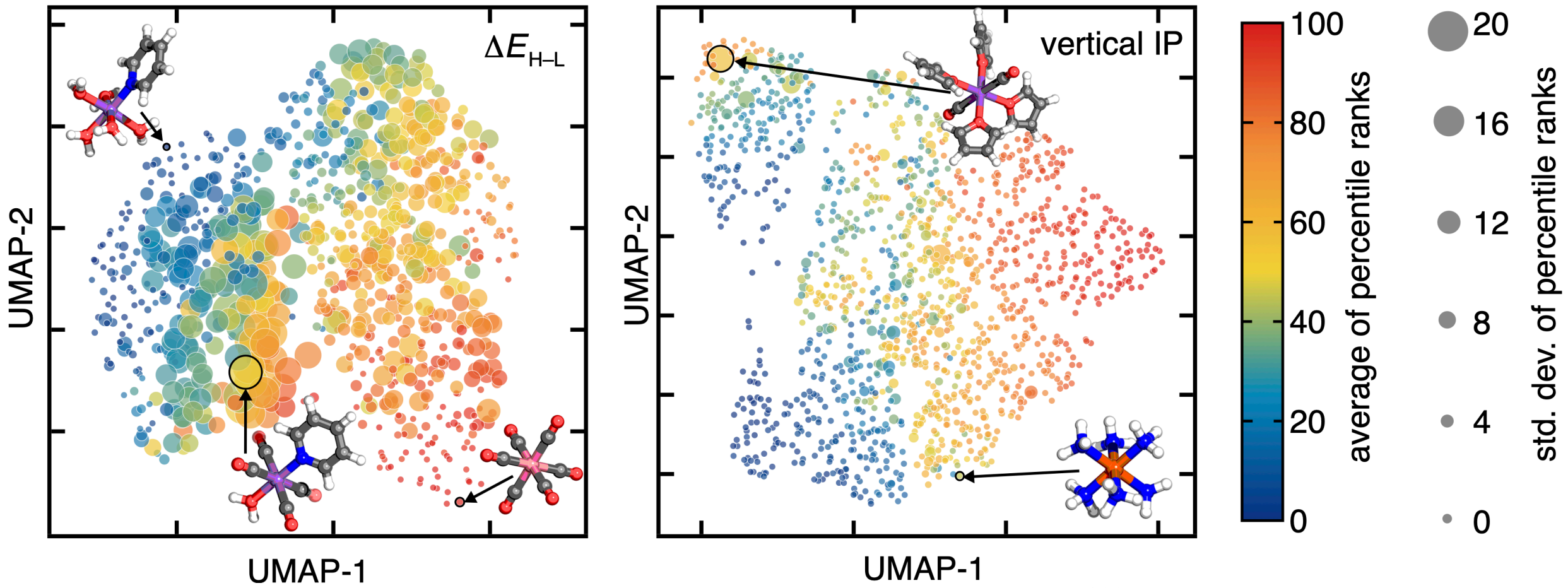

**Fig. 3** Uniform manifold approximation and projection (UMAP)[100] 2D visualization of $\Delta E_{\text{H-L}}$ (left) and vertical IP (right). Each TMC is shown as a circle that is colored by the average of the percentile ranks of the property ($\Delta E_{\text{H-L}}$ (left) and vertical IP (right)) obtained over all 23 DFAs and is scaled by the std. dev. of the percentile ranks over the DFAs. Representative TMCs from left to right in the left pane: Mn(II)$(H_2O)_5(C_5H_5N)$, Mn(II)$(CO)_4(H_2O)(C_5H_5N)$, Co(III)$(CO)_6$, and in the right pane: HS Mn(II)$(C_4H_4O)_4(CO)_2$ and LS Fe(II)$(NH_3)_6$. Atoms are colored as follows: orange for Fe, purple for Mn, pink for Co, blue for N, red for O, gray for C, and white for H.

In contrast to $\Delta E_{\text{H-L}}$, the percentile rank of vertical IP properties for TMCs remains nearly constant across all 23 DFAs, including those that have intermediate values (e.g., LS $Fe(II)(NH_3)_6$ in Fig. 3, ESI Table S4). This result is expected since variation in the functional was qualitatively observed to rigidly shift the vertical IP distribution (Fig. 2). Still, there are a few exceptions with large percentile rank standard deviation among the DFAs for vertical IP. In one extreme example, HS $Mn(II)(C_4H_4O)_4(CO)_2$ has a low percentile rank (i.e., < 40) for most pure GGAs (e.g., 36 for BLYP) and meta-GGAs but a higher percentile rank (i.e., > 60) for hybrids (e.g., 66 for SCAN0) and double hybrids (ESI Table S4). Nevertheless, these variations for vertical IP percentile rank are still far less than were observed for $\Delta E_{\text{H-L}}$.

Basis set dependence could also be expected[101-104] to influence our observations so we repeated our analysis with a larger triple-ζ (i.e., def2-TZVP) basis set than the double-ζ LACVP* basis set that is more amenable to high-throughput screening. Over each of the three properties and 23 DFAs, we observe that properties computed using the small and large basis sets display both high linear correlation (Pearson's $r > 0.98$) and absolute property prediction agreement (ESI Table S5 and Fig. S7). This observation holds independent of the DFA or property compared. Although one may expect vertical IP to be more basis set dependent, e.g., for complexes with strongly negatively charged ligands[34, 105], we observe little basis set dependence even for this property (ESI Fig. S8). Therefore, we conclude that DFA dependence outweighs basis dependence for property evaluation, and subsequent discussions focus on results obtained with the LACVP* basis set.

**2b. Universal design rules invariant of DFA choices**

Feature analysis of ML models provides valuable abstractions of learned design principles that can be used to guide for materials design.[36] For transition metal chemistry, a series of revised autocorrelations (RAC-155)[34] that are products and differences on the molecular graph of heuristic properties (e.g., electronegativity, $\chi$; nuclear charge, $Z$; topology, $T$; covalent radius, $S$; and identity, $I$) have been used to train predictive ML (e.g., kernel ridge regression, KRR, or artificial neural network, ANN) models. Feature selection to determine the relative importance of individual RACs in terms of their distance to the metal on the molecular graph as well as their electronic (i.e., Z or $\chi$) vs geometric (i.e., $S$, $T$, or $I$) nature has been used to reveal design principles for individual properties.[34] For example, $\Delta E_{\text{H-L}}$ has been shown[34] to be strongly metal-local and electronic in nature, whereas frontier orbital and vertical IP-related properties are known[106] to depend much more on the overall size and shape of the TMC. Nevertheless, such feature selection-derived design principles have been exclusively obtained with a single DFA. To identify sensitivity of design principles to the DFA used for generating ML model training data, we performed random forest-ranked recursive feature addition (RF-RFA) from RAC-155 with KRR models following our previously established procedure[59, 106] for all 23 DFAs (see Sec. 4).

Across this wide set of DFAs, the RF-RFA/KRR-selected features are insensitive to the functional choice for each of the three properties studied (Fig. 4 and ESI Fig. S9–S14). This observation holds both for the DFA-sensitive $\Delta E_{\text{H-L}}$ and DFA-insensitive vertical IP (Fig. 4). Even when small quantitative differences are observed with changing DFA, the magnitude of differences of the selected features between each of the three properties is significantly larger than the dependence of the selected features on the DFAs. Consistent with the linear correlation analysis, we also observe weak dependence of the selected features by basis set for a given property-DFA combination, reinforcing the value of small basis sets for computational high throughput screening (ESI Fig. S15). Even for pairs of DFAs where we observed the poorest linear correlations, the selected feature sets are still strikingly similar (Fig. 4).

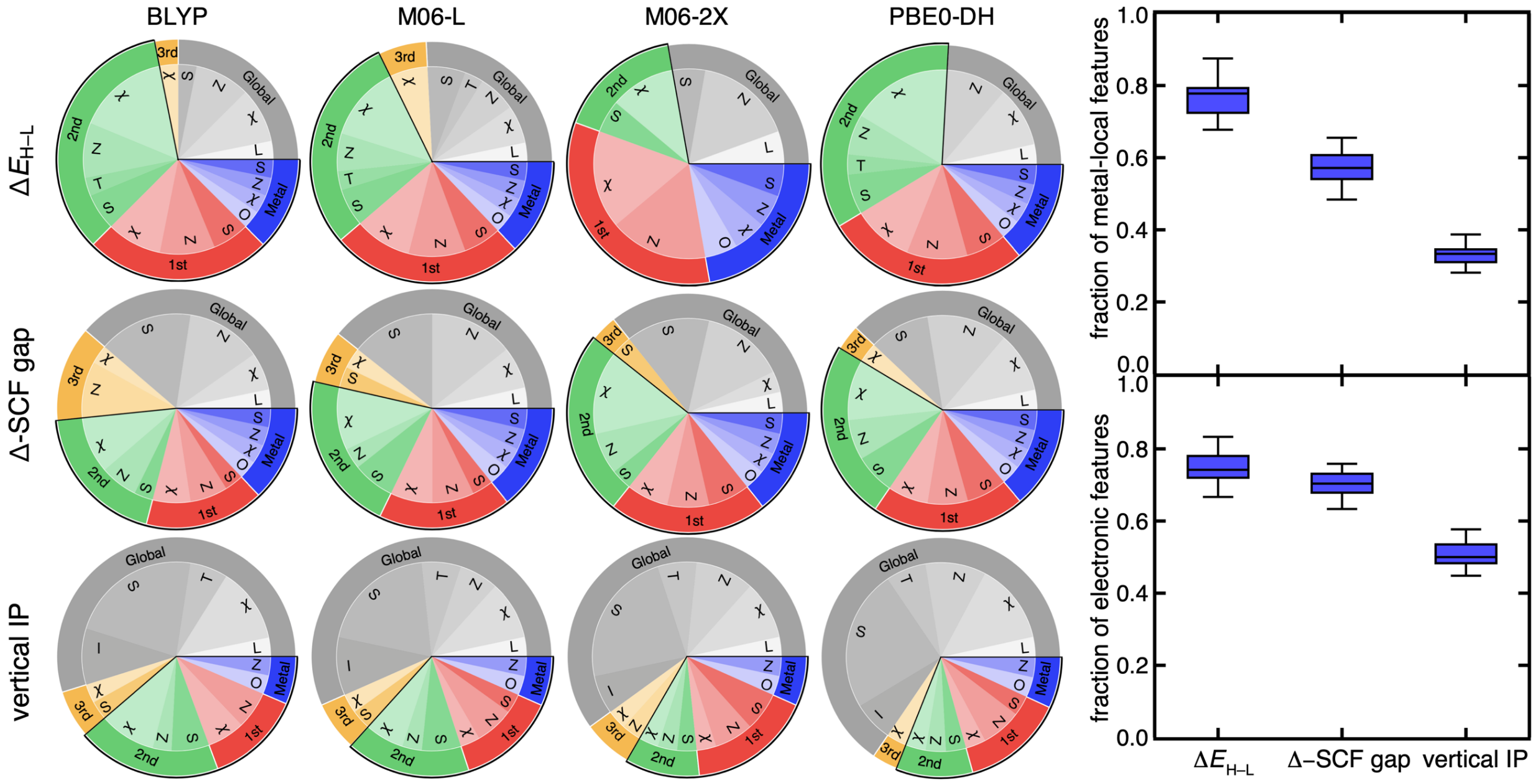

**Fig. 4** (left) Pie charts of the RF-RFA/KRR-selected RAC-155 features for $\Delta E_{\text{H-L}}$ (top), Δ-SCF gap (middle), and vertical IP (bottom) for the DFAs indicated (top). Features are grouped by the most metal-distal atoms: metal in blue, first coordination sphere in red, second coordination sphere in green, third coordination sphere in orange, and more distant, global features in gray. A black outline groups the first three categories (i.e., within two bond paths to the metal) as metal-local features. Within each connectivity distance category, the property (i.e., χ, S, T, Z, or I) is also indicated, with the oxidation/spin state (O) assigned as metal-local and the ligand charge (L) assigned as global. (right) Box plot for the fraction of metal-local features (top) and the fraction of electronic features (bottom) for all 23 DFAs at each property. Following our previous work[59, 106], we have categorized χ, Z, O, and L as electronic features, with all remaining features categorized as geometric.

For example, RF-RFA/KRR on $\Delta E_{\text{H-L}}$ from either of the poorly correlated pair of GGA BLYP and meta-GGA hybrid M06-2X DFAs produce selected feature sets with comparably high fractions of metal-local features (BLYP: 0.72, M06-2X: 0.72) and electronic (i.e., electronegativity, nuclear charge, oxidation state, see Sec. 4) features (BLYP: 0.78, M06-2X: 0.83, Fig. 4). The observation of the invariance of selected features for $\Delta E_{\text{H-L}}$ also holds for M06-L and M06-2X (Fig. 4). Thus, the feature-derived design rules are insensitive to the significant differences in the distributions of $\Delta E_{\text{H-L}}$ obtained with each of the DFAs, despite this difference leading to variations in percentile rank or low correlations among functionals (Fig. 2).

Although higher-rung double hybrids have been shown[95] in some cases to yield more accurate property prediction for spin state ordering, feature selection for all three properties on the PBE0-DH double hybrid also yields very similar selected features to DFAs from lower rungs (Fig. 4 and ESI Fig. S9–S14). Notably, semi-local DFAs that often yield unphysically small or closed HOMO-LUMO gaps give nearly the same selected features relative to range-separated functionals that contain asymptotically correct, non-local (i.e., $1/r$) exchange, even for the vertical IP and Δ-SCF gap (ESI Fig. S9–S14). Taken together, these observations provide powerful support for the design rules revealed through RF-RFA/KRR; such design features are robust to DFA choice and basis set beyond what could be achieved for absolute or even relative (i.e., rank ordering) property prediction across diverse properties.

A related, open question is the extent to which observations we have made are sensitive to the data set on which the models are trained. Consistent with prior work using B3LYP on modest data sets[34, 106], the $\Delta E_{\mathrm{H\text{-}L}}$, Δ-SCF gap, and vertical IP demonstrate decreasing dependence on metal-local and electronic features across representative examples from our broader 23 DFA set (Fig. 4). Importantly, the RF-RFA/KRR-selected features are quantitatively comparable, even when considering distinct data sets. The current set contains both smaller complexes with considerably more diverse metal-local chemistry (i.e., both P/S/Cl- and C/N/O-coordinating) and a greater number of ligand types and sizes relative to the set in previous work[34, 106] that had only five unique ligand types with a narrow range (i.e., C/N/O) of metal-coordinating atoms (Fig. 5 and ESI Fig. S16). Thus, not only is the RF-RFA/KRR feature map insensitive to method choice but the design rules are likely insensitive to data set choice as long as sufficient variation (e.g., metal identity and ligand field strength) is included in the set.[107]

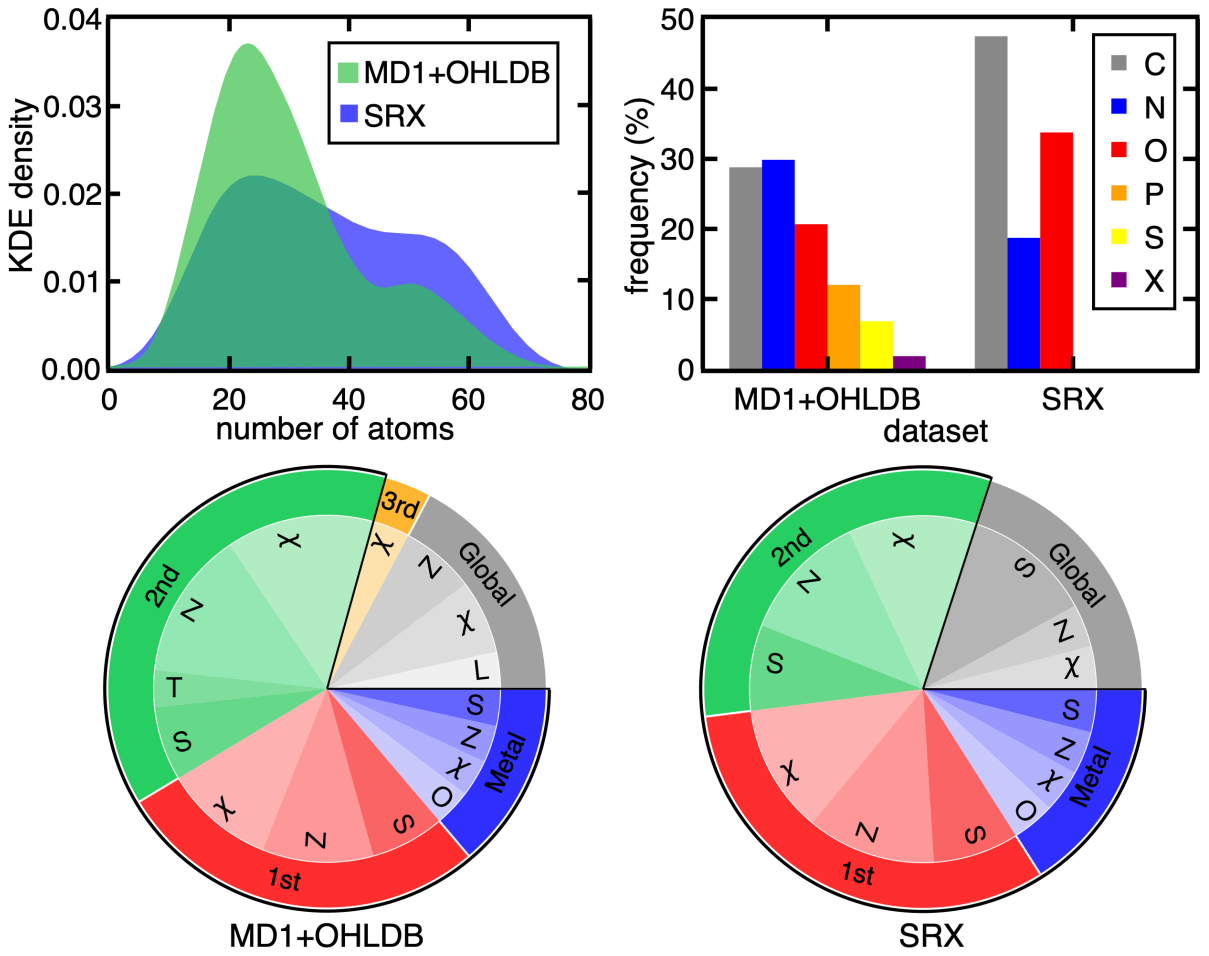

**Fig. 5** (top, left) Kernel density estimation (KDE) of the size distribution of complexes in two subsets of data from prior work (*MD1+OHLDB*) used in this study and the small redox set (*SRX*) of only five small ligand types from previous work.[34, 106] (top, right) Clustered bar graph for the connecting atom identity (X indicates any halide) in the two sets. (bottom) Pie charts of the features selected by RF-RFA KRR for $\Delta E_{\text{H-L}}$ with B3LYP/LACVP* for *MD1+OHLDB* (left) and *SRX* (right). The pie chart labels follow the format of those in Fig. 4.

Overall, the superior robustness of RF-RFA/KRR-selected features both to data set and to DFA suggests an efficient approach for materials design. To quickly reveal design rules for new properties in materials spaces that have twin combinatorial and method accuracy challenges such as open shell transition metal chemistry, low cost DFAs (e.g., GGAs) and small basis sets (i.e., double-ζ) on modest data sets of small, representative complexes can be used to efficiently reveal design principles even when they would be insufficient for individual property predictions.

To enable the exploration of a large chemical space for discovery, we also trained artificial neural network (ANN) models. ANNs have been shown to generalize better than kernel-based models[37] on data sets (e.g., hundreds of open-shell TMCs) such as those studied here. Given the large space of hyperparameters involved in training ANN models, independently training an ANN for each DFA could lead to differences due to the training procedure. Indeed, we observe that small differences in weight initialization and the stochastic nature of model optimization lead to distinct architectures even when the essential features from RF-RFA/KRR indicate the structure-property mapping should be similar (Fig. 6).

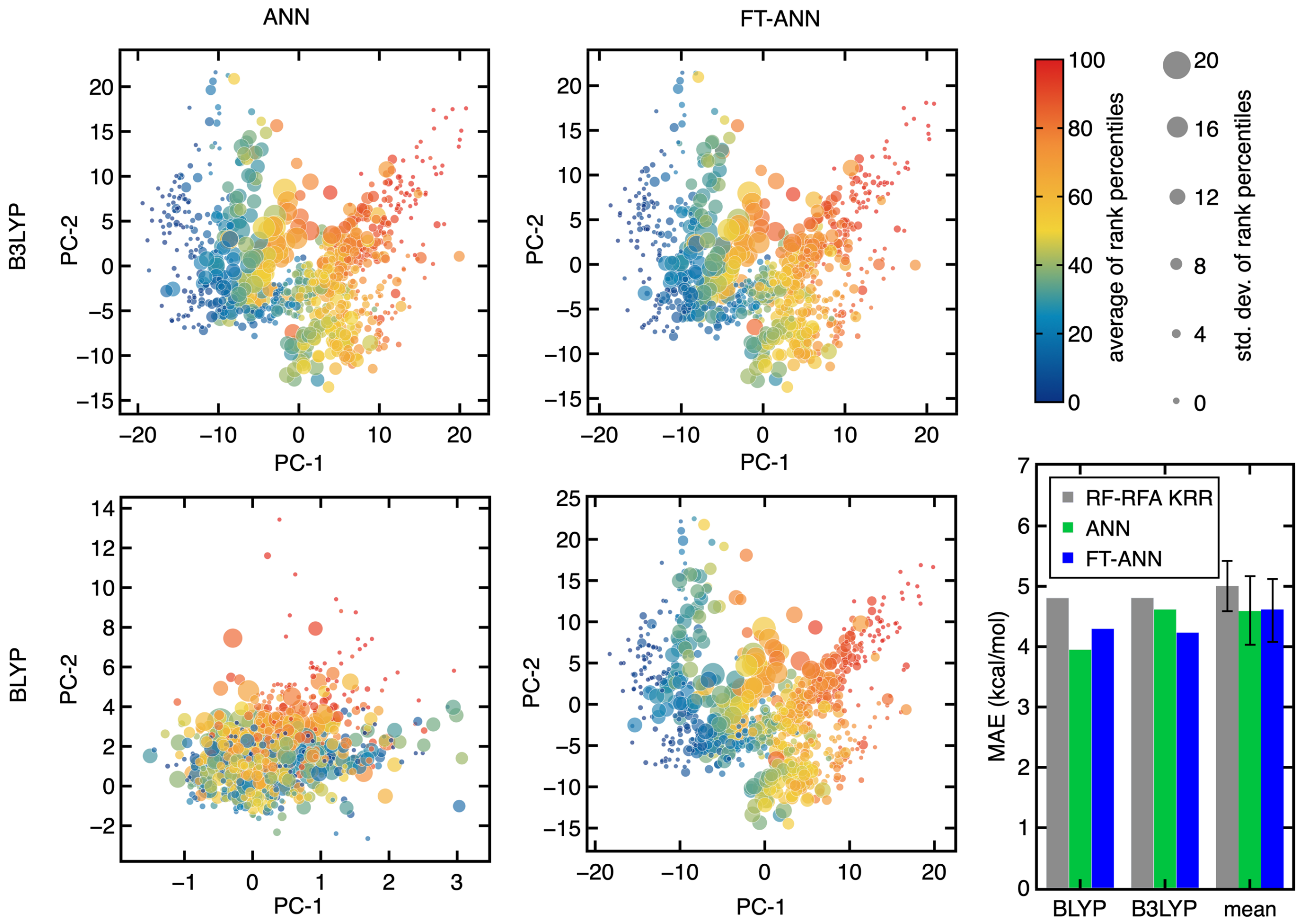

**Fig. 6** Principal component analysis (PCA) of the latent space for ANNs (left) and FT-ANNs (middle) for $\Delta E_{\text{H-L}}$ obtained with B3LYP (top) and BLYP (bottom). The PCA is fit to the latent space of the B3LYP FT-ANN and applied to the latent space of other three models. For each PCA plot, data are colored by their average $\Delta E_{\text{H-L}}$ percentile rank across all 23 DFAs and scaled by the std. dev. of the percentile ranks over all 23 DFAs, as indicated in the legend (upper right). The mean absolute error (MAE) of $\Delta E_{\text{H-L}}$ on the set-aside test data for the three different ML models (bottom right): RF-RFA KRR (gray), ANN (green), and FT-ANN (blue). The average MAE (labeled mean) of all model types for each of the 23 DFAs is also shown, with the std. dev. of the MAEs of 23 DFAs as the error bar.

To overcome the challenge of inequivalent ANN latent spaces, we use the weights of the ANN trained with B3LYP data as a starting point to train fine-tuned ANNs (FT-ANNs) on properties obtained with each of the other 22 DFAs (Fig. 6 and see Sec. 4 and ESI Table S6). The FT-ANNs trained through this procedure have comparable latent spaces for different DFAs without sacrificing prediction accuracy in comparison to alternative (i.e., RF-RFA/KRR or standard ANN) models (Fig. 6). These observations hold regardless of whether the DFAs had poor linear correlations with each other or with the parent B3LYP DFA used to obtain the initial ANN model weights and architecture (ESI Fig. S17–S19 and Tables S7–S9). Despite having similar latent spaces, each of the 23 ANNs predicts distinct properties approximating each DFA, enabling us to understand in the context of large-scale chemical discovery how ML models differing in DFA data sources will influence absolute or relative property prediction performance in ML-driven chemical space exploration.

**2c. Robust chemical discovery using the consensus among multiple DFAs**

Leveraging the B3LYP ANN along with the 22 FT-ANN models trained on the other DFAs, we next identified how ML-approximated knowledge of all DFA predictions will influence the design of lead compounds. We first define a target property and then identify consensus lead TMCs as the set of materials in which a majority (i.e., $\geqslant$12) of the ML models trained on distinct DFAs are in agreement about the target property value. Since we have selected a wide-ranging set of DFAs that include semi-local functionals, meta-GGAs, and range-separated hybrids along with varied HF exchange and MP2 correlation fractions, the consensus lead TMCs cannot be selected just due to the dominance of a single family of closely related functionals (ESI Tables S1 and S10). In our procedure, we also perform discovery using uncertainty quantification[108], restricting our chemical discovery task to regions of chemical space with high ML model confidence (ESI Fig. S20 and Table S11).

We apply our consensus-based approach in the exploration of a large space of 187,200 TMCs obtained from Ref. 59. This enumerated space consists of HS and LS M(II/III) midrow metals (M = Cr, Mn, Fe, or Co) with 36 unique ligands in heteroleptic and homoleptic mononuclear octahedral TMCs (ESI Tables S12–S14). The diverse chemistry, metal-coordinating atom types, symmetry of the complexes, and sizes of the ligands produces a large space of smaller TMCs along with those with up to 200 atoms (ESI Fig. S21 and Tables S12–S13). We first identify leads with small, targeted frontier orbital gaps (i.e., Δ-SCF gaps < 3 eV) and next search for those that have near degenerate spin states (i.e., $|\Delta E_{\text{H-L}}| < 5$ kcal/mol) as candidate spin crossover (SCO) complexes.

Exploring complexes with targeted Δ-SCF gaps < 3 eV in this design space is motivated by their relative rarity (ca. 0.1%) in the original set of data from the 23 DFAs (ESI Fig. S22). We find that lead TMCs for the targeted Δ-SCF gap are robust to the choice of DFAs, producing similar results independent of the DFA (Fig. 7 and ESI Table S15). This observation holds even for DFAs with weak linear correlations among our 23-DFA set, such as BLYP and M06-2X (Δ-SCF gap Pearson's $r = 0.80$), which still recommend similar (38% overlap) lead complexes (ESI Fig. S23). Although the original DFA data contain few complexes with targeted (i.e., < 3 eV) Δ-SCF gaps, the ML models generalize well on the 187,200 complex design space, yielding fruitful candidate lead complexes for this design objective. As would be observed with a single DFA, the consensus targeted Δ-SCF gap lead complexes favor large or bidentate N- or O-coordinating ligands with no significant metal preference, an observation that follows the expected trend of smaller Δ-SCF gap with increasing system size (Fig. 8 and ESI Fig. S24). Because of the DFA robustness of the lead targeted Δ-SCF gap complexes, this observation does not change if we only employed a single DFA of any kind or single family of DFAs (ESI Fig. S25).

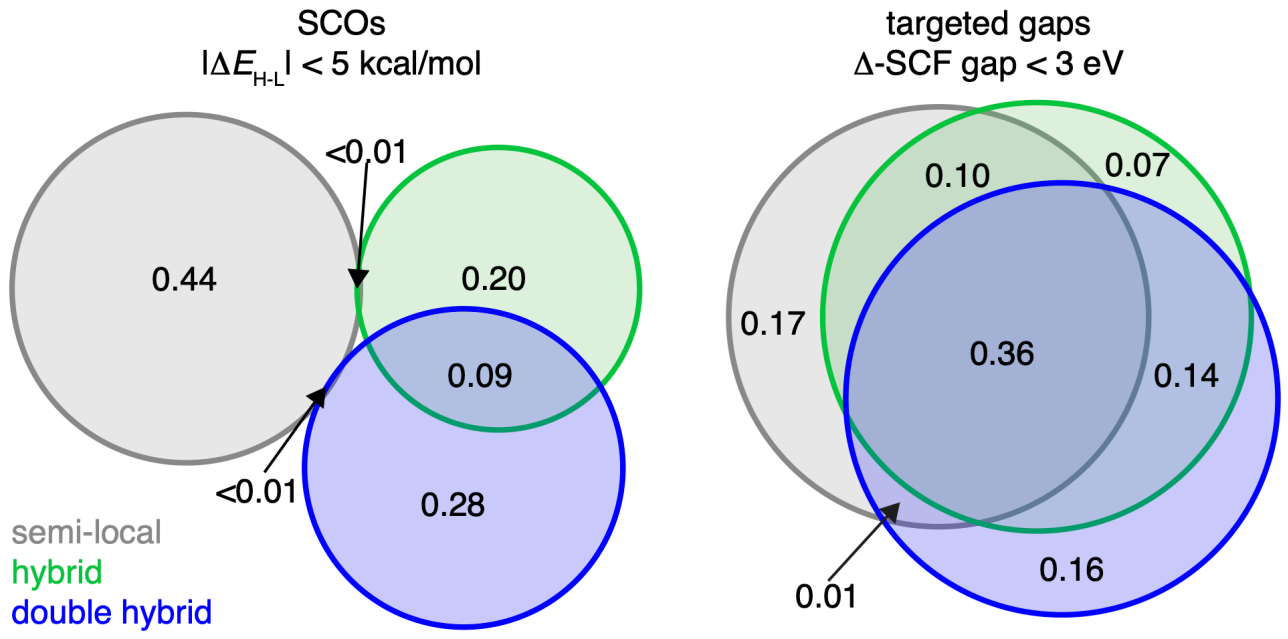

**Fig. 7** Venn diagrams of lead spin-crossover (SCO) complexes (left) and targeted gap complexes (right) favored by different groupings of DFAs: semi-local (gray, GGAs and meta-GGAs), hybrid (green, GGA hybrids, range-separated hybrids, and meta-GGA hybrids), and double-hybrid (blue). The number within each subset shows the fraction of the complexes in each relevant intersection with respect to the union of all subsets.

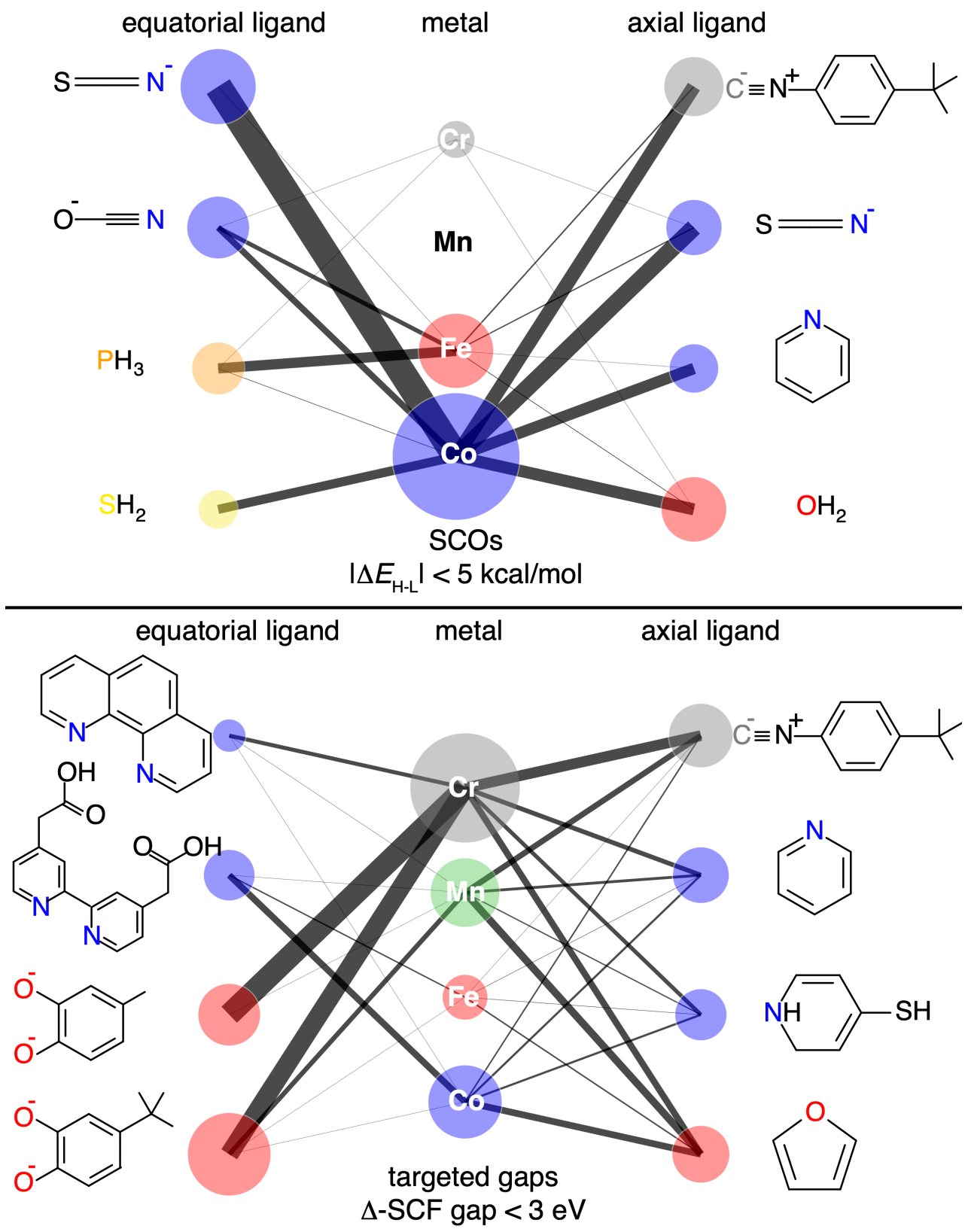

**Fig. 8** Network graph that illustrates the statistics of consensus ML lead SCO complexes (top) and targeted Δ-SCF gap complexes (bottom). The size of the sphere represents the relative abundance of the metal or equatorial/axial ligand appearing in the lead TMCs, and the width of a line connecting a metal and a ligand shows the relative abundance of this metal-ligand combination in the leads. Metals are colored as: gray for Cr, green for Mn, red for Fe, and blue for Co, and coordinating-atom types are colored as: gray for C, blue for N, red for O, orange for P, and yellow for S.

To validate our approach on a more challenging property, we next target discovery of SCOs with $|\Delta E_{\text{H-L}}| < 5$ kcal/mol because identified SCO complex chemistry is known to be strongly sensitive to DFA choice (e.g., HF exchange fraction[34, 36]). We find that SCO complexes are indeed very sensitive to HF exchange fraction, resulting in limited overlap (< 1%) between the leads discovered by pure semi-local DFAs and the HF-exchange-containing hybrid DFAs or double-hybrid DFAs (Fig. 7 and ESI Fig. S26). Although both GGA-hybrid and double-hybrid DFAs are expected to be more similar to each other than to a pure GGA, their suggested lead SCO complexes also differ significantly, with only 9% overlap (Fig. 7). Additionally, lead SCO complexes can differ even when we compare DFAs within the same rung of "Jacob's ladder" that were observed to have strong linear correlation with each other for $\Delta E_{\text{H-L}}$. For example, the two meta-GGA hybrids SCAN0 and TPSSh ($\Delta E_{\text{H-L}}$ Pearson's $r = 0.97$) recommend vastly different lead SCO complexes (i.e., only 3% in common), likely due to a rigid shift of $\Delta E_{\text{H-L}}$ values between the two DFAs (ESI Fig. S26). This divergent behavior of DFA sensitivity depending on the design objective suggests that the conventional workflow of only considering a single DFA for chemical discovery may work for some design targets but not others.

We considered specific examples of how the large DFA sensitivity of $\Delta E_{\text{H-L}}$ values results in DFA-dependent chemistry for the SCO candidates. As expected[36, 47-48, 54], we find that GGAs (e.g., BLYP) have a low spin bias and favor O-coordinating weak field ligands, whereas hybrid functionals (e.g. B3LYP) favor N-coordinating intermediate field ligands in SCO candidates (ESI Fig. S27). When taking a majority of all 23 DFAs into account, consensus lead SCO complexes are mostly Fe/Co complexes with weak/moderate field ligands, matching expectations from experimentally characterized SCOs[109] (Fig. 8). Specifically, the consensus lead SCO complexes rule out C-coordinating strong field ligands or extremely weak field ligands such as small anions (e.g., $S^{2-}$, $F^-$, and $I^-$). Despite the large variation in ligand chemistry in the original design space, we observe few Cr or Mn SCO complexes, indicating no consensus designs for Cr- and Mn-containing SCOs (ESI Table S12). Importantly, both sets of discrete lead complexes (i.e., SCO and targeted Δ-SCF gap) identified by the ANNs follow the design rules revealed by RF-RFA/KRR, i.e., that $\Delta E_{\text{H-L}}$ depends much more on metal-local features than the Δ-SCF gap (Fig. 5).

To demonstrate the distinct advantages of our consensus-based workflow for chemical discovery, we mined experimentally observed SCO complexes from the Cambridge Structural Database (CSD)[110] following slight modifications to the procedure used in prior work[94, 111] and compared them to our ML lead complexes (ESI Text S1). We observe significant overlap in the B3LYP ANN latent space between the experimentally identified SCO complexes and those discovered by our consensus ML approach (Fig. 9). For example, $Co(II)(NCS^-)_4(NCO^-)(C_2H_3N)$, a SCO complex predicted by our consensus ML approach, is close to an experimentally observed Co(II), N-coordinating SCO complex (CSD refcode: JUMPEO, Fig. 9). When small differences between the experimentally observed SCO complexes and our consensus ML leads are observed, they likely result from differences between our design space and the experimentally studied SCO compounds[112]. For example, a hexadentate Mn(II) complex with mixed N- and O-coordinating atoms is an experimentally-observed SCO complex (CSD refcode: YANLAC), but we do not have any hexadentate TMCs in our design space and therefore do not predict any similar Mn(II) compounds to be SCOs.

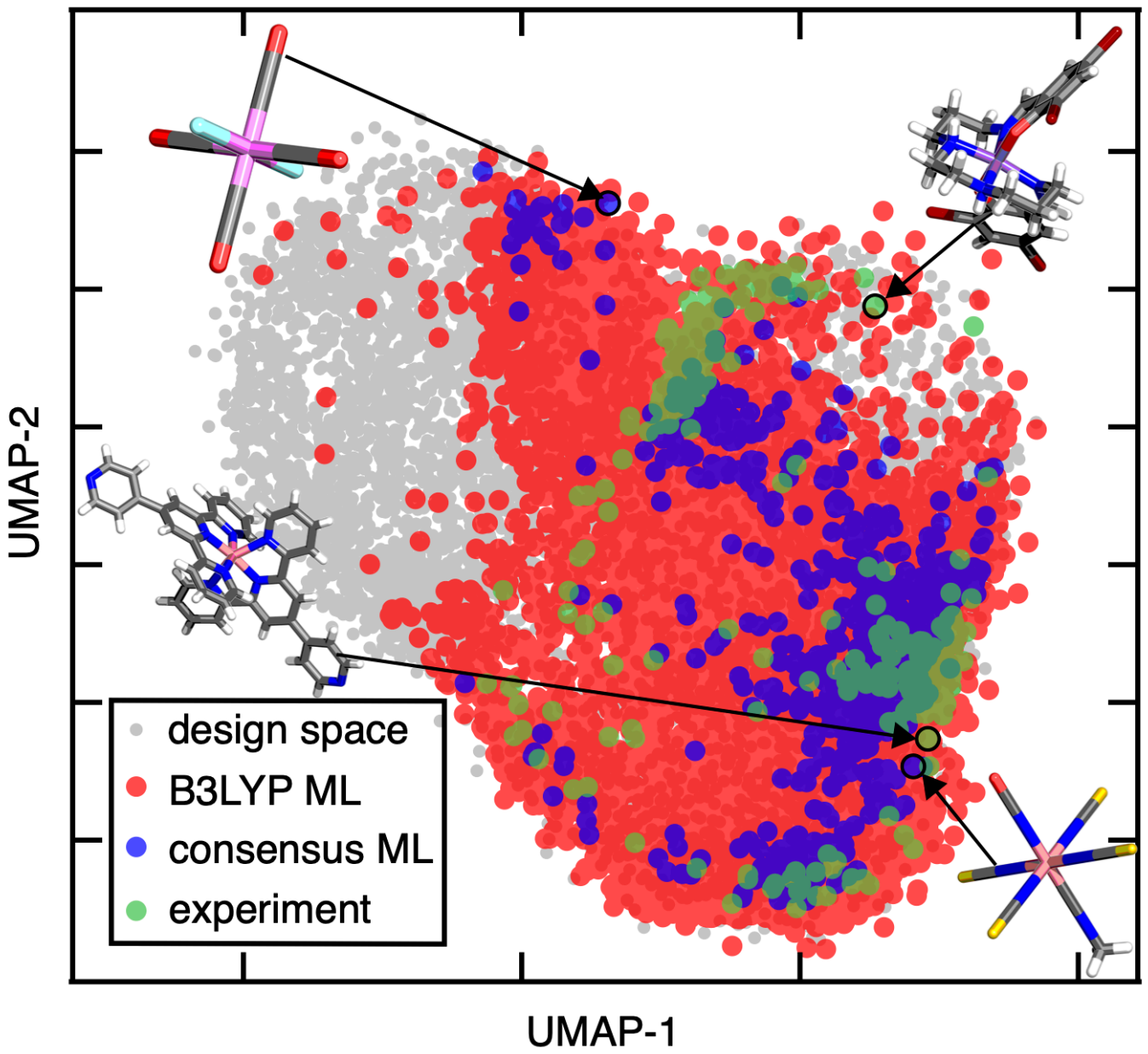

**Fig. 9** UMAP 2D visualization of $\Delta E_{H\text{-}L}$ for the design space of 187,200 TMCs (gray), lead SCO complexes predicted by a single ANN trained on B3LYP data (red), by the consensus approach of 23 (FT)-ANNs trained on all 23 DFAs (blue), and experimentally observed SCO complexes (green). The B3LYP-only leads cover 1/6 of the design space, although it appears to cover more due to the way density is represented on the plot. Representative complexes from left to right: JUMPEO (experiment), Cr(III)$(CO)_4(F^-)_2$ (design space), Co(II)$(NCS^-)_4$(NCO-)$(C_2H_3N)$ (design space), and YANLAC (experiment). Atoms are colored as follows: orange for Fe, purple for Mn, pink for Cr, blue for N, red for O, gray for C, cyan for F, dark red for Br, and white for H.

In comparison to the consensus-based approach, when we apply our conventional workflow of using a single DFA (e.g., B3LYP) to discover lead SCO complexes, we find that the candidate SCOs occupy a much larger region in the B3LYP ANN latent space than the experimental SCO complexes (Fig. 9). This suggests that many of the lead compounds obtained from an ANN trained on the single B3LYP DFA are likely false positives (Fig. 9). This comparison demonstrates the power of using DFA consensus to constrain ML-identified lead complexes to reasonable chemical spaces during chemical discovery. Notably, use of converged wavefunctions and structures from the parent DFA to derive properties with other DFAs as well as model weights not just improves consistency in the consensus-based approach but also limits the computational overhead as well in comparison to best available alternatives (e.g., correlated wavefunction theory or experiments).

Due to their widespread study, Fe metal centers with N-coordinating ligands dominate our set of experimentally-studied SCO complexes[109] (ESI Table S16). The robustness of our consensus approach motivates us to make additional recommendations beyond these well studied Fe/N SCOs. Since Co(II) and Co(III) complexes appear frequently in the consensus leads and lie near the experimental Fe SCO complexes in the latent space of B3LYP ANN model, our studies suggest their potential for further experimental or theoretical validation of such materials in SCO complex design (Fig. 8, ESI Fig. S28 and Table S14). The most likely candidate ligand chemistry suggested by the consensus screen would be to explore Co compounds with N-coordinating intermediate field ligands such as isothiocyanate, cyanate, and pyridine or higher denticity analogues.

## 3. Conclusions

While the limited accuracy of density functional approximations in challenging materials spaces is well established, the use of a single DFA in virtual high-throughput screening and machine learning has remained a necessity. To understand the potential biases a single DFA choice introduces in discovery campaigns, we computed three properties, $\Delta E_{\text{H-L}}$, Δ-SCF gap, and vertical IP, for over 2,000 open-shell TMCs with 23 DFAs. For DFAs distributed over multiple rungs of "Jacob's ladder" (e.g., semi-local to double-hybrids), absolute properties differed, but linear correlations were high. Over the three studied properties, distinct DFA-dependence was observed from low (i.e., vertical IP) to intermediate (i.e., frontier orbital Δ-SCF gap) or high (i.e., $\Delta E_{\text{H-L}}$) sensitivity. Sensitivity to DFA choice was observed to be greater than that with varying basis set from an affordable double-ζ (i.e., LACVP*) to a larger, triple-ζ (i.e., def2-TZVP) basis set.

RF-RFA/KRR feature selection revealed invariant design rules for all 23 DFAs at each property, even for those DFAs that have poor linear and rank correlations. In addition, the selected features were strikingly similar between data sets that contained different metal-coordinating chemistry and system sizes. The robustness of RF-RFA/KRR selected features suggest that universal design rules for new properties can be readily uncovered with low-cost DFAs and small basis sets on small, representative complexes. Such design rules can then guide more limited exploration for refinement of properties at higher levels of theory.

To enable large-scale chemical discovery informed by the 23 DFAs, we developed a fine-tuning procedure to obtain 22 comparably-performing ANNs trained on data from the other 22 DFAs. When using these models to explore a large space of 187,200 TMCs, we obtained design principles for lead SCO complexes and complexes with a targeted Δ-SCF gap. Lead targeted-gap complexes were robust to the choice of DFAs, producing similar leads regardless of DFA. Conversely, lead SCO complexes were very sensitive to both the choice of DFA family and HF exchange fraction. This observation suggests that the conventional use of a single DFA in VHTS and ML workflows is robust for only specific types of properties.

In order to overcome the limitations of a single DFA, we required consensus among more than half of the ML-model-predicted DFA properties. While the single-DFA approach and consensus approach were consistent with each other and recapitulated RF-RFA/KRR design principles, the consensus-based approach was critical to identifying lead SCO complexes. These compounds overlapped significantly in the ANN model latent space with experimentally observed SCO complexes in CSD. In contrast, lead SCO complexes identified by a single DFA (e.g., B3LYP) occupied a much larger region compared to those experimentally observed SCO complexes, indicating many B3LYP ANN leads would be false positives. Thus, DFA consensus with the approach described here to constrain ML-identified leads during chemical discovery is a promising method to improve prediction robustness without increasing computational cost.

## 4. Computational details

### 4a. Data sets and calculation details

We employ two subsets of data, as curated in prior work[59] from five prior studies[32-34, 106, 113] that originally corresponded to a total of 2,828 mononuclear octahedral transition-metal complexes in equilibrium geometries obtained with gas-phase density functional theory (DFT). In comparison to the prior curation[59] (i.e., where the sets were referred to as *MD1* and *OHLDB*), we refined the data further by de-duplicating structures with identical molecular graph, charge, and spin state across the two sets. This final filtering step followed the procedure for molecular graph identification described in Ref. 94 and resulted in a data set of 2,639 unique complexes. Details of all complexes are provided in the ESI. As in the original studies[32-34, 106, 113], the complexes contain M(III)/M(II) (M = Cr, Mn, Fe, or Co) centers with high spin (HS) and low spin (LS) multiplicities defined as: quintet-singlet for $d^4$ Mn(III)/Cr(II) and $d^6$ Co(III)/Fe(II), sextet-doublet for $d^5$ Fe(III)/Mn(II), and quartet-doublet $d^3$ Cr(III) and $d^7$ Co(II).

For all DFT geometry optimizations carried out in the original work, TeraChem[114-115], as automated by molSimplify[116-117] and molSimplify automatic design (mAD)[106], was employed. These calculations used the B3LYP[69-71] hybrid functional with the LACVP* basis set, which corresponds to the LANL2DZ[118] effective core potential for transition metals (i.e., Cr, Mn, Fe, Co) and heavier elements (i.e., I or Br) and the 6-31G* basis for all remaining elements. All non-singlet states were calculated with an unrestricted formalism and singlet states with a restricted formalism. In all these calculations, level shifting of 1.0 Ha on majority virtual spin orbitals and 0.1 Ha on minority virtual spin orbitals was employed.

We developed an approach to maximize correspondence between B3LYP and the 22

additionally studied DFAs that also reduced computational cost (ESI Table S1). We carried out single point energy evaluations on B3LYP/LACVP* structures both with a larger, triple-ζ def2-TZVP basis set for all 23 DFAs as well as the 22 additional DFAs with LACVP*. To automate these single-point energy calculations, we interfaced molSimplify with a developer version of Psi4[15] (1.4a2.dev723). This step was necessary in part because meta-GGAs, double-hybrids, and Minnesota functionals and the larger basis set were unavailable in TeraChem. The 23 functionals included in our study are described in Sec. 2, and they were used with all default definitions applied in Psi4, as indicated in ESI Table S1.

In our accelerated workflow, we used the previously converged[32-34, 106, 113] B3LYP/LACVP* TeraChem wavefunction molecular orbital coefficients as an initial guess for a Psi4 B3LYP/LACVP* SCF calculation. The converged Psi4 B3LYP/LACVP* wavefunction was used as an initial guess to obtain self-consistent single point energies in Psi4 for all other functionals with LACVP* basis set using the recommended default grid size[119-120] with 99 radial points and 590 spherical points (ESI Fig. S29 and Table S17). We also carried out single point energy calculations with the larger def2-TZVP[121] basis set for all combinations of functionals and complexes (ESI Fig. S29). For these larger basis calculations, we first carried out basis set projection to obtain the B3LYP/def2-TZVP converged wavefunction from the B3LYP/LACVP* result in Psi4. These calculations were run with a maximum of 50 SCF iterations to reach the SCF convergence with both the energy and density convergence thresholds being $3.0 \times 10^{-5}$ Ha (ESI Fig. S30 and Table S17). Select pure GGA and meta-GGA calculations did not initially converge. In these cases, we performed calculations with a hybrid form of the pure GGA or meta-GGA, sequentially reducing the percentage of HF exchange and extrapolating to the 0% HF (i.e., pure GGA) total energy (ESI Fig. S31–S32 and Text S2).

The three primary properties computed (i.e., $\Delta E_{\text{H-L}}$, IP, and Δ-SCF gap) were all evaluated on B3LYP/LACVP* geometries obtained with TeraChem. For vertical IP and Δ-SCF gap evaluated on the *N*-electron reference system, we adopted a consistent spin state convention: we removed a majority spin electron *N*-1-electron calculation (i.e., for IP) and added a minority spin electron for the *N*+1-electron case (i.e., for EA), in each case starting the Psi4 calculation with an initial guess for the (*N*-1/*N*+1)-electron calculation from the converged B3LYP/LACVP* TeraChem result (ESI Fig. S29 and Fig. S33).

The filtering procedure applied in previous work[59] required that all B3LYP/LACVP* wavefunctions had limited spin contamination (i.e., $\langle S^2 \rangle$ deviations from the expected $S(S+1)$ value < 1.0 $\mu_{\mathrm{B}}^2$), the number of calculations excluded by this filtering threshold is sensitive to the choice of functional (ESI Fig. S34). In this work, we increased the cutoff for inclusion to 1.1 $\mu_{\mathrm{B}}^2$ (ESI Fig. S34). Finally, complexes were retained in the data set only if properties could be converged below this cutoff from all 23 functionals (ESI Tables S18–S19).

**4b. ML models**

As in prior work[59], we use revised auto-correlations[34] (RACs) as descriptors for all our machine learning models. RACs are sums of products and differences of five atom-wise heuristic properties (i.e., topology, identity, electronegativity, covalent radius, and nuclear charge) on the 2D molecular graph. As motivated previously[34], we applied the maximum bond depth of three and eliminated RACs that were invariant over the mononuclear octahedral transition metal complexes, leaving 151 RACs in total (ESI Text S3). Along with three overall complex features[32, 34] (i.e., oxidation state, spin multiplicity, and total ligand charge), we obtained a feature set of 154 descriptors in total. For both kernel ridge regression (KRR) and artificial neural network (ANN) models, the hyperparameters were selected using Hyperopt[122] with 200 evaluations on a range of hyperparameters, using a random 80%/20% train/test split and 20% of the training data (i.e., 16% overall) used as the validation set (ESI Tables S20–S21). As in prior work[59, 106], recursive feature addition (RFA) was carried out on random-forest-ranked features (i.e., RF-RFA) to obtain the selected feature set that gives the best-performing KRR model with the lowest mean absolute error. All KRR models were implemented in scikit-learn[123] with a radial-basis function kernel. Details of all models and selected features are provided in the ESI.

All ANN models were trained using Keras[124] with Tensorflow[125] as the backend and Hyperopt[122] for hyperparameter selection (ESI Table S20). To avoid randomness in the weight initialization and to increase the consistency between ANN models trained with DFT data derived from different functionals, we also obtained a fine-tuned B3LYP ANN model with a reduced (i.e., $1x10^{-5}$) learning rate for each of the other 22 functionals, which produced 22 fine-tuned ANN (FT-ANN) models at each property and basis set combination (ESI Table S6). All ANN models were trained with the Adam optimizer up to 2,000 epochs, and dropout, batch normalization, and early stopping were applied to avoid over-fitting (ESI Table S20).

ASSOCIATED CONTENT

**Electronic Supplementary Information**.

Summary of 23 DFAs; Statistics of properties obtained by 23 DFAs; Pearson's *r* matrix and parity plots for $\Delta E_{\text{H-L}}$, vertical IP, and Δ-SCF gap; Distributions of $\Delta E_{\text{H-L}}$, vertical IP, and Δ-SCF gap at different DFAs; UMAP 2D visualization of Δ-SCF gap data; Percentile ranks at each DFA for example complexes; Statistics and parity plots for basis set comparison; Pie charts of RF-RFA KRR selected features at different DFAs; RF-RFA KRR selected features for vertical IP for difference datasets; Hyperparameters for FT-ANN models; MAEs, $R^2$, and scaled MAEs of all 23 functionals for three properties; Uncertainty quantification metric and its cutoff; Possible metal, oxidation, spin state, and ligand combinations in the 187,000 complex space; Size distribution of the 187,200 complexes design space; Histograms of Δ-SCF gap grouped by system size; Venn diagrams of lead Δ-SCF gap complexes and lead SCO complexes; Network graph of lead Δ-SCF gap and lead SCO complexes; Procedure of isolating of candidate SCO complexes; Statistics of experimental SCO complexes; UMAP visualization of selected lead SCO complexes; Computational workflow and DFT parameters used therein; Hartree-Fock linear extrapolation scheme; Statistics of SCF iterations for convergence and failed calculations before and after HF extrapolation; Electron configuration diagram of electron addition/removal convention; Extended description of RAC featurization; Range of hyperparameters sampled during Hyperopt for KRR and ANN models.

**Data Availability**

The datasets supporting this article have been uploaded as part of the electronic supplementary information.

AUTHOR INFORMATION

**Corresponding Author**

*email: hjkulik@mit.edu phone: 617-253-4584

**Notes**

The authors declare no competing financial interest.

ACKNOWLEDGMENT

The authors acknowledge primary support by the Office of Naval Research under Grant

Numbers N00014-17-1-2956, N00014-18-1-2434, and N00014-20-1-2150 (to C.D., M.G.T, and H.J.K). The machine learning effort as also supported by DARPA (grant number DE18AP00039). F.L. was partially supported by the Department of Energy under Grant Number DE-SC0018096 and a MolSSI fellowship (Grant No. ACI-1547580). M.G.T. was supported by the Department of Energy under Grant Number DE-SC0012702. S. C. was supported by an MIT Energy Initiative Fund for Undergraduate Research. This work made use of Department of Defense HPCMP computing resources. This work was also carried out in part using computational resources from the Extreme Science and Engineering Discovery Environment (XSEDE), which is supported by National Science Foundation Grant Number ACI-1548562. H.J.K. holds a Career Award at the Scientific Interface from the Burroughs Wellcome Fund, an AAAS Marion Milligan Mason Award, and an Alfred P. Sloan Fellowship in Chemistry, which supported this work. The authors thank Adam H. Steeves, Aditya Nandy, and Vyshnavi Vennelakanti for providing a critical reading of the manuscript.

**Supporting Information for**
***Machine learning to tame divergent density functional approximations: a new path to consensus materials design principles***

Chenru Duan[1,2], Shuxin Chen[1], Michael G. Taylor[1], Fang Liu[1], and Heather J. Kulik[1]
[1]Department of Chemical Engineering, Massachusetts Institute of Technology, Cambridge, MA 02139
[2]Department of Chemistry, Massachusetts Institute of Technology, Cambridge, MA 02139

**Contents**

**Table S1**. Summary of 23 functionals studied in this work, including their rungs on "Jacob's ladder" of DFT, Hartree-Fock (HF) exchange fraction, long-range correction (LRC) range-separation parameter ($bohr^{-1}$), MP2 correlation fraction, and whether empirical (i.e., D3) dispersion correction is included.

| Functional | Type | Exchange type | HF exchange fraction | LRC fraction | MP2 correlation | D3 dispersion |
|---|---|---|---|---|---|---|
| BP86 | GGA | GGA | -- | -- | -- | No |
| BLYP | GGA | GGA | -- | -- | -- | No |
| PBE | GGA | GGA | -- | -- | -- | No |
| TPSS | meta-GGA | meta-GGA | -- | -- | -- | No |
| SCAN | meta-GGA | meta-GGA | -- | -- | -- | No |
| M06-L | meta-GGA | meta-GGA | -- | -- | -- | No |
| MN15-L | meta-GGA | meta-GGA | -- | -- | -- | No |
| B3LYP | GGA hybrid | GGA | 0.200 | -- | -- | No |
| B3P86 | GGA hybrid | GGA | 0.200 | -- | -- | No |
| B3PW91 | GGA hybrid | GGA | 0.200 | -- | -- | No |
| PBE0 | GGA hybrid | GGA | 0.250 | -- | -- | No |
| ωB97X | RS hybrid | GGA | 0.158 | 0.300 | -- | No |
| LRC-ωPBEh | RS hybrid | GGA | 0.200 | 0.200 | -- | No |
| TPSSh | meta-GGA hybrid | meta-GGA | 0.100 | -- | -- | No |
| SCAN0 | meta-GGA hybrid | meta-GGA | 0.250 | -- | -- | No |
| M06 | meta-GGA hybrid | meta-GGA | 0.270 | -- | -- | No |
| M06-2X | meta-GGA hybrid | meta-GGA | 0.540 | -- | -- | No |
| MN15 | meta-GGA hybrid | meta-GGA | 0.440 | -- | -- | No |
| B2GP-PLYP | double hybrid | GGA | 0.650 | -- | 0.360 | No |
| PBE0-DH | double hybrid | GGA | 0.500 | -- | 0.125 | No |
| DSD-BLYP-D3BJ | double hybrid | GGA | 0.710 | -- | 1.000 | Yes |
| DSD-PBEB95-D3BJ | double hybrid | GGA | 0.660 | -- | 1.000 | Yes |
| DSD-PBEP6-D3BJ | double hybrid | GGA | 0.690 | -- | 1.000 | Yes |

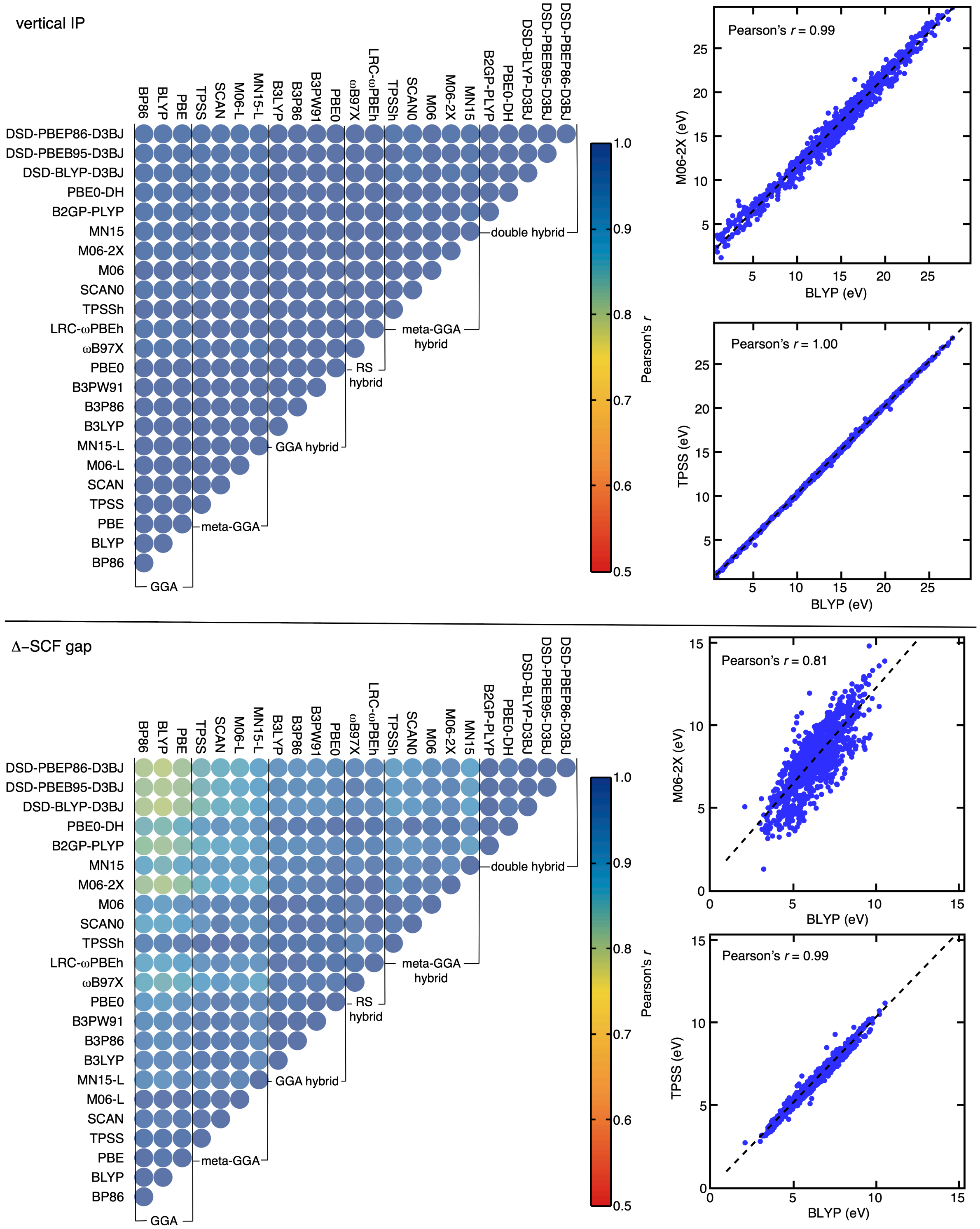

**Figure S1**. (top) An upper triangular matrix colored by Pearson's $r$ for pairs of 23 functionals for vertical IP (shown at left) along with a parity plot of vertical IP for M06-2X and BLYP (top right) and for TPSS and BLYP (bottom right). (bottom) An upper triangular matrix colored by Pearson's r for pairs of 23 functionals for Δ-SCF gap (shown at left) along with a parity plot of Δ-SCF gap for M06-2X and BLYP (top right), and for TPSS and BLYP (bottom right).

**Table S2.** Summary of the mean value, standard deviation (std. dev.), minimum (min.), and maximum (max.) with each DFA for three properties considered in this work. Units of each measurement are indicated at top.

| | $\Delta E_{H-L}$ (kcal/mol) | | | | Δ-SCF gap (eV) | | | | vertical IP (eV) | | | |
|---|---|---|---|---|---|---|---|---|---|---|---|---|
| | mean | std. dev. | min. | max. | mean | std. dev. | min. | max. | mean | std. dev. | min. | max. |
| BP86 | 11.4 | 30.8 | -52.0 | 132.8 | 6.3 | 1.3 | 2.1 | 10.8 | 14.7 | 4.8 | 0.9 | 28.0 |
| BLYP | 7.3 | 26.5 | -49.4 | 113.2 | 6.3 | 1.3 | 2.1 | 10.5 | 14.5 | 4.8 | 0.7 | 27.6 |
| PBE | 8.8 | 30.9 | -56.5 | 133.3 | 6.3 | 1.3 | 2.2 | 10.8 | 14.6 | 4.8 | 0.7 | 27.8 |
| TPSS | 4.4 | 28.0 | -52.8 | 116.0 | 6.5 | 1.4 | 2.7 | 11.2 | 14.8 | 4.8 | 0.7 | 28.0 |
| SCAN | 1.5 | 30.6 | -66.0 | 110.4 | 6.8 | 1.5 | 2.0 | 11.8 | 15.0 | 4.8 | 0.7 | 28.6 |
| M06-L | -2.2 | 26.4 | -59.2 | 100.4 | 6.7 | 1.4 | 2.5 | 11.8 | 15.0 | 4.8 | 0.9 | 28.4 |
| MN15-L | -27.5 | 33.8 | -98.7 | 82.5 | 7.0 | 1.5 | 2.2 | 12.2 | 15.1 | 4.8 | 1.1 | 28.6 |
| B3LYP | -8.6 | 24.5 | -60.5 | 80.2 | 7.1 | 1.5 | 2.5 | 13.8 | 15.3 | 4.9 | 1.2 | 28.7 |
| B3P86 | -6.0 | 26.8 | -61.5 | 90.4 | 7.1 | 1.5 | 2.2 | 14.0 | 15.6 | 4.8 | 1.2 | 29.0 |
| B3PW91 | -10.0 | 27.1 | -66.5 | 87.1 | 7.2 | 1.5 | 2.1 | 14.0 | 15.4 | 4.8 | 1.0 | 28.8 |
| PBE0 | -14.6 | 27.0 | -72.3 | 82.2 | 7.3 | 1.6 | 1.8 | 14.4 | 15.5 | 4.9 | 1.0 | 29.0 |
| ωB97X | -12.3 | 24.4 | -60.6 | 79.3 | 7.8 | 1.7 | 0.9 | 14.6 | 15.9 | 4.9 | 1.5 | 29.2 |
| LRC-ωPBEh | -13.5 | 27.8 | -69.1 | 79.3 | 7.6 | 1.7 | 1.8 | 14.5 | 15.7 | 4.9 | 1.1 | 29.1 |
| TPSSh | -5.0 | 26.5 | -60.3 | 94.1 | 6.5 | 1.9 | 2.6 | 11.9 | 15.1 | 4.8 | 0.7 | 28.4 |
| SCAN0 | -19.0 | 27.3 | -82.0 | 74.7 | 7.6 | 1.8 | 0.9 | 15.3 | 15.7 | 4.9 | 0.9 | 29.6 |
| M06 | -21.2 | 29.8 | -83.3 | 80.4 | 7.2 | 1.5 | 1.9 | 13.8 | 15.5 | 4.8 | 1.3 | 29.0 |
| M06-2X | -40.9 | 26.3 | -122.8 | 50.7 | 8.0 | 1.9 | 1.3 | 14.8 | 16.2 | 4.9 | 1.2 | 29.6 |
| MN15 | -7.9 | 29.6 | -75.6 | 90.6 | 7.0 | 1.6 | 2.3 | 13.0 | 15.6 | 4.9 | 1.3 | 29.1 |
| B2GP-PLYP | -20.5 | 28.3 | -71.9 | 85.1 | 7.8 | 1.8 | 1.9 | 14.5 | 15.6 | 4.9 | 1.0 | 28.5 |
| PBE0-DH | -22.5 | 28.0 | -80.3 | 80.6 | 7.8 | 1.8 | 1.6 | 15.0 | 15.8 | 4.9 | 1.2 | 29.2 |
| DSD-BLYP-D3BJ | -20.4 | 28.7 | -69.9 | 89.1 | 7.9 | 1.9 | 1.9 | 14.5 | 15.6 | 5.0 | 0.9 | 28.5 |
| DSD-PBEB95-D3BJ | -19.7 | 28.1 | -69.2 | 87.1 | 7.9 | 1.8 | 2.0 | 14.4 | 15.6 | 4.9 | 1.1 | 28.6 |
| DSD-PBEP6-D3BJ | -20.5 | 29.0 | -70.7 | 90.6 | 7.9 | 1.8 | 1.9 | 14.5 | 15.6 | 5.0 | 0.8 | 28.5 |

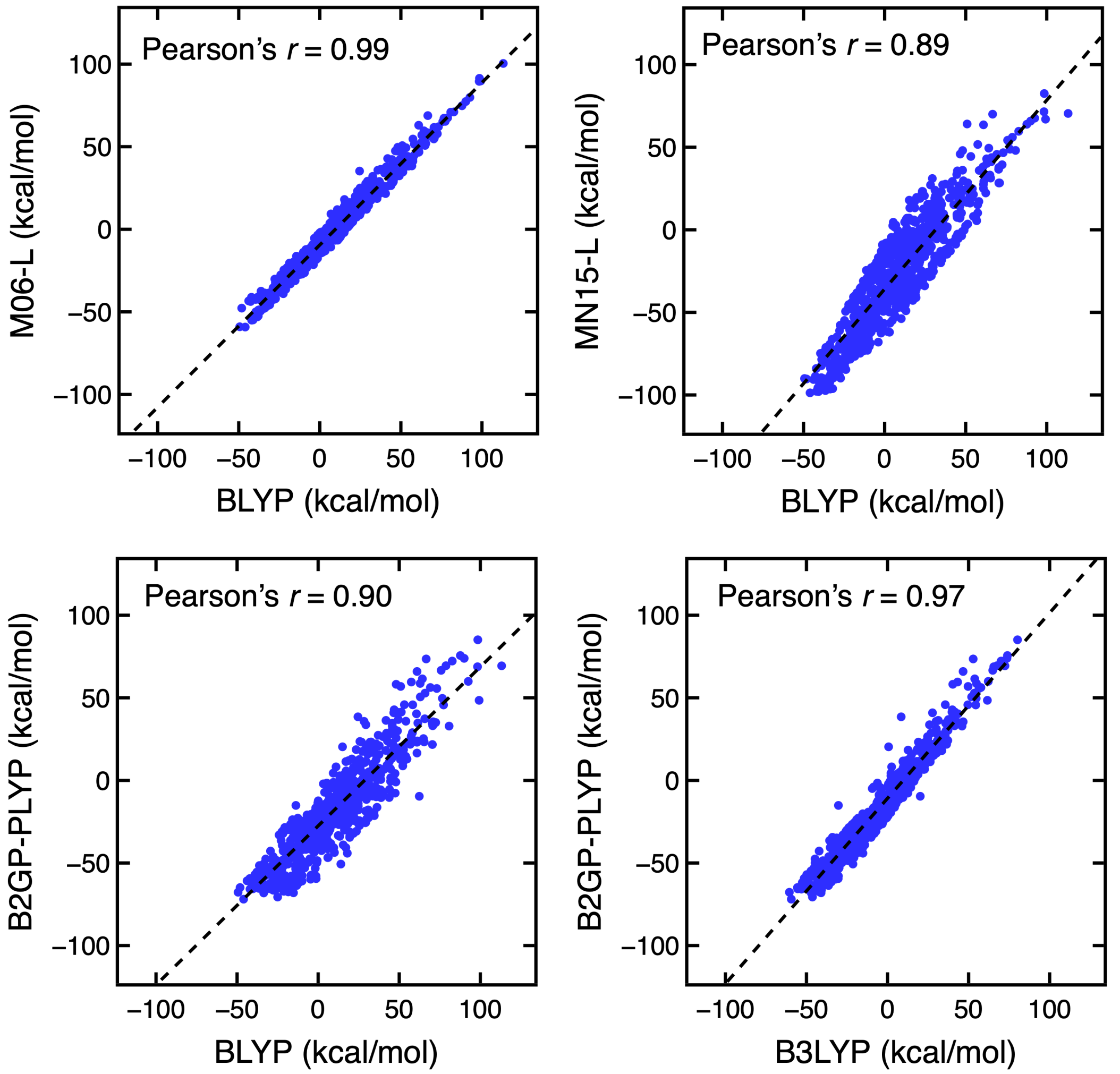

**Figure S2.** Parity plots of $\Delta E_{\text{H-L}}$ for select functionals: BLYP vs. M06-L (top left), BLYP vs. MN15-L (top right), BLYP vs B2GP-PLYP (bottom left), and B3LYP vs B2GP-PLYP (bottom right). A linear regression fit is shown as black dashed line in each parity plot, and the Pearson's *r* is shown in inset.

**Table S3.** Pearson's *r* of $\Delta E_{\text{H-L}}$ among DFAs from the M06 family using the LACVP* basis.

| | M06-L | M06 | M06-2X |
|---|---|---|---|
| M06-L | 1.00 | 0.95 | 0.72 |
| M06 | 0.95 | 1.00 | 0.86 |
| M06-2X | 0.72 | 0.86 | 1.00 |

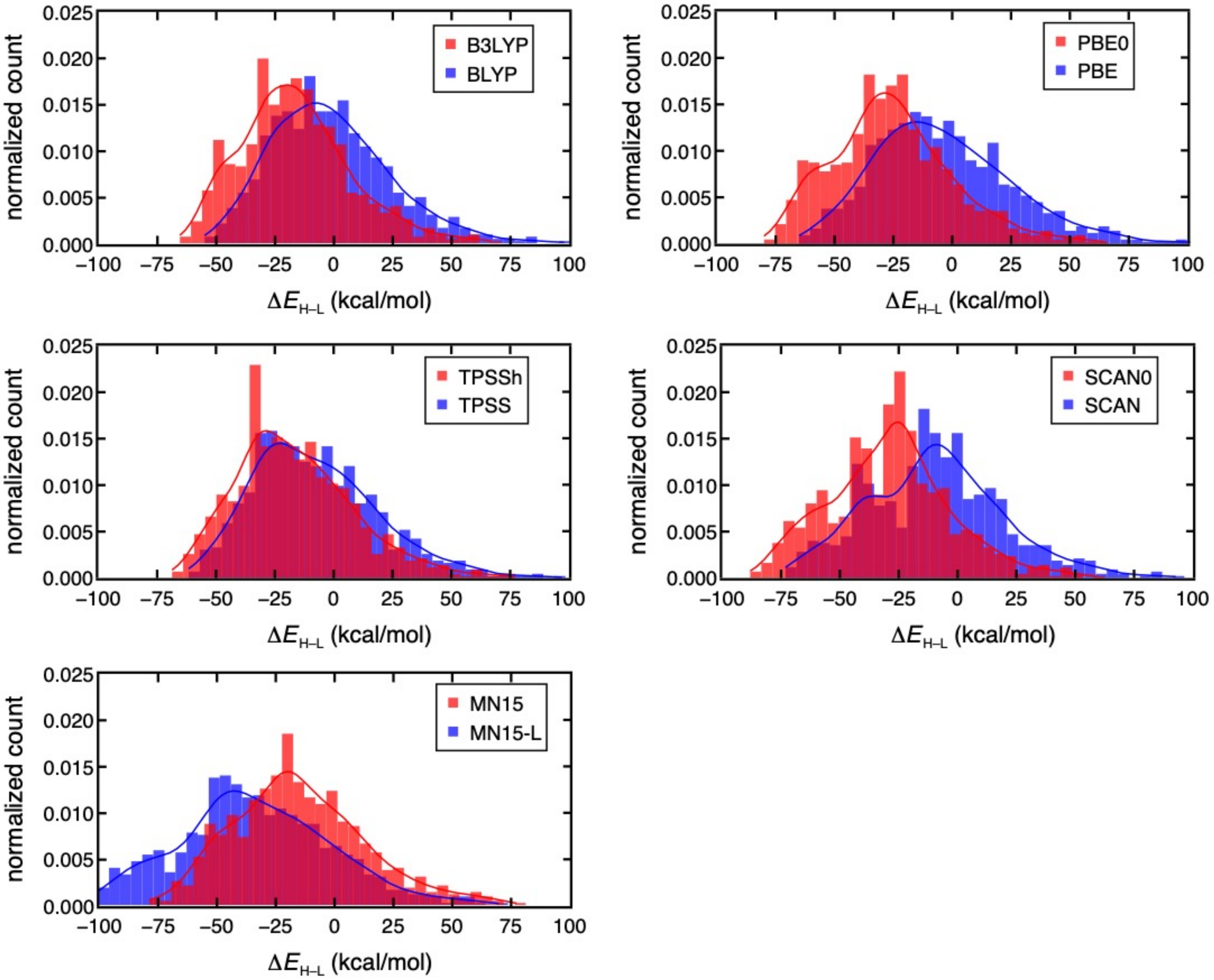

**Figure S3.** Distribution of $\Delta E_{\text{H-L}}$ (in kcal/mol) for select DFAs (from left to right and top to bottom, as indicated in inset legend): BLYP and B3LYP (top left), PBE and PBE0 (top right), TPSS and TPSSh (middle left), SCAN and SCAN0 (middle right), and MN15-L and MN15 (bottom left). The bin size is 5 kcal/mol for all distributions.

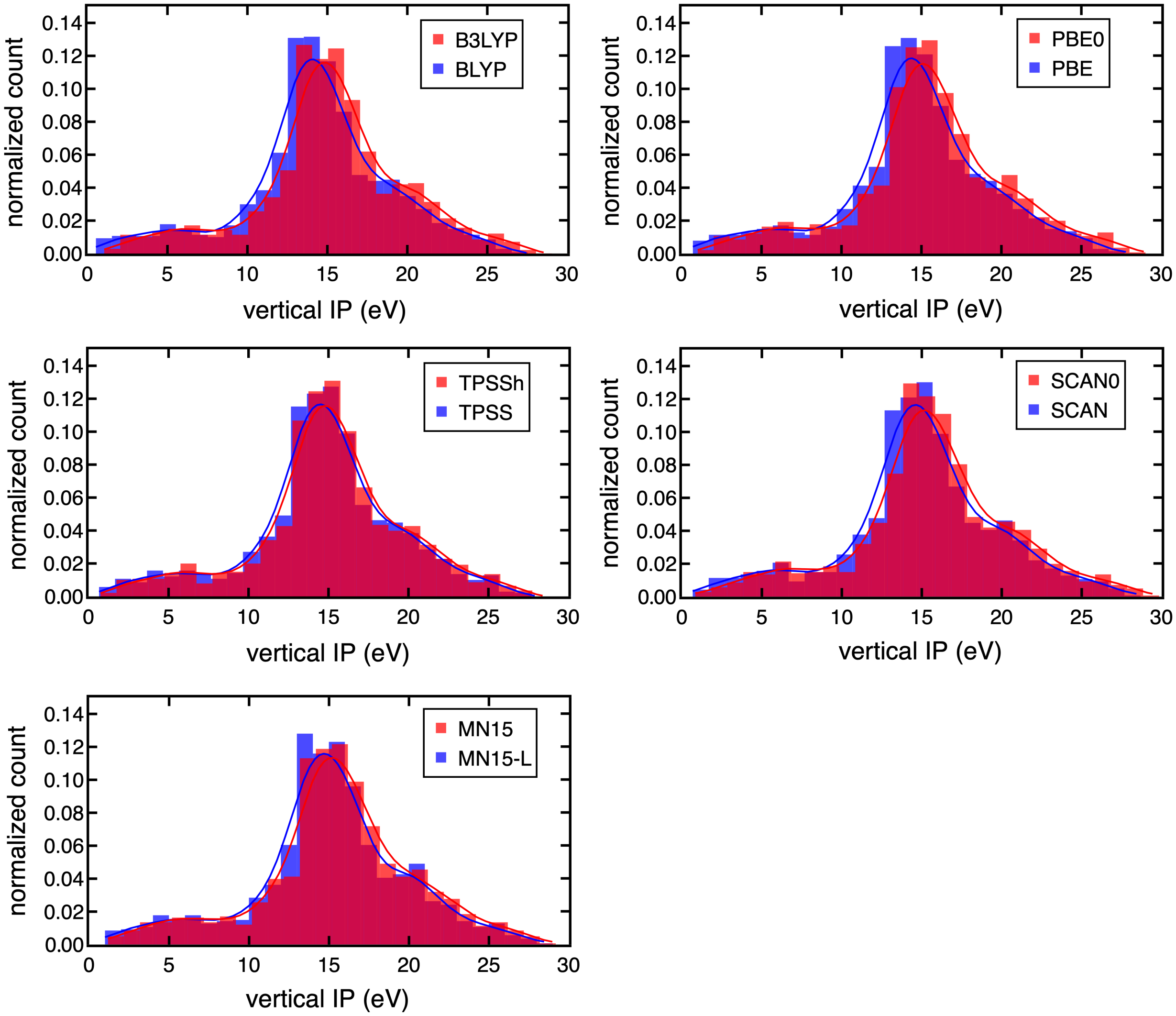

**Figure S4.** Distribution of vertical IP (in eV) for select DFAs (from left to right and top to bottom, as indicated in inset legend): BLYP and B3LYP (top left), PBE and PBE0 (top right), TPSS and TPSSh (middle left), SCAN and. SCAN0 (middle right), and MN15-L and MN15 (bottom left). The bin size is 1 eV for all distributions.

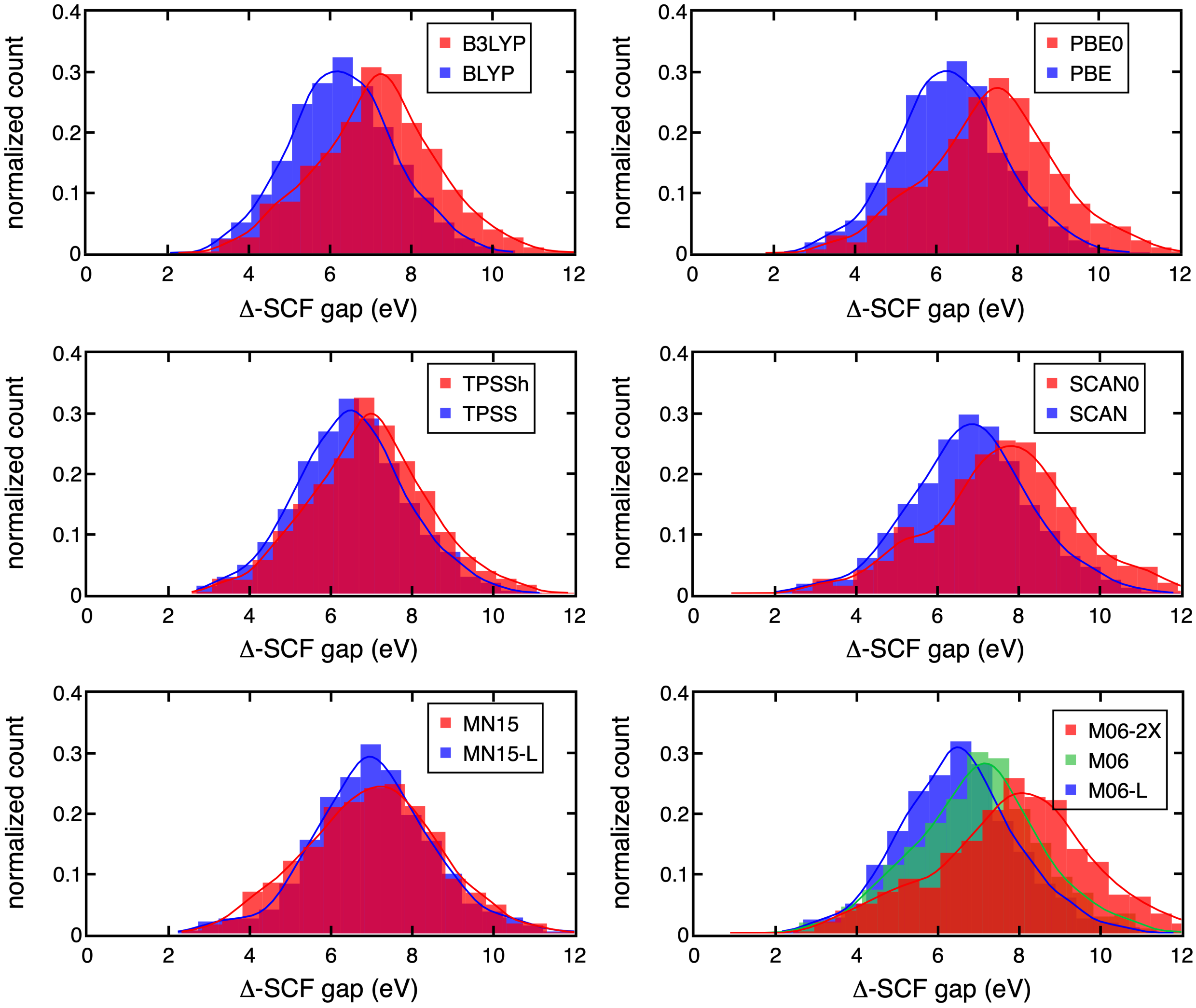

**Figure S5.** Distribution of Δ-SCF gap (in eV) for select DFAs (from left to right and top to bottom, as indicated in inset legend): BLYP and B3LYP (top left), PBE and PBE0 (top right), TPSS and TPSSh (middle left), SCAN and SCAN0 (middle right), MN15-L and MN15 (bottom left), and M06-L, M06, and M06-2X (bottom right). The bin size is 0.5 eV for all distributions.

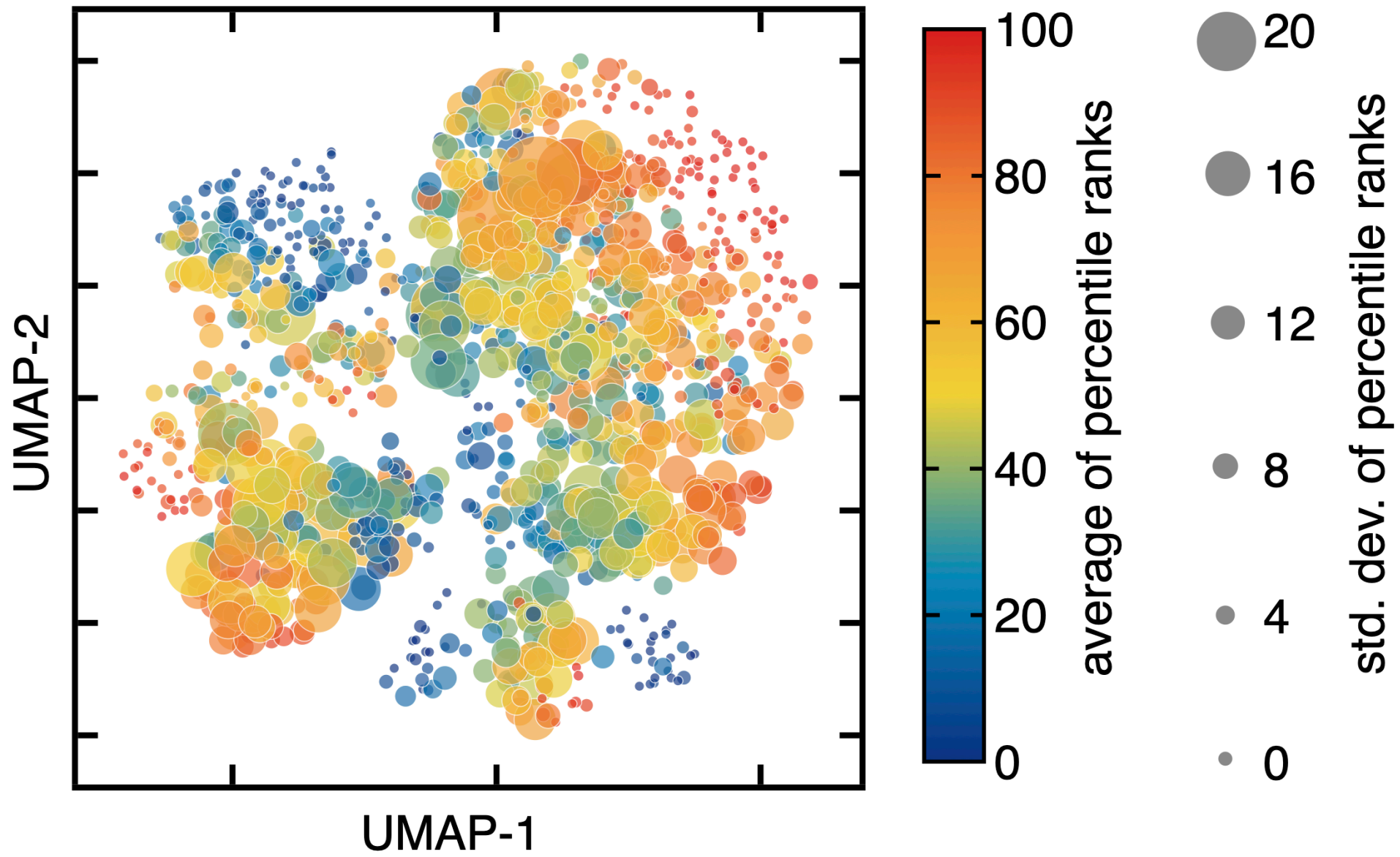

**Figure S6.** UMAP 2D visualization of Δ-SCF gap for the data set. Each complex is represented as a circle that is colored by the average (red for high and blue for low as in inset legend) Δ-SCF gap and scaled by the std. dev. of the percentile ranks of the Δ-SCF gap from the 23 DFAs.

**Table S4.** Percentile ranks for each DFA and their mean value and std. dev. for five representative complexes in Figure 3 in the main text.

| | $\Delta E_{H\text{-}L}$ | | | vertical IP | |
|---|---|---|---|---|---|
| | $Co(III)(CO)_6$ | $Mn(II)(H_2O)_5(pyr)$ | $Mn(II)(CO)_4(H_2O)(pyr)$ | $Fe(II)(NH_3)_6$ | $Mn(II)(furan)_4(CO)_2$ |
| BP86 | 97 | 3 | 73 | 48 | 37 |
| BLYP | 98 | 3 | 70 | 50 | 36 |
| PBE | 97 | 3 | 72 | 48 | 37 |
| TPSS | 98 | 3 | 70 | 46 | 36 |
| SCAN | 98 | 6 | 62 | 46 | 64 |
| M06-L | 98 | 4 | 67 | 46 | 35 |
| MN15-L | 97 | 0 | 45 | 45 | 35 |
| B3LYP | 98 | 3 | 50 | 47 | 59 |
| B3P86 | 98 | 3 | 53 | 45 | 61 |
| B3PW91 | 98 | 2 | 57 | 45 | 62 |
| PBE0 | 98 | 2 | 49 | 44 | 63 |
| ωB97X | 98 | 3 | 37 | 44 | 67 |
| LRC-ωPBEh | 98 | 3 | 48 | 42 | 67 |
| TPSSh | 98 | 3 | 62 | 44 | 61 |
| SCAN0 | 98 | 6 | 39 | 43 | 66 |
| M06 | 98 | 1 | 50 | 44 | 64 |
| M06-2X | 97 | 1 | 20 | 44 | 66 |
| MN15 | 97 | 0 | 55 | 33 | 59 |
| B2GP-PLYP | 97 | 0 | 30 | 44 | 67 |
| PBE0-DH | 98 | 1 | 34 | 43 | 67 |
| DSD-BLYP-D3BJ | 97 | 0 | 26 | 44 | 67 |
| DSD-PBEB95-D3BJ | 97 | 0 | 26 | 43 | 67 |
| DSD-PBEP6-D3BJ | 97 | 0 | 28 | 43 | 68 |
| mean | 97.6 | 2.3 | 48.8 | 44.3 | 57.0 |
| std. dev. | 0.5 | 1.9 | 16.1 | 3.1 | 12.7 |

**Table S5**. Summary of Pearson's *r* (labeled as *r*), coefficient of determination ($R^2$), and mean absolute difference (MAD) between a small (LACVP*) and large (def2-TZVP) basis set evaluated at each functional for three properties considered in this work.

| | $\Delta E_{H-L}$ (kcal/mol) | | | Δ-SCF gap (eV) | | | vertical IP (eV) | | |
|---|---|---|---|---|---|---|---|---|---|
| | *r* | $R^2$ | MAD | *r* | $R^2$ | MAD | *r* | $R^2$ | MAD |
| BP86 | 0.99 | 0.97 | 4.2 | 0.99 | 0.95 | 0.24 | 1.00 | 1.00 | 0.15 |
| BLYP | 0.99 | 0.97 | 3.6 | 0.99 | 0.94 | 0.25 | 1.00 | 1.00 | 0.14 |
| PBE | 0.99 | 0.96 | 4.7 | 0.99 | 0.95 | 0.24 | 1.00 | 1.00 | 0.15 |
| TPSS | 0.99 | 0.89 | 8.4 | 0.97 | 0.91 | 0.30 | 1.00 | 0.99 | 0.20 |
| SCAN | 0.99 | 0.93 | 7.2 | 0.97 | 0.91 | 0.33 | 1.00 | 1.00 | 0.22 |
| M06-L | 0.97 | 0.88 | 7.4 | 0.97 | 0.93 | 0.24 | 1.00 | 1.00 | 0.18 |
| MN15-L | 0.99 | 0.98 | 3.5 | 0.98 | 0.96 | 0.21 | 1.00 | 1.00 | 0.16 |
| B3LYP | 0.98 | 0.95 | 3.5 | 0.96 | 0.91 | 0.31 | 1.00 | 1.00 | 0.16 |
| B3P86 | 0.98 | 0.96 | 3.8 | 0.96 | 0.91 | 0.32 | 1.00 | 1.00 | 0.23 |
| B3PW91 | 0.98 | 0.95 | 4.1 | 0.96 | 0.91 | 0.32 | 1.00 | 0.99 | 0.24 |
| PBE0 | 0.98 | 0.95 | 4.2 | 0.96 | 0.91 | 0.33 | 1.00 | 0.99 | 0.25 |
| ωB97X | 0.98 | 0.93 | 4.6 | 0.96 | 0.90 | 0.39 | 1.00 | 0.99 | 0.33 |
| LRC-ωPBEh | 0.98 | 0.95 | 4.6 | 0.96 | 0.91 | 0.35 | 1.00 | 0.99 | 0.29 |
| TPSSh | 0.98 | 0.88 | 7.8 | 0.96 | 0.90 | 0.34 | 1.00 | 1.00 | 0.24 |
| SCAN0 | 0.97 | 0.92 | 5.8 | 0.96 | 0.91 | 0.37 | 1.00 | 0.99 | 0.29 |
| M06 | 0.97 | 0.93 | 6.2 | 0.95 | 0.91 | 0.32 | 1.00 | 0.99 | 0.22 |
| M06-2X | 0.97 | 0.90 | 6.5 | 0.95 | 0.91 | 0.41 | 1.00 | 0.99 | 0.34 |
| MN15 | 0.95 | 0.87 | 8.7 | 0.95 | 0.88 | 0.41 | 1.00 | 0.99 | 0.27 |
| B2GP-PLYP | 0.97 | 0.90 | 7.1 | 0.96 | 0.87 | 0.50 | 1.00 | 0.99 | 0.23 |
| PBE0-DH | 0.98 | 0.93 | 8.0 | 0.96 | 0.91 | 0.40 | 1.00 | 0.99 | 0.27 |
| DSD-BLYP-D3BJ | 0.97 | 0.88 | 8.0 | 0.96 | 0.87 | 0.52 | 1.00 | 0.99 | 0.23 |
| DSD-PBEB95-D3BJ | 0.97 | 0.89 | 8.0 | 0.96 | 0.89 | 0.45 | 1.00 | 0.99 | 0.23 |
| DSD-PBEP6-D3BJ | 0.97 | 0.85 | 8.5 | 0.96 | 0.88 | 0.50 | 1.00 | 0.99 | 0.23 |

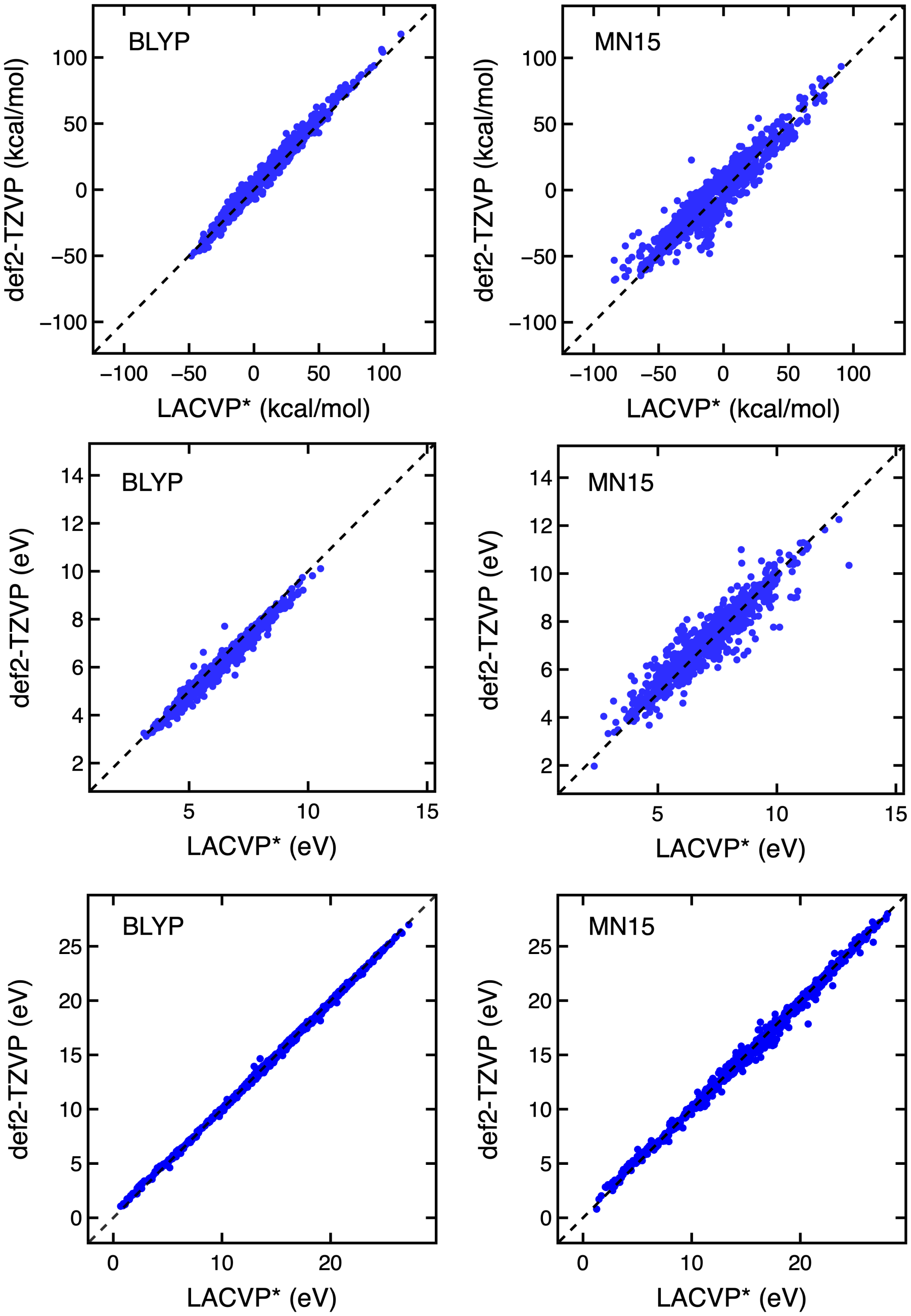

**Figure S7.** Parity plots of three properties, $\Delta E_{\text{H-L}}$ (top), Δ-SCF gap (middle), and vertical IP (bottom), obtained with LACVP* and def2-TZVP for two representative functionals: BLYP (left) and MN15 (right).

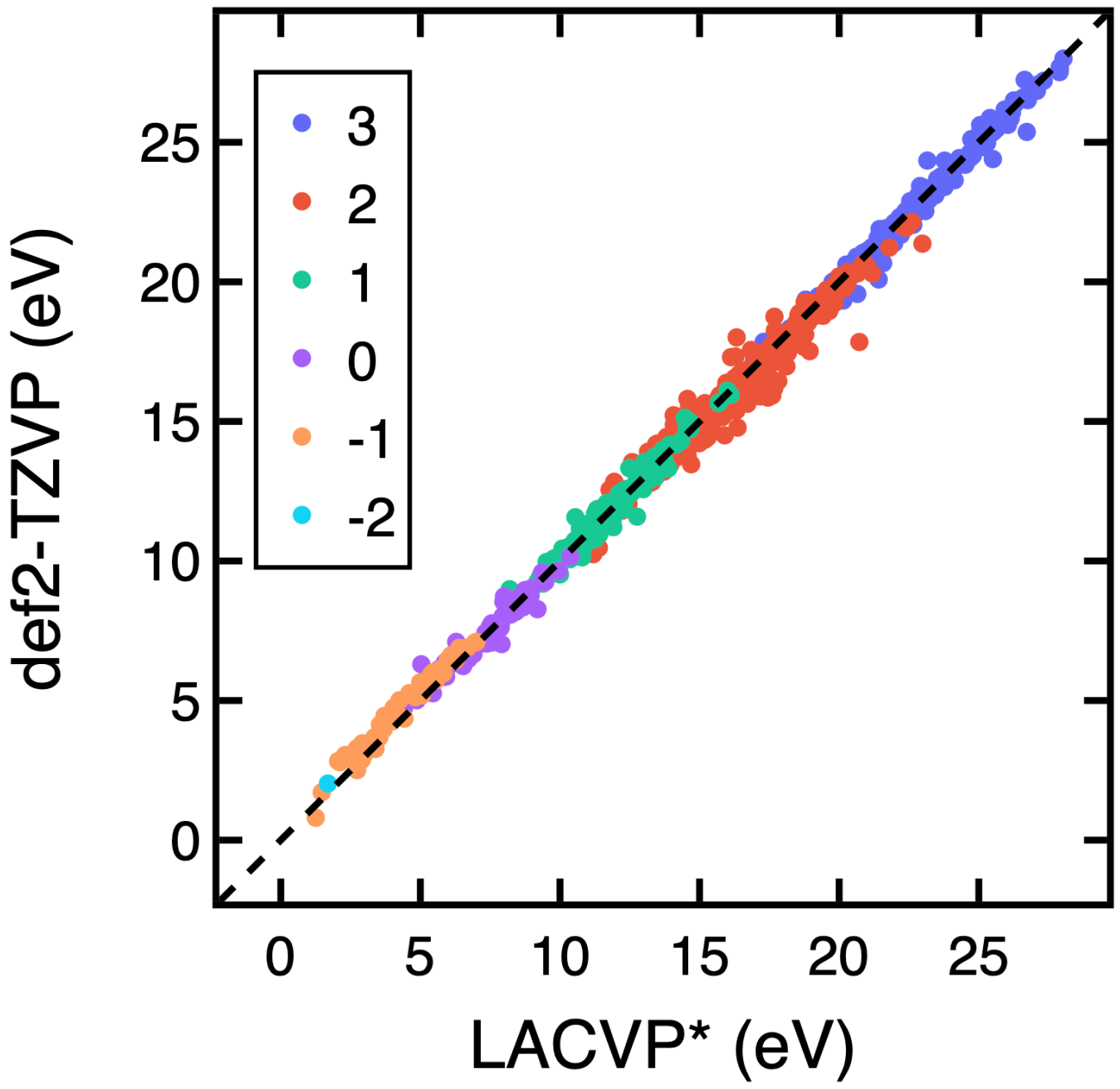

**Figure S8.** Parity plot of the vertical IP obtained with MN15 using LACVP* vs the def2-TZVP basis set grouped by the total charge of the TMC as indicated in inset colorbar.

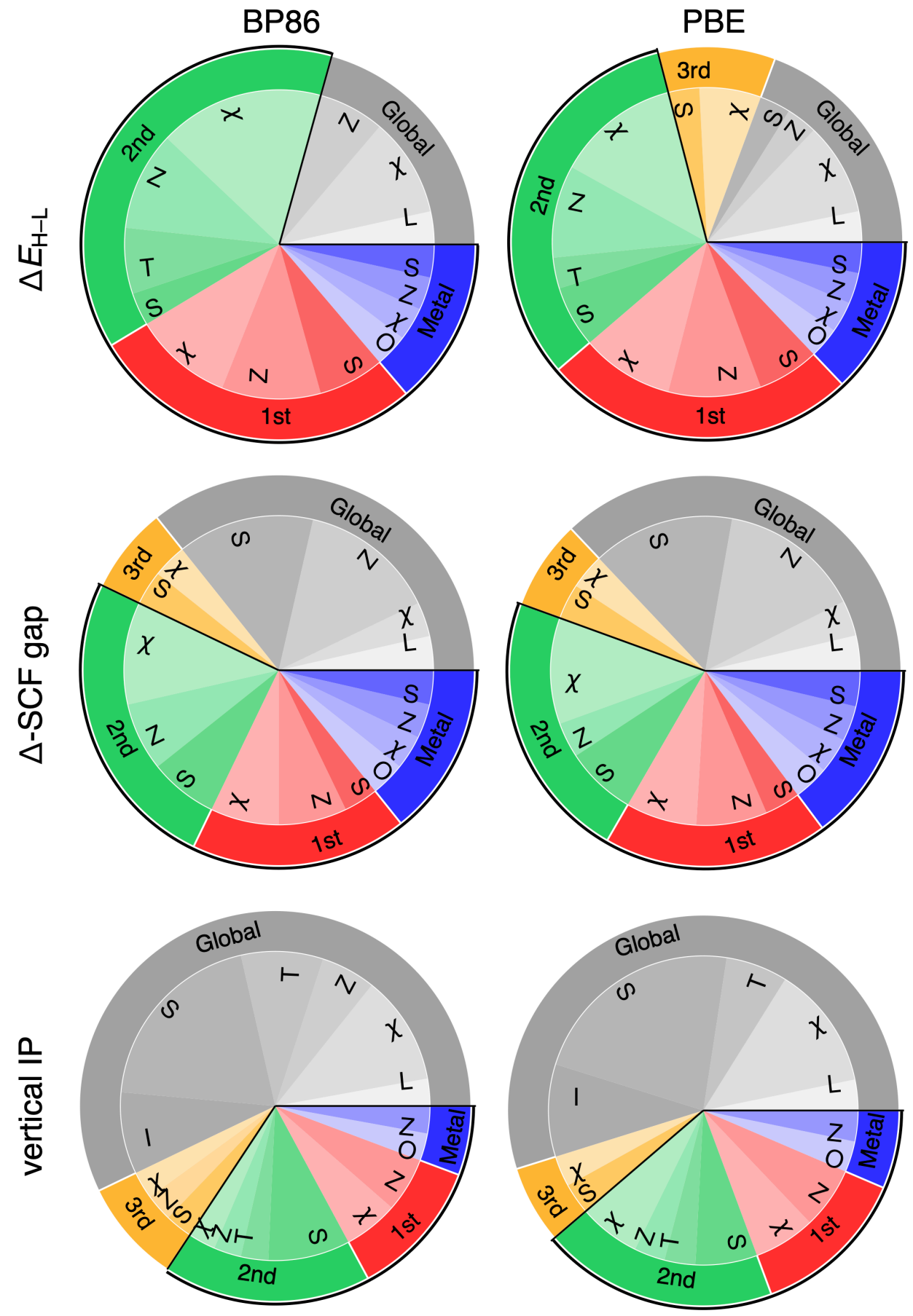

**Figure S9.** Pie charts of the RF-RFA/KRR-selected features for $\Delta E_{\text{H-L}}$ (top), Δ-SCF gap (middle), and vertical IP (bottom) for two additional GGAs that are not shown in Figure 4 in the main text. The format of the pie charts is the same as that in Figure 4.

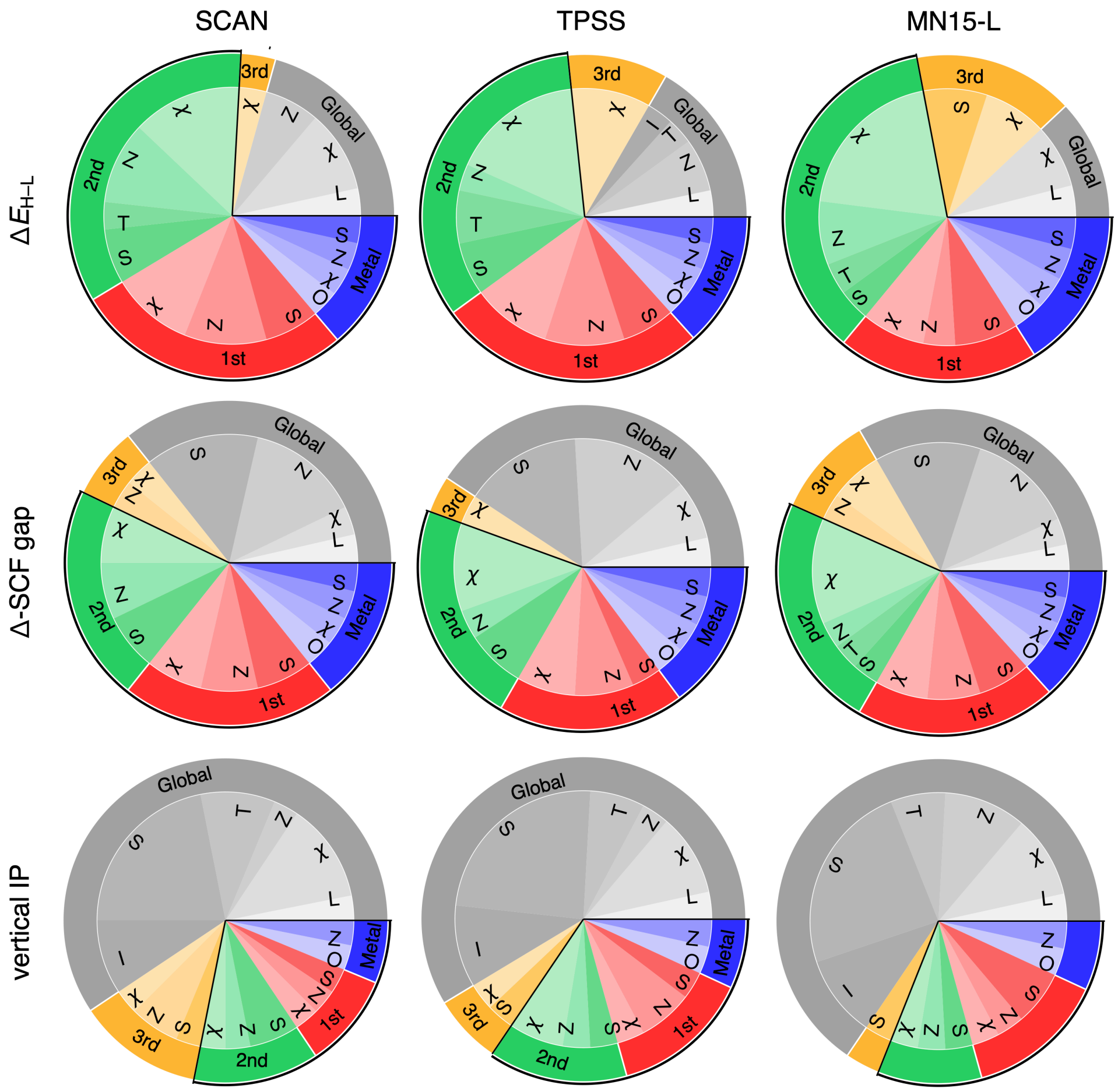

**Figure S10.** Pie charts of the RF-RFA/KRR-selected features for $\Delta E_{\text{H-L}}$ (top), Δ-SCF gap (middle), and vertical IP (bottom) for three additional meta-GGAs that are not shown in Figure 4 in the main text. The format of the pie charts is the same as that in Figure 4.

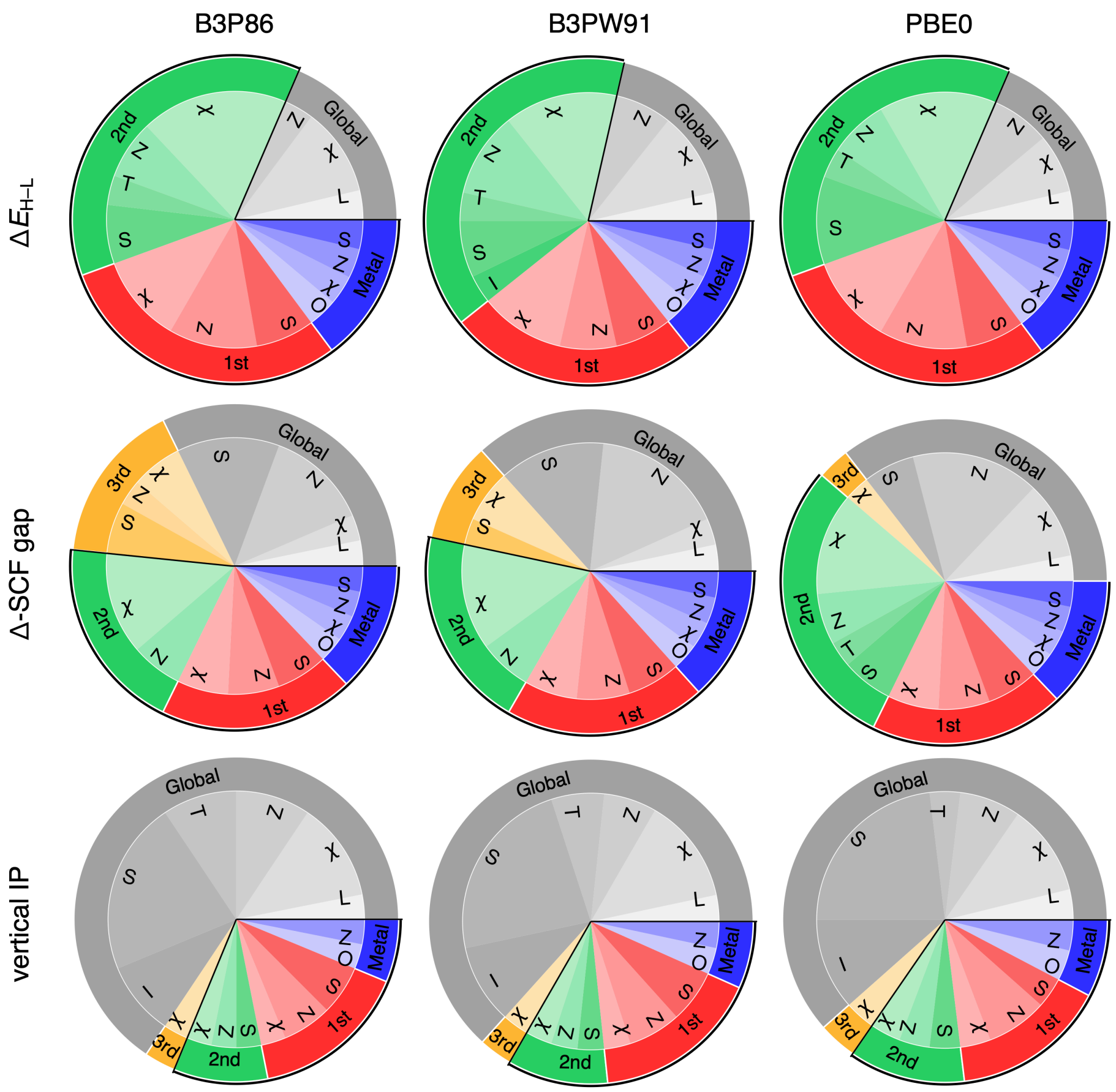

**Figure S11.** Pie charts of the RF-RFA/KRR-selected features for $\Delta E_{\text{H-L}}$ (top), Δ-SCF gap (middle), and vertical IP (bottom) for three additional GGA hybrids that are not shown in Figure 4 in the main text and Figure S15. The format of the pie charts is the same as that in Figure 4.

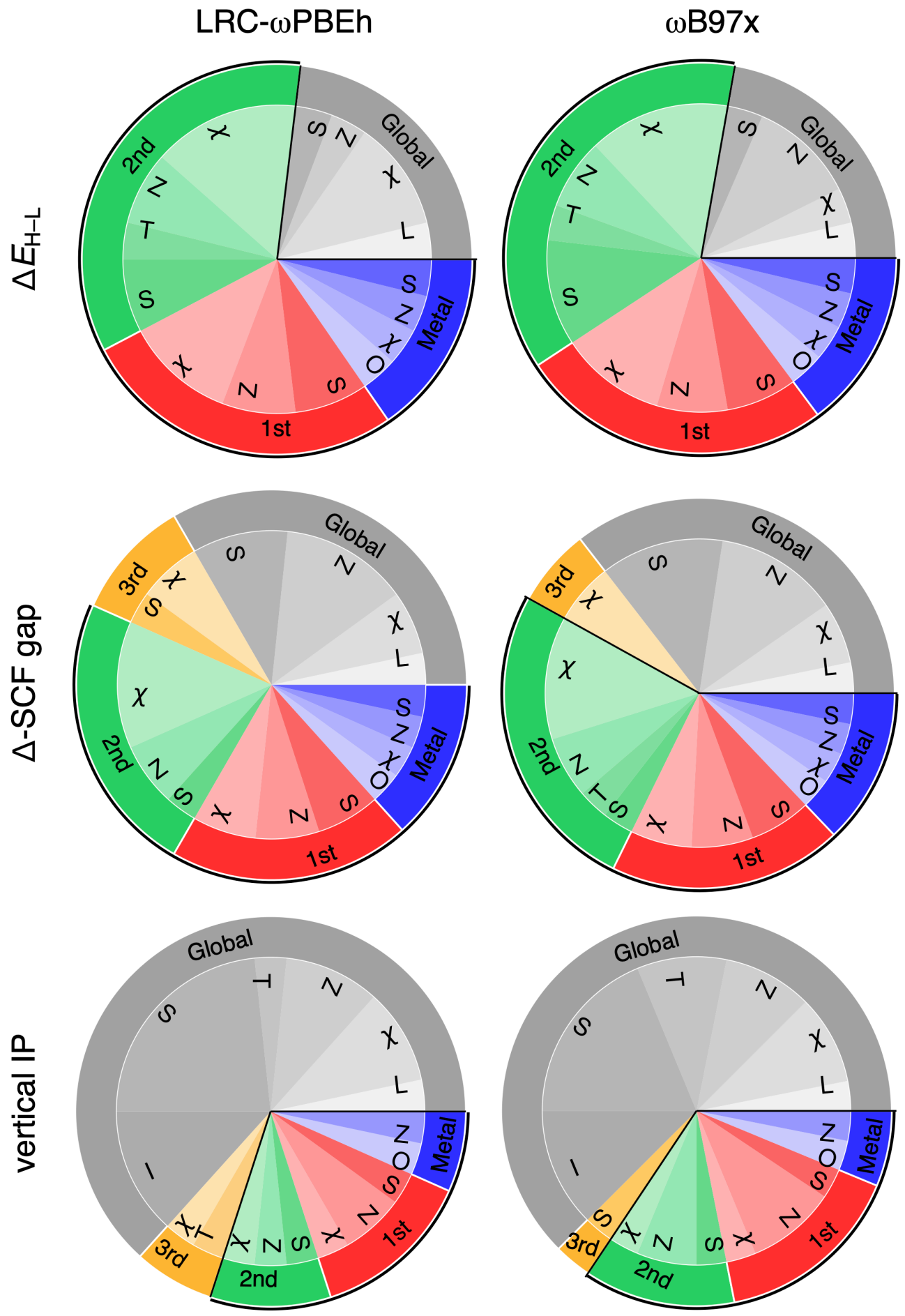

**Figure S12.** Pie charts of the RF-RFA/KRR-selected features for $\Delta E_{\text{H-L}}$ (top), Δ-SCF gap (middle), and vertical IP (bottom) for the two range-separated hybrids that are not shown in Figure 4 in the main text. The format of the pie charts is the same as that in Figure 4.

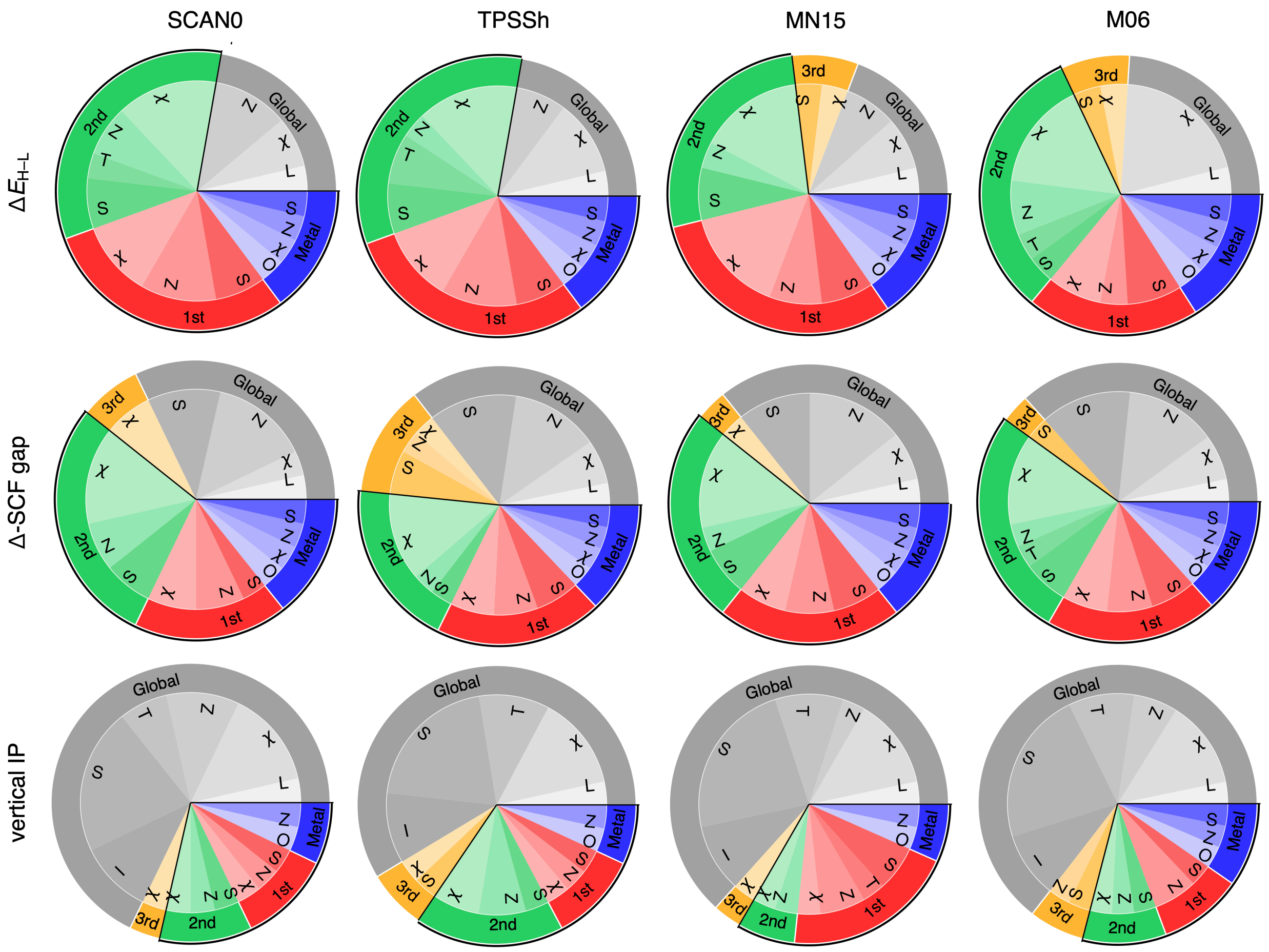

**Figure S13.** Pie charts of the RF-RFA/KRR-selected features for $\Delta E_{\text{H-L}}$ (top), Δ-SCF gap (middle), and vertical IP (bottom) for the four additional meta-GGA hybrids that are not shown in Figure 4 in the main text. The format of the pie charts is the same as that in Figure 4.

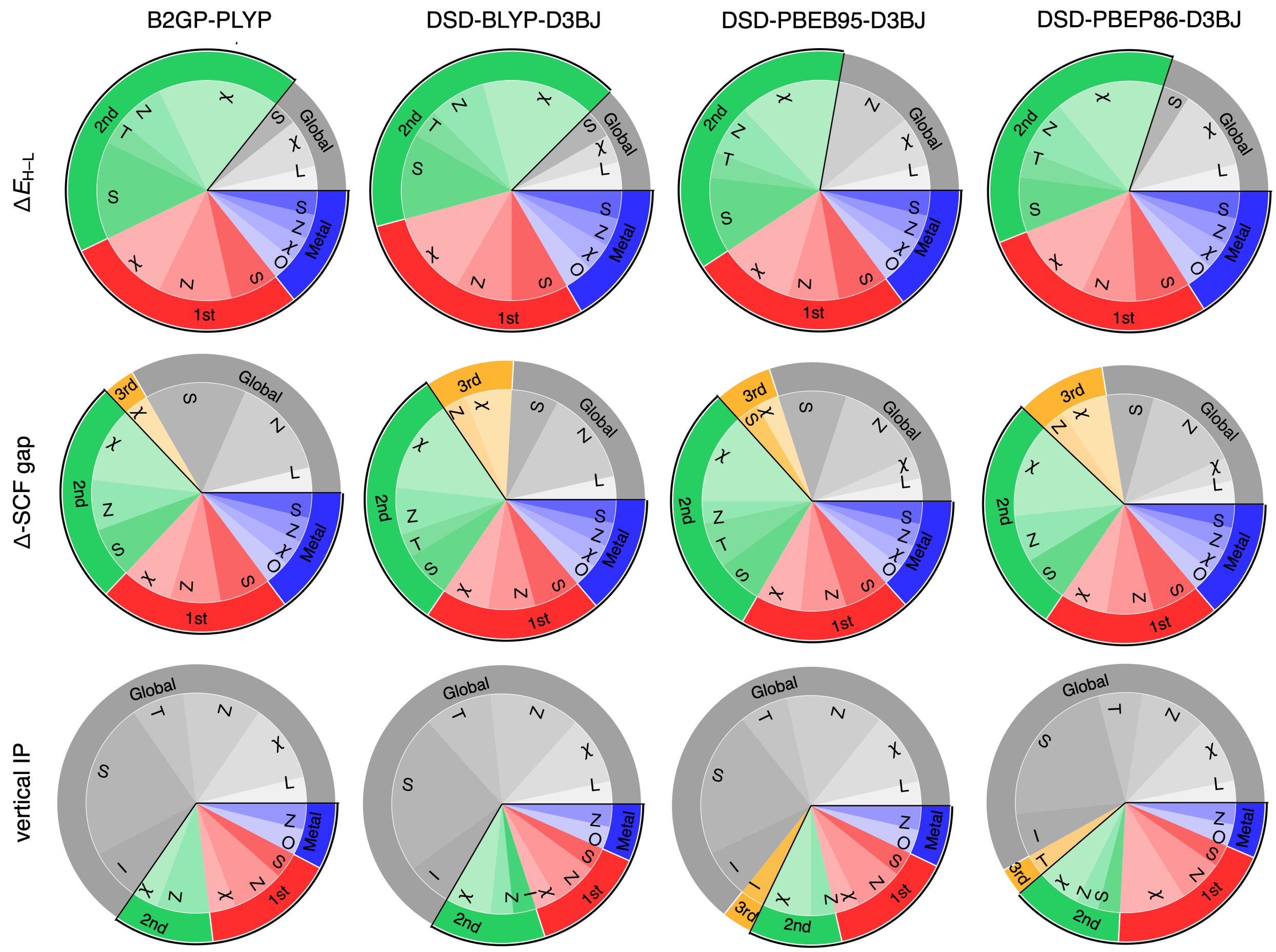

**Figure S14.** Pie charts of the RF-RFA/KRR-selected features for $\Delta E_{\text{H-L}}$ (top), Δ-SCF gap (middle), and vertical IP (bottom) for the four double hybrids that are not shown in Figure 4 in the main text. The format of the pie charts is the same as that in Figure 4.

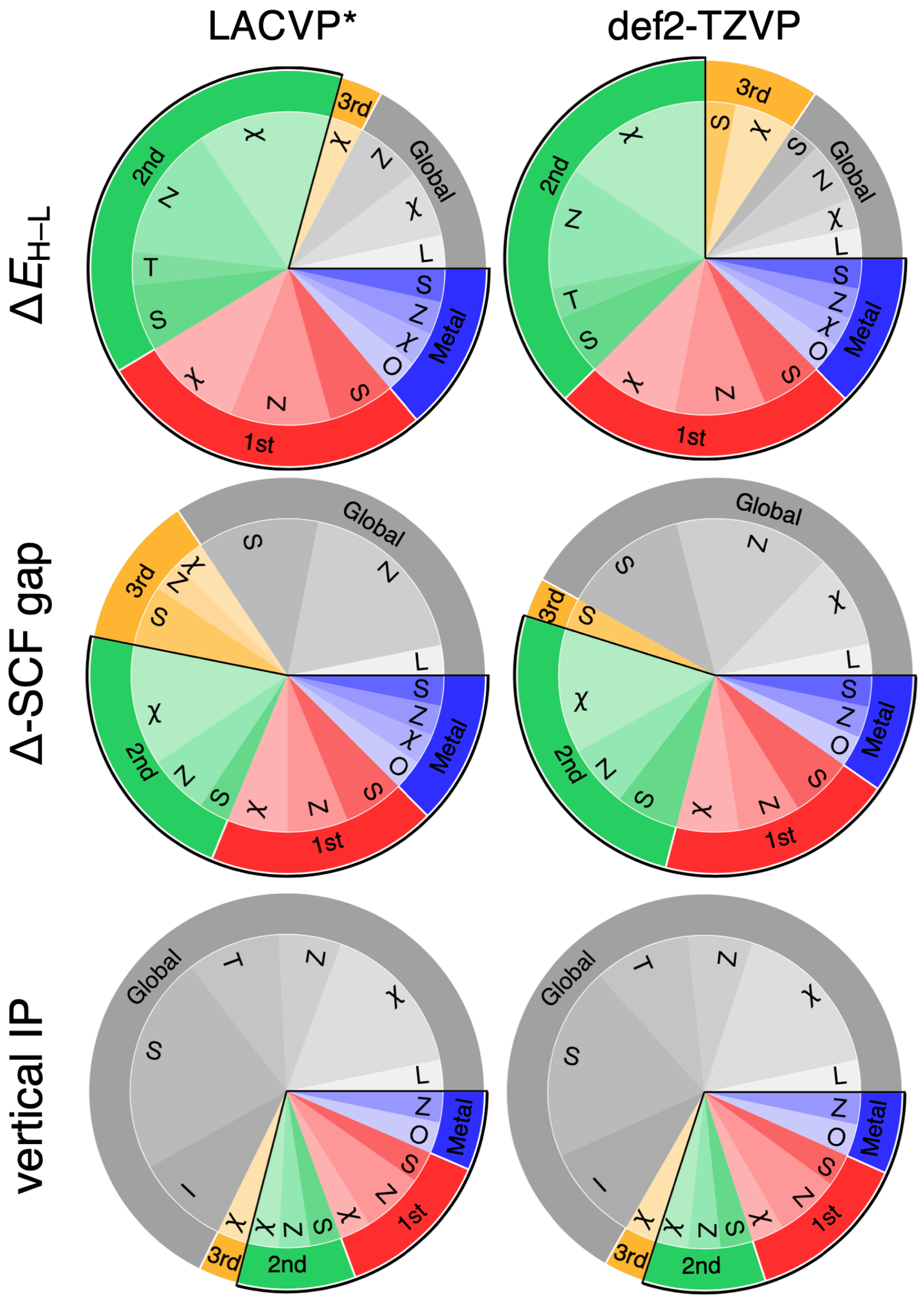

**Figure S15.** Pie charts of the RF-RFA/KRR-selected features selected features of RF-RFA/KRR for B3LYP $\Delta E_{\text{H-L}}$ (top), Δ-SCF gap (middle), and vertical IP (bottom) with the LACVP* (left) and def2-TZVP (right) basis sets. The format of the pie charts is the same as that in Figure 4.

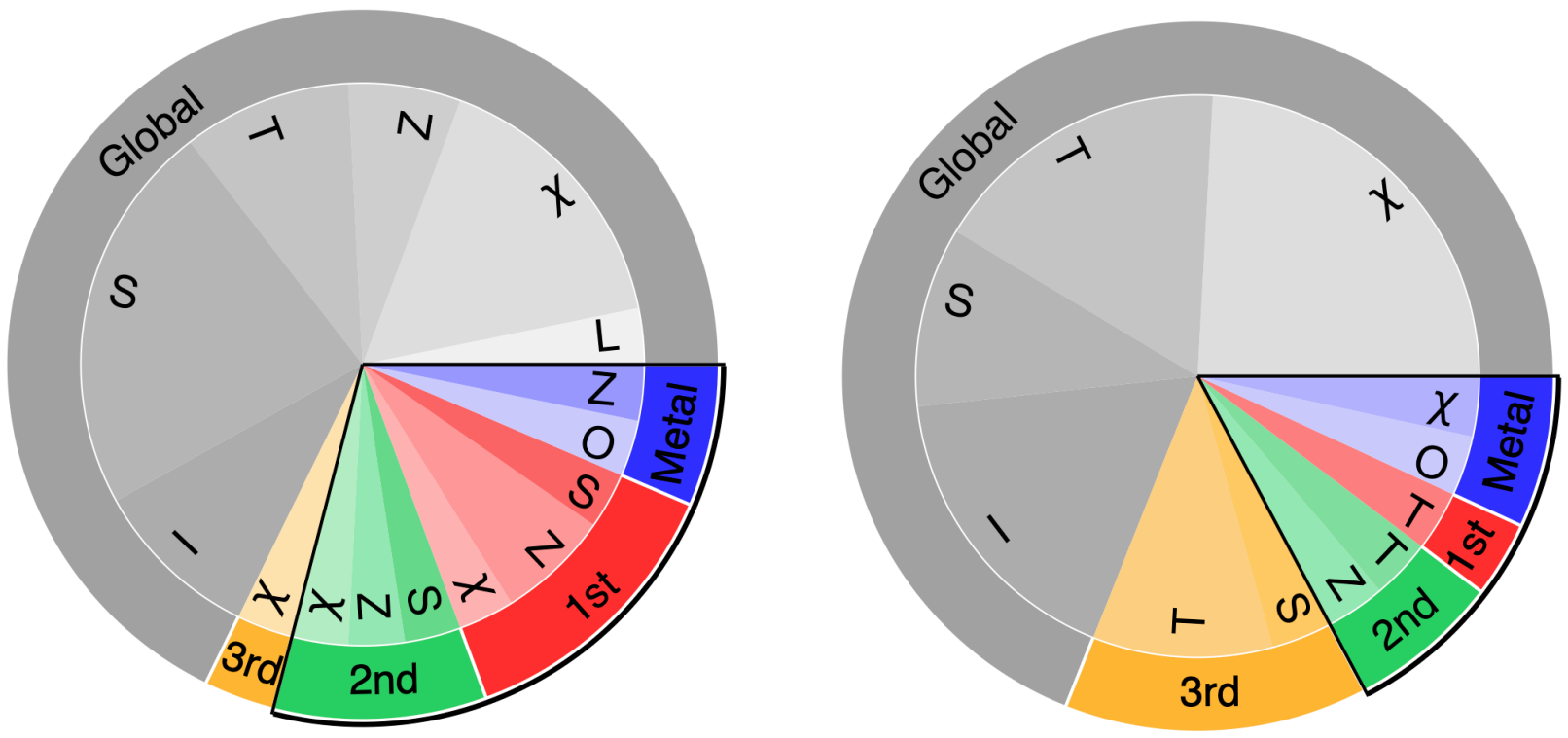

**Figure S16.** Pie charts of the RF-RFA/KRR-selected features for vertical IP with B3LYP/LACVP* applied to a large dataset in this work (i.e., *MD1+OHLDB*, left) and the smaller set of complexes (i.e., *SRX*) from previous work[1-2] (right).

**Table S6**. Hyperparameters for fine-tuned ANN (FT-ANN) models obtained for each property and basis set combination. To obtain these models, we re-optimized the model weights of a B3LYP ANN model with a reduced (i.e., 1e-5) learning rate for each of the non-B3LYP functionals at each basis set and property combination. The hyperparameters of the FT-ANN models at each basis set and property combination are the same as those in the original B3LYP ANN model.

| | LACVP* | | | def2-TZVP | | |
|---|---|---|---|---|---|---|
| | $\Delta E_{\text{H-L}}$ | vertical IP | Δ-SCF gap | $\Delta E_{\text{H-L}}$ | vertical IP | Δ-SCF gap |
| Architecture | [512,512,512] | [256,256,256] | [512,512,512] | [256,256,256] | [256,256,256] | [256,256,256] |
| L2 regularization | 2.2e-4 | 1.0e-4 | 4.8e-4 | 2.0e-3 | 3.0e-3 | 5.6e-2 |
| Dropout rate | 0.41 | 0.30 | 0.02 | 0.08 | 0.25 | 0.28 |
| Learning rate | 3.3e-4 | 7.3e-4 | 4.8e-4 | 7.5e-4 | 9.0e-4 | 8.8e-4 |
| Beta1 | 0.88 | 0.85 | 0.88 | 0.94 | 0.88 | 0.89 |
| Batch size | 128 | 256 | 32 | 256 | 128 | 32 |

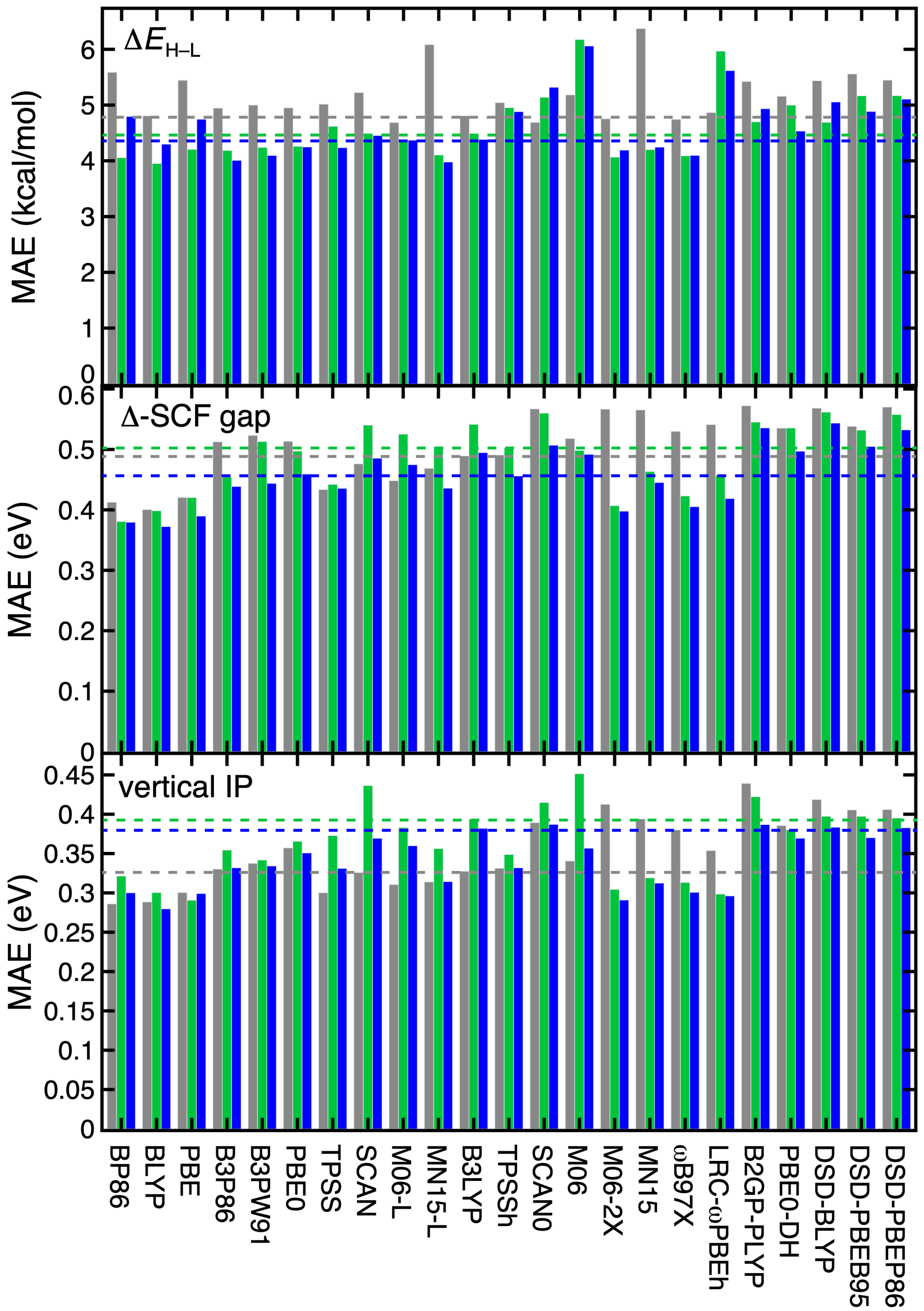

**Figure S17.** Mean absolute error (MAE) for three types of models: RF-RFA/KRR (gray), ANN (green), and FT-ANN (blue) of all 23 DFAs with the LACVP* basis set for three properties: $\Delta E_{\text{H-L}}$ (top), Δ-SCF gap (middle), and vertical IP (bottom). A horizontal dashed line is shown for the B3LYP MAE for each type of model as a reference. The D3BJ dispersion correction is included in all three DSD double hybrids.

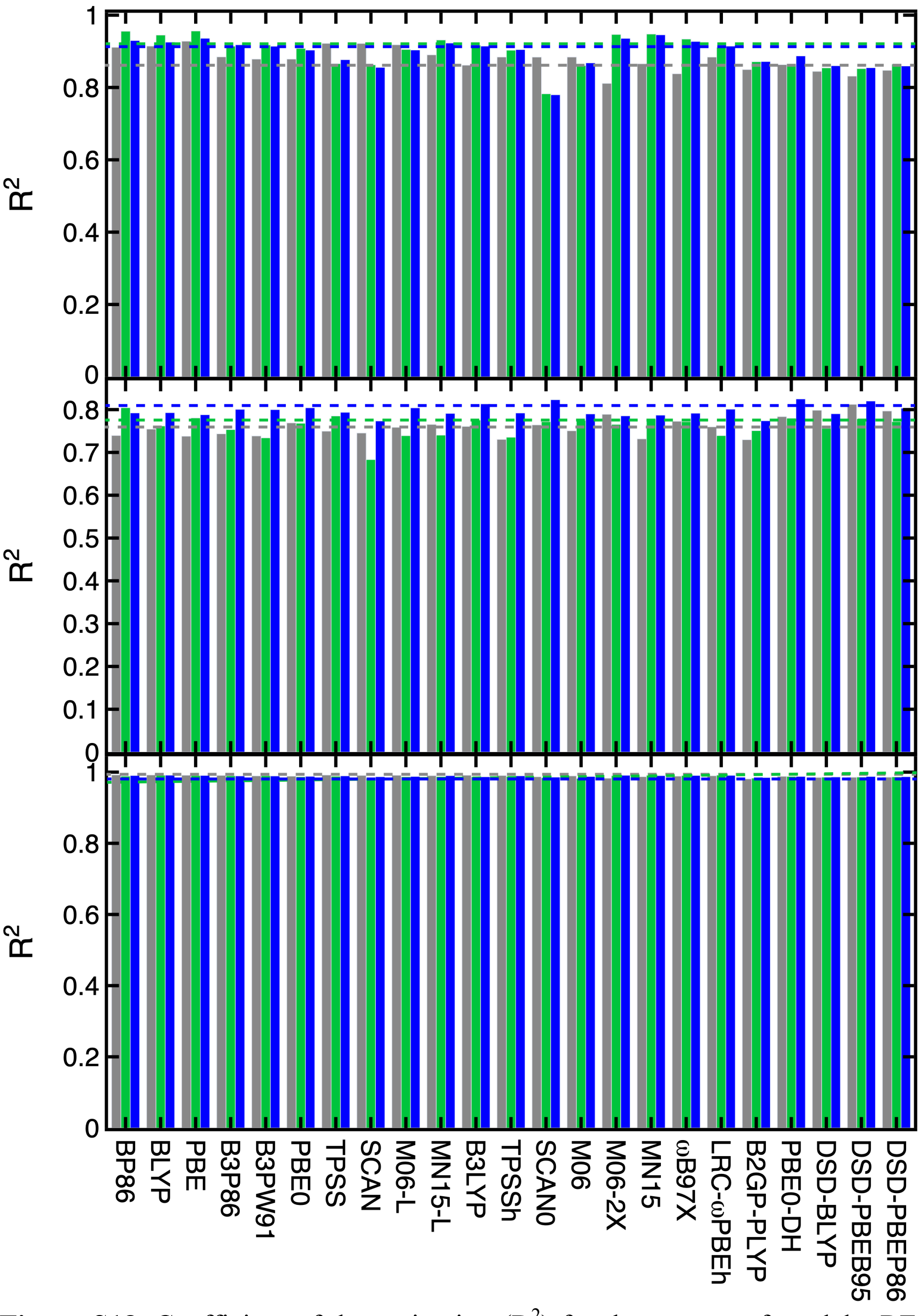

**Figure S18.** Coefficient of determination ($R^2$) for three types of models: RF-RFA/KRR (gray), ANN (green), and FT-ANN (blue) of all 23 DFAs with the LACVP* basis set for three properties: $\Delta E_{\text{H-L}}$ (top), Δ-SCF gap (middle), and vertical IP (bottom). A horizontal dashed line is shown for B3LYP $R^2$ for each type of model as a reference. The D3BJ dispersion correction is applied in all three DSD double hybrids.

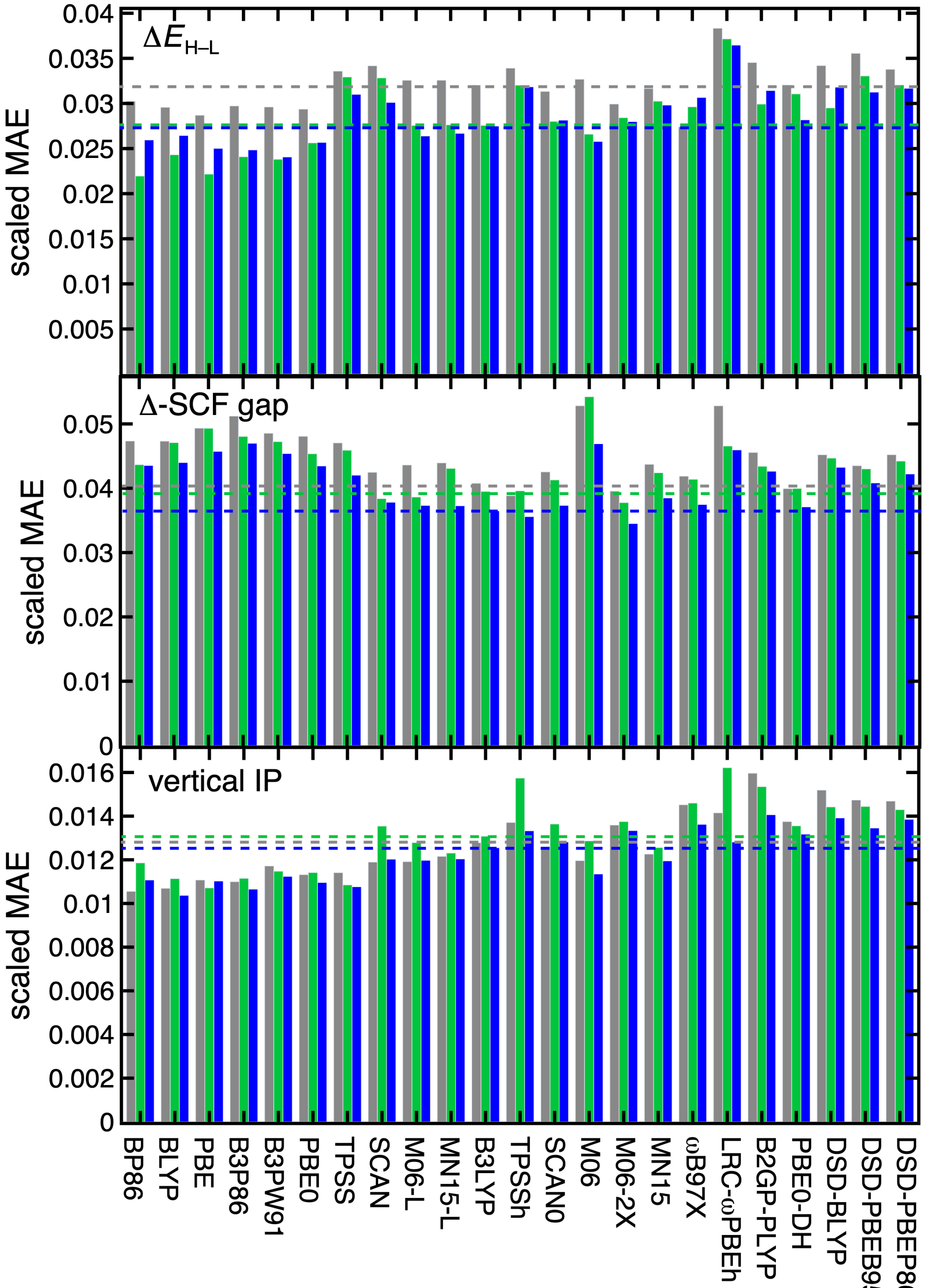

**Figure S19.** Scaled MAE for three types of models: RF-RFA/KRR (gray), ANN (green), and FT-ANN (blue) of all 23 DFAs with the LACVP* basis set for three properties: $\Delta E_{\text{H-L}}$ (top), Δ-SCF gap (middle), and vertical IP (bottom). A horizontal dashed line is shown for the B3LYP scaled MAE for reference. The D3BJ dispersion correction is applied in all three DSD double hybrids.

**Table S7.** Average value (mean) and standard deviation (std. dev.) of the MAE of model performance of 23 functionals with the LACVP* basis set.

| | $\Delta E_{H-L}$ (kcal/mol) | | | Δ-SCF gap (eV) | | | vertical IP (eV) | | |
|---|---|---|---|---|---|---|---|---|---|
| | RF-RFA KRR | ANN | FT-ANN | RF-RFA KRR | ANN | FT-ANN | RF-RFA KRR | ANN | FT-ANN |
| mean | 5.2 | 4.6 | 4.6 | 0.51 | 0.49 | 0.46 | 0.35 | 0.36 | 0.34 |
| std. dev | 0.4 | 0.6 | 0.5 | 0.05 | 0.05 | 0.05 | 0.04 | 0.05 | 0.03 |

**Table S8.** Average value (mean) and standard deviation (std. dev.) of the $R^2$ of model performance of 23 functionals with the LACVP* basis set.

| | $\Delta E_{H-L}$ | | | Δ-SCF gap | | | vertical IP | | |
|---|---|---|---|---|---|---|---|---|---|
| | RF-RFA KRR | ANN | FT-ANN | RF-RFA KRR | ANN | FT-ANN | RF-RFA KRR | ANN | FT-ANN |
| mean | 0.88 | 0.90 | 0.90 | 0.76 | 0.76 | 0.79 | 0.99 | 0.99 | 0.99 |
| std. dev | 0.03 | 0.04 | 0.04 | 0.02 | 0.02 | 0.01 | 0.00 | 0.00 | 0.00 |

**Table S9.** Average value (mean) and standard deviation (std. dev.) of the scaled MAE of model performance of 23 functionals with the LACVP* basis set.

| | $\Delta E_{H-L}$ | | | Δ-SCF gap | | | vertical IP | | |
|---|---|---|---|---|---|---|---|---|---|
| | RF-RFA KRR | ANN | FT-ANN | RF-RFA KRR | ANN | FT-ANN | RF-RFA KRR | ANN | FT-ANN |
| mean | 0.032 | 0.029 | 0.029 | 0.045 | 0.044 | 0.041 | 0.013 | 0.013 | 0.012 |
| std. dev | 0.002 | 0.004 | 0.003 | 0.004 | 0.004 | 0.004 | 0.002 | 0.002 | 0.001 |

**Table S10.** Categories of DFAs studied in this work on each rung of "Jacob's ladder" or by inclusions of HF exchange.

| semi-local (7) | | hybrid (11) | | | double hybrid (5) |
|---|---|---|---|---|---|
| GGA | meta-GGA | GGA hybrid | meta-GGA hybrid | range-separated hybrid | double hybrid |
| 3 | 4 | 4 | 5 | 2 | 5 |

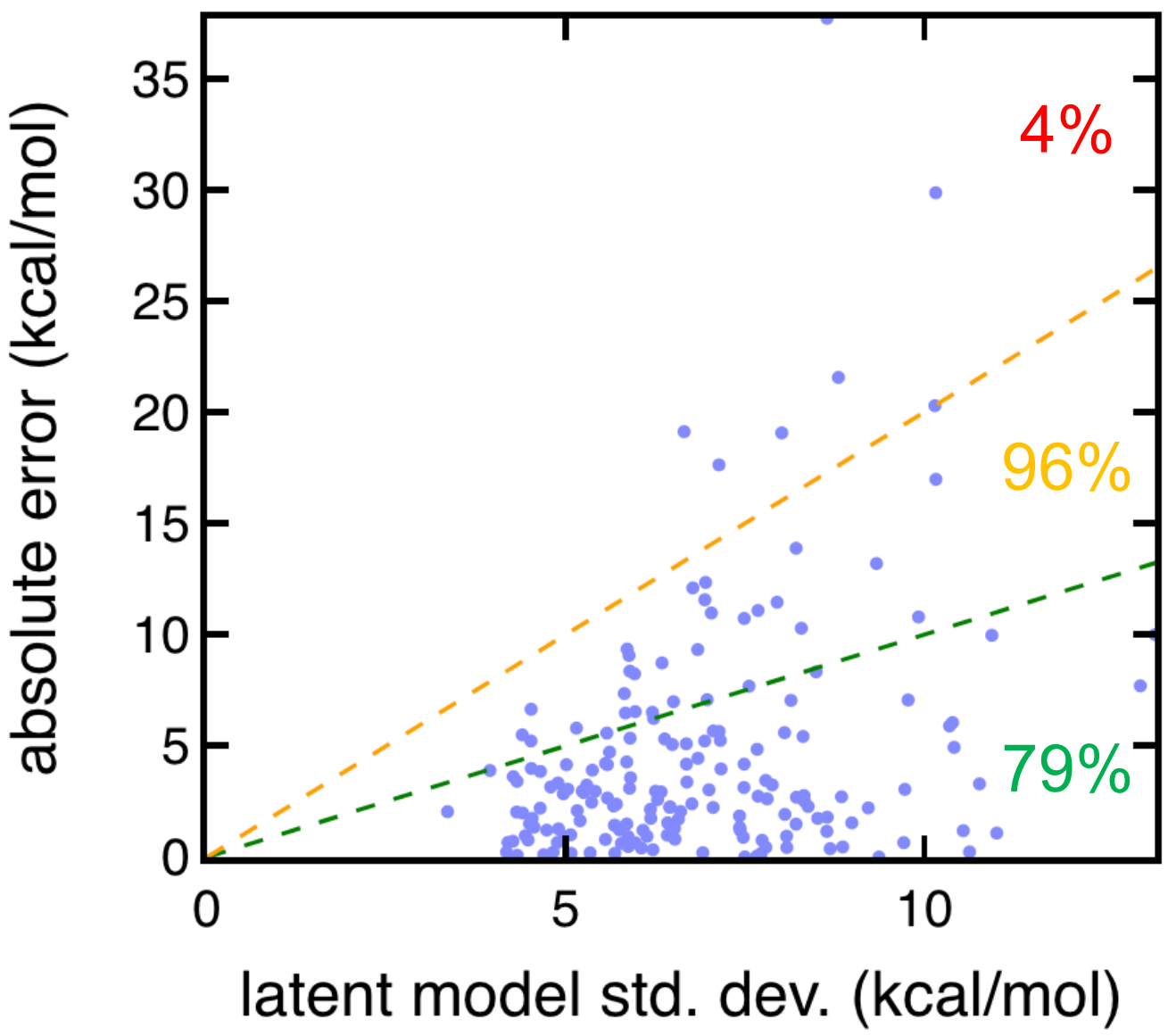

**Figure S20.** Spin-splitting ANN model trained on B3LYP/LACVP* data with absolute errors vs. the uncertainty metric[3] (labeled as latent model std. dev.,), with both in kcal/mol. The model uncertainty is calibrated against the set-aside test data with maximum likelihood following our established procedure[3]. The green and yellow dashed lines corresponds to one and two std. dev., respectively. 79% of the complexes are within one std. dev. and 96% of the complexes are within two std. dev.

**Table S11.** Cutoffs for latent model std. dev. (i.e., calibrated distance in latent space) used during the exploration of SCO complexes ($|\Delta E_{\text{H-L}}| < 5$ kcal/mol) and complexes at a targeted Δ-SCF gap (i.e., < 3 eV).

| | |
|---|---|
| SCO complexes | 20 kcal/mol |
| Targeted-gap complexes | 2 eV |

**Table S12.** Design space ligands with the net charge (charge), denticity (dent), number of atoms ($n_{at}$), connecting atom type (CA), and SMILES string. These ligands are the unique ligands in the dataset from "merged dataset", *MD1*, introduced in Ref. 4.

| id | name | charge | dent. | $n_{at}$ | CA | SMILES |
|---|---|---|---|---|---|---|
| 1 | acac | -1 | 2 | 14 | O | O=C(C)C=[CH-](O)C |
| 2 | aceticacidbipyridine | 0 | 2 | 32 | N | n1ccc(cc1c1nccc(c1)CC(=O)O)CC(=O)O |
| 3 | acetonitrile | 0 | 1 | 6 | N | N#CC |
| 4 | ammonia | 0 | 1 | 4 | N | N |
| 5 | benzisc | 0 | 1 | 16 | C | [C-]#[N+]Cc1ccccc |
| 6 | bipy | 0 | 2 | 20 | O | n1ccccc1c1ncccc1 |
| 7 | carbonyl | 0 | 1 | 2 | N | C#[O] |
| 8 | cyanide | -1 | 1 | 2 | C | [C-]#N |
| 9 | cyanoaceticporphyrin | -2 | 4 | 52 | N | N1=C2C=[CH2-][CH-]1=C(c1[nH]c(cc1)/C=C/1\N=C(/C(=c/3\[nH]/c(=C\2)/cc3)/C=C(/C(=O)O)\C#N)C=C1)/C=C(/C(=O)O)\C#N |
| 10 | cyanopyridine | 0 | 1 | 12 | N | C1(=CCNC=C1)C#N |
| 11 | en | 0 | 2 | 12 | N | NCCN |
| 12 | formaldehyde | 0 | 1 | 4 | O | C=O |
| 13 | furan | 0 | 1 | 9 | O | o1cccc1 |
| 14 | isothiocyanate | -1 | 1 | 3 | N | [N-]=C=S |
| 15 | mebipyridine | 0 | 2 | 26 | N | n1ccc(cc1c1nccc(c1)C)C |
| 16 | mec | -2 | 2 | 15 | O | [O-]c1c(cc(cc1)C)[O-] |
| 17 | methylamine | 0 | 1 | 7 | N | CN |
| 18 | misc | 0 | 1 | 6 | C | [C-]#[N+]C |
| 19 | ox | -2 | 2 | 6 | O | C(=O)(C(=O)[O-])[O-] |
| 20 | phen | 0 | 2 | 22 | N | c1cc2ccc3cccnc3c2nc1 |
| 21 | phenisc | 0 | 1 | 13 | C | [C-]#[N+]c1ccccc1 |
| 22 | pisc | 0 | 1 | 25 | C | [C-]#[N+]c1ccc(C(C)(C)C)cc1 |
| 23 | porphyrin | -2 | 4 | 36 | N | N1=C2C=[CH2-][CH-]1=Cc1[nH]c(cc1)/C=C/1\N=C(/C=c/3\[nH]/c(=C\2)/cc3)C=C1 |
| 24 | pph3 | 0 | 1 | 34 | P | c1c(P(c2ccccc2)c2ccccc2)cccc1 |
| 25 | py | 0 | 1 | 11 | C | C1=CCNC=C1 |
| 26 | tbuc | -2 | 2 | 24 | O | [O-]c1c(cc(C(C)(C)C)cc1)[O-] |
| 27 | thiopyridine | 0 | 1 | 12 | N | C1(=CCNC=C1)S |
| 28 | water | 0 | 1 | 3 | O | O |
| 29 | fluoride | -1 | 1 | 1 | F | [F-] |
| 30 | iodide | -1 | 1 | 1 | I | [I-] |
| 31 | [O-][O-] | -2 | 1 | 2 | O | [O-][O-] |
| 32 | hydroxyl | -1 | 1 | 2 | O | [OH-] |
| 33 | phosphine | 0 | 1 | 4 | P | [PH3] |
| 34 | [S--] | -2 | 1 | 1 | S | [S--] |
| 35 | hydrogen sulfide | 0 | 1 | 3 | S | [SH2] |
| 36 | cyanate | -1 | 1 | 3 | N | N#C[O-] |

**Table S13.** Allowed ligand combinations in the theoretical complex space according to the symmetry and allowed equatorial or axial ligand type. The 11,700 complexes are combined with 16 possible metal/oxidation/spin state combinations to produce a theoretical space of 187,200 complexes.

| class | Allowed eq | Allowed ax | total |
|---|---|---|---|
| homoleptic | monodentate (25) | -- | 25 |
| heteroleptic, ax1 = ax2 | monodentate (25) | monodentate different from eq (24) | 25×24 = 600 |
| | bidentate (9) | monodentate (25) | 9×25 = 225 |
| | tetradentate (2) | monodentate (25) | 2×25 = 50 |
| heteroleptic, ax1 ≠ ax2 | monodentate (25) | two monodentate ligands $\binom{25}{2}$ | 25×300 = 7500 |
| | bidentate (9) | | 9×300 = 2700 |
| | tetradentate (2) | | 2×300 = 600 |
| Total | | | 11700 |

**Table S14.** Definitions of spin multiplicities (2*S*+1) for each metal and oxidation state in the theoretical complex space.

| | | M(II) multiplicity | M(III) multiplicity |
|---|---|---|---|
| Cr | LS | 1 | 2 |
| | HS | 5 | 4 |
| Mn | LS | 2 | 1 |
| | HS | 6 | 5 |
| Fe | LS | 1 | 2 |
| | HS | 5 | 6 |
| Co | LS | 2 | 1 |
| | HS | 4 | 5 |

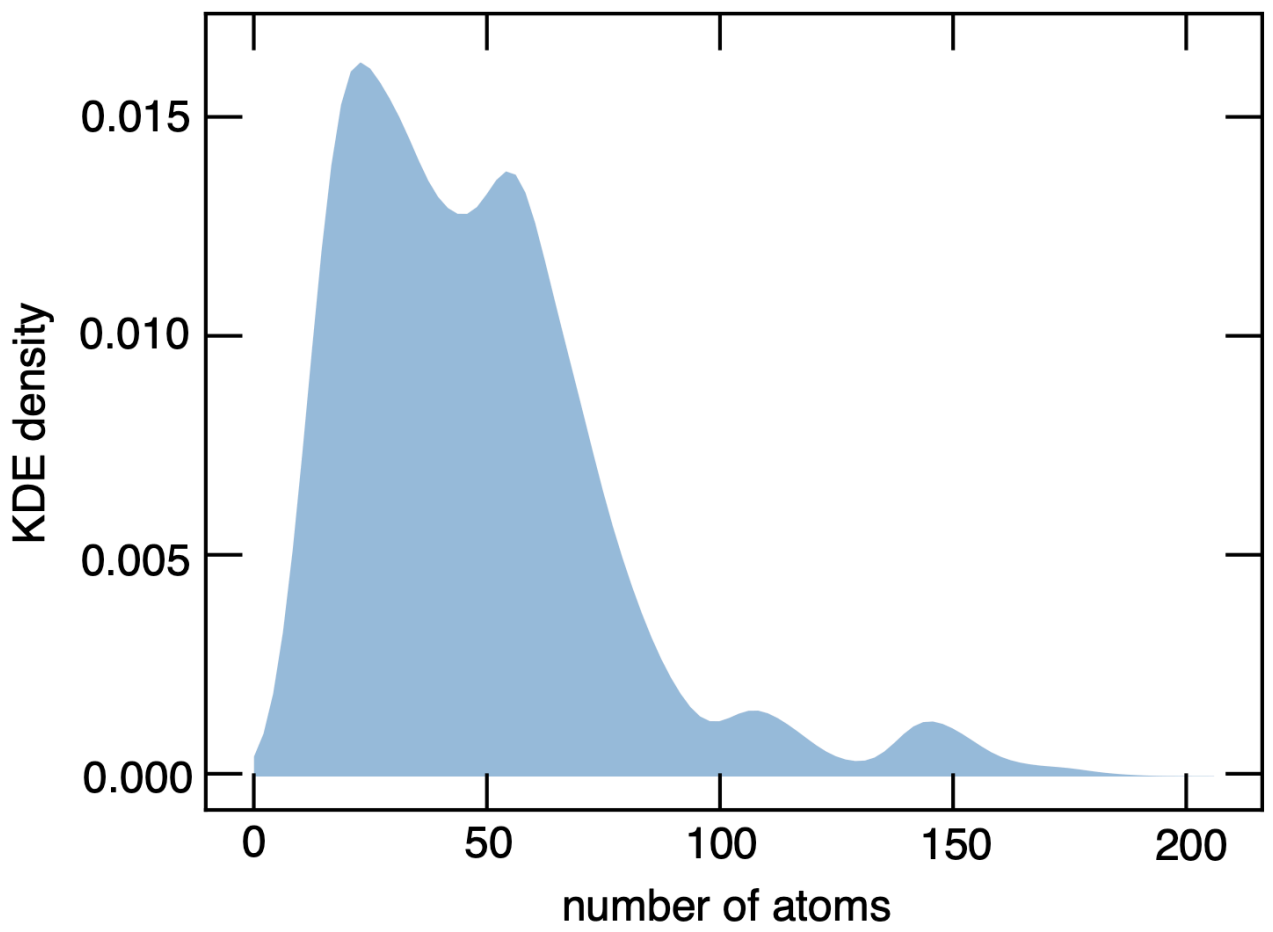

**Figure S21.** Kernel density estimation (KDE) of the size distribution of the 187,200 complexes design space.

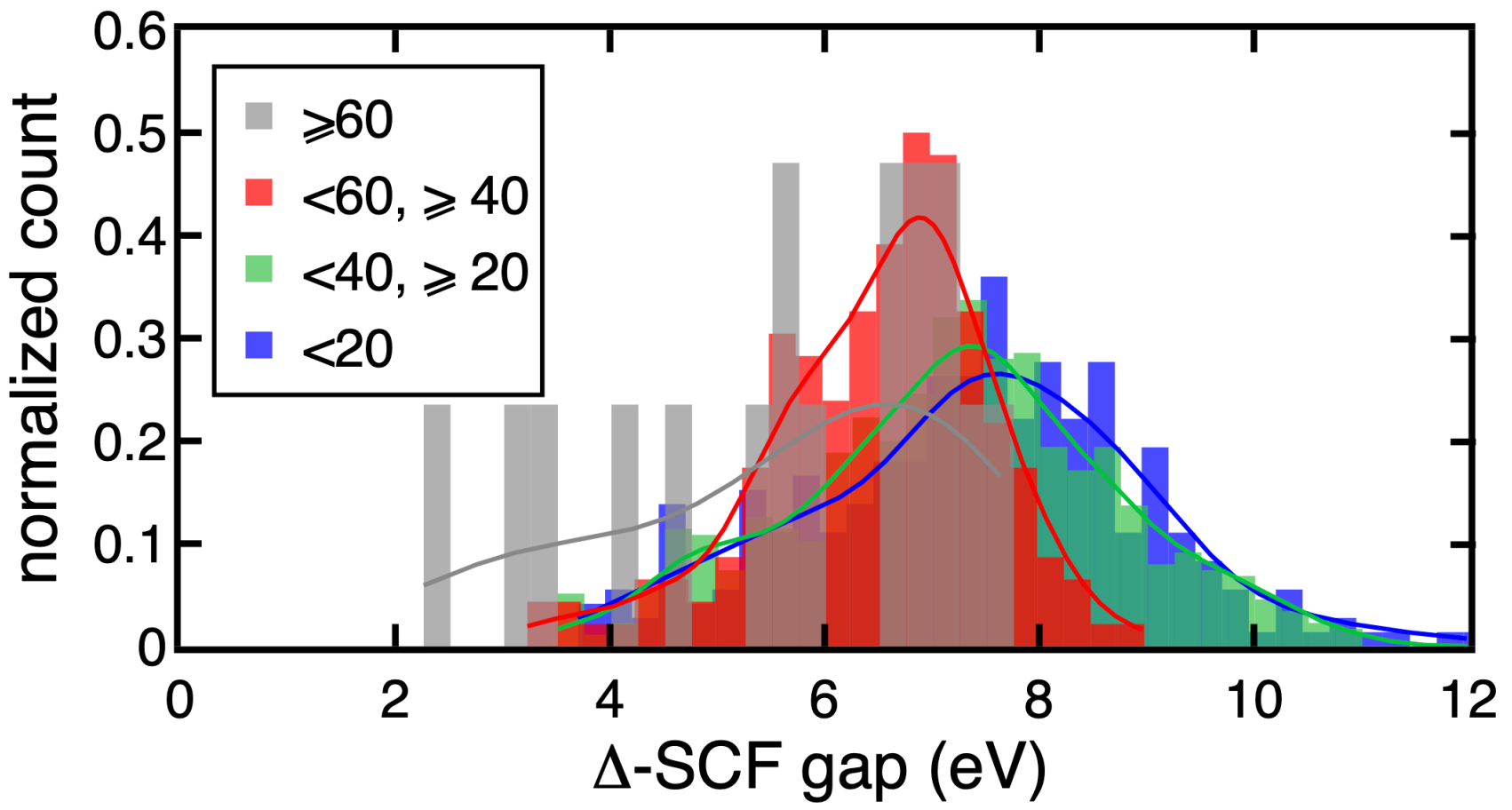

**Figure S22.** Normalized histograms of Δ-SCF gap over the B3LYP/LACVP* data grouped by the system size: < 20 (blue), between 20 and 40 (green), between 40 and 60 (red), and > 60 (gray).

**Table S15.** Density functional categorization used in Figure 8 in the main text.

| | |
|---|---|
| semi-local (GGA and meta-GGA) | BLYP, BP86, PBE, SCAN, TPSS, M06-L, MN15-L |
| hybrid (range-separated and global) | B3LYP, B3P86, B3PW91, PBE0, SCAN0, TPSSh, M06, M06-2X, MN15, LRC-ωPBEh, ωB97x |
| double hybrid | B2GP-PLYP, PBE0-DH, DSD-BLYP-D3BJ, DSD-PBEB95-D3BJ, DSD-PBEP86-D3BJ |

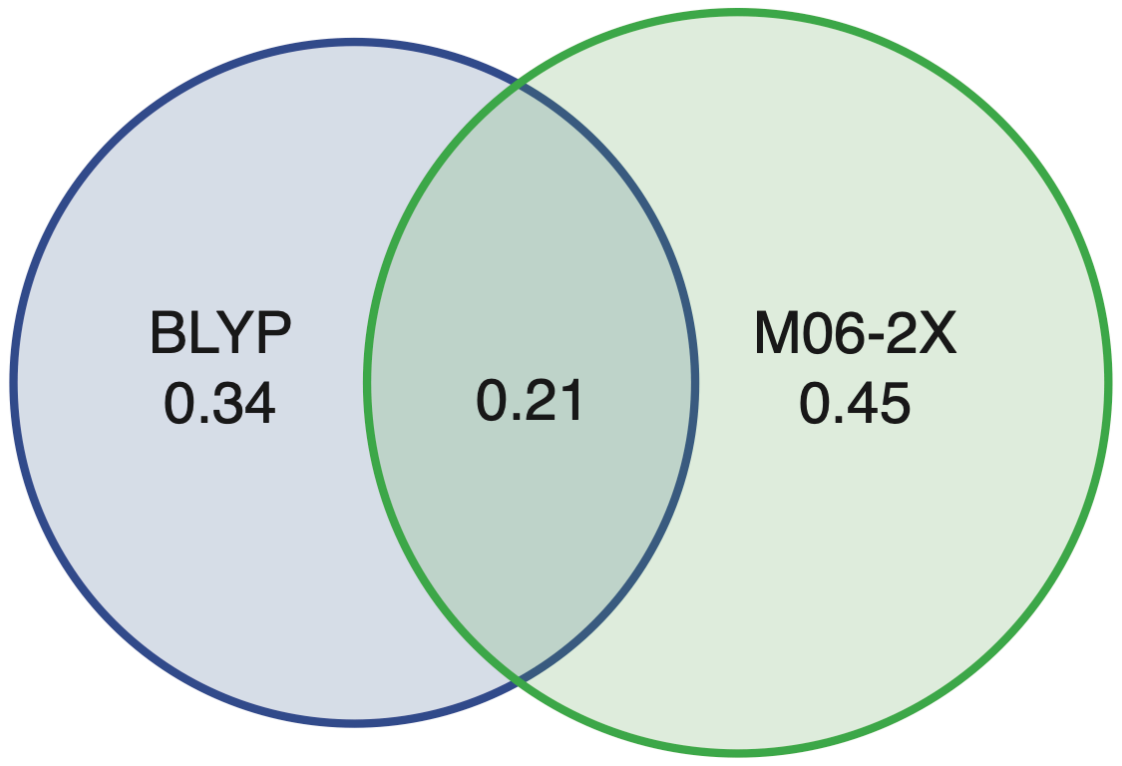

**Figure 23.** Venn diagram of lead targeted gap (Δ-SCF gap < 3 eV) complexes discovered by BLYP and M06-2X.

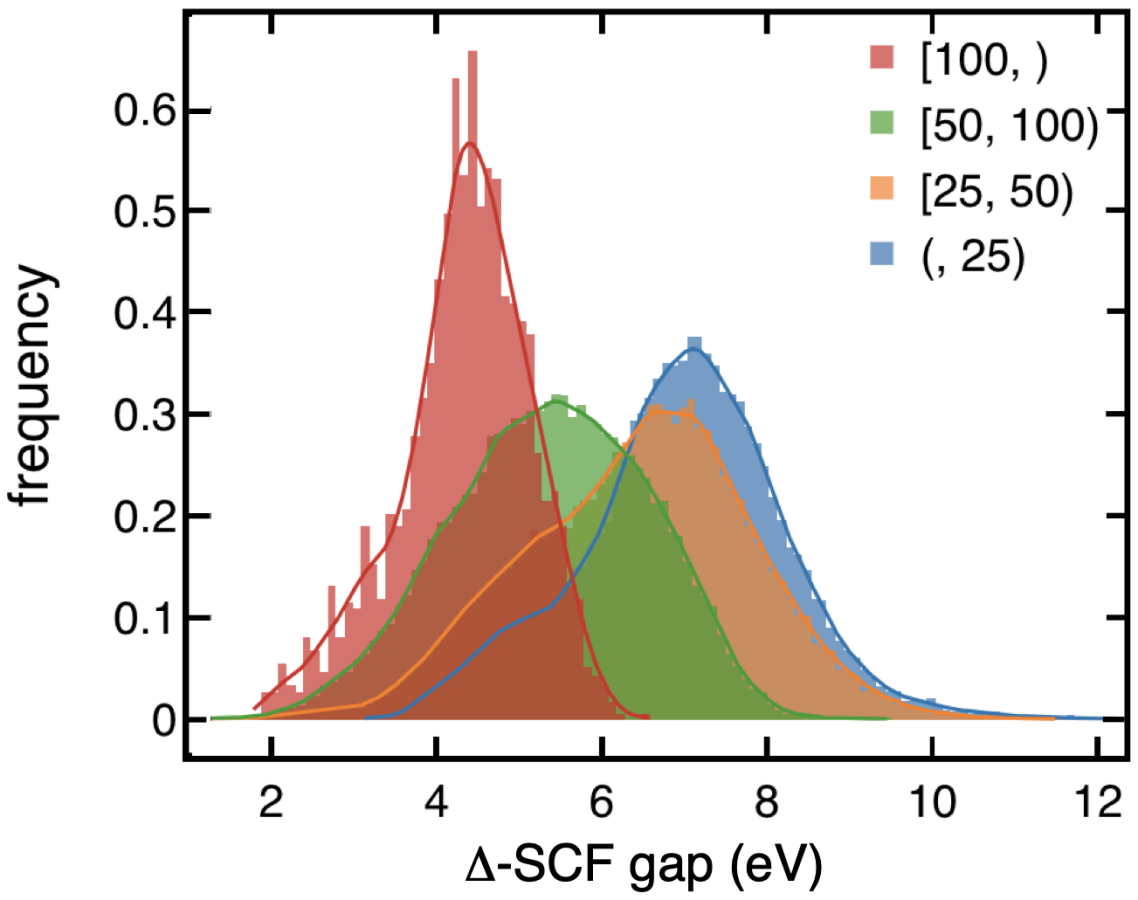

**Figure S24.** Distribution of predicted Δ-SCF gap from the B3LYP ANN model with different complex size ranges in the 187,200 complexes design space: below 25 atoms (blue), between 25 and 50 atoms (orange), between 50 and 100 atoms (green), above 100 atoms (red).

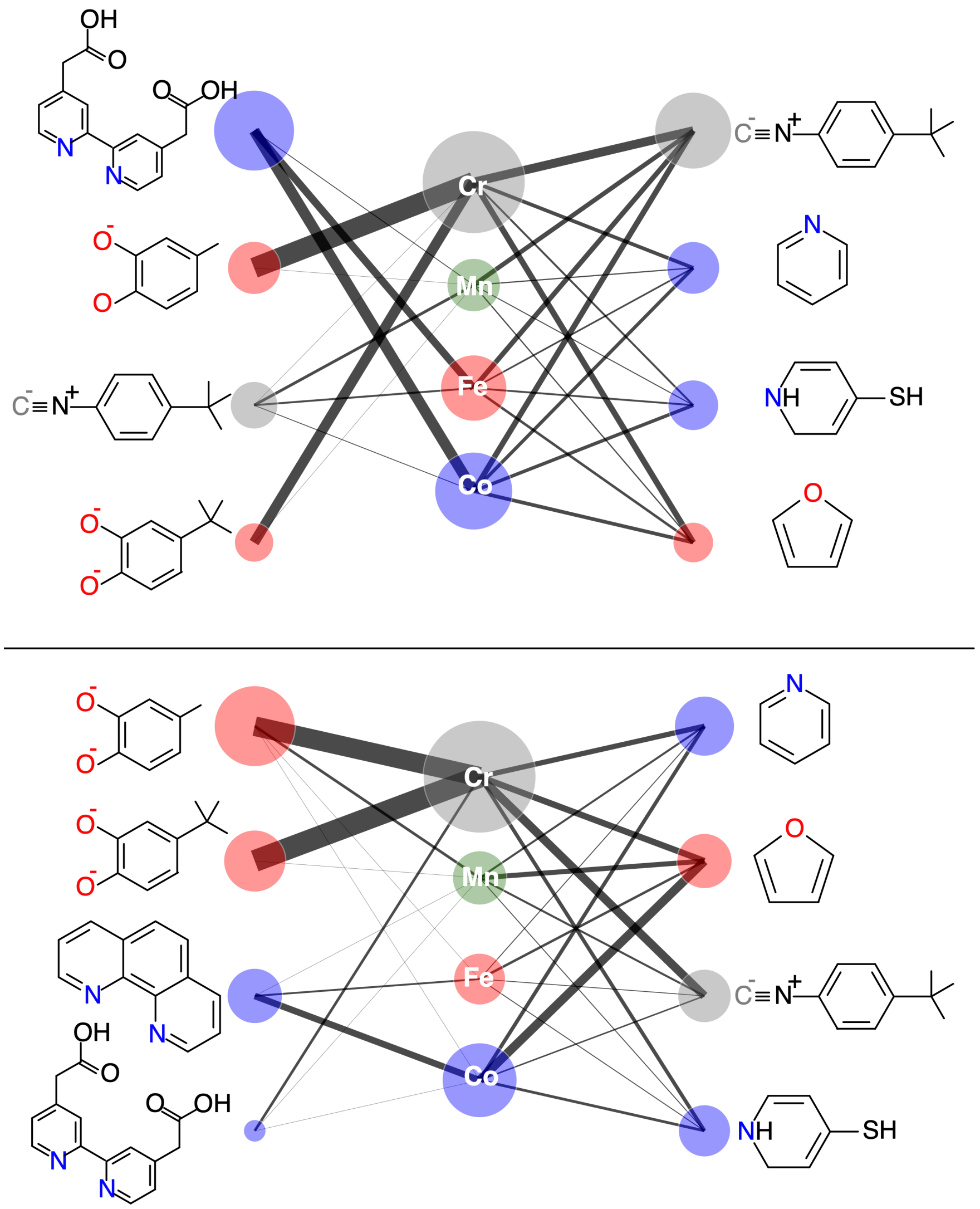

**Figure S25.** Network graph of the statistics for lead targeted Δ-SCF gap complexes from the BLYP FT-ANN (top) and B3LYP ANN (bottom). The scale of the circle indicates the relative abundance of the metal or equatorial/axial ligand appearing in the leads, and the width of a line connecting a metal and a ligand shows the relative abundance of this metal-ligand combination appearing in the leads. Metals are colored as the following: gray for Cr, green for Mn, red for Fe, and blue for Co. Coordinating atom types are colored as the following: gray for C, blue for N, and red for O.

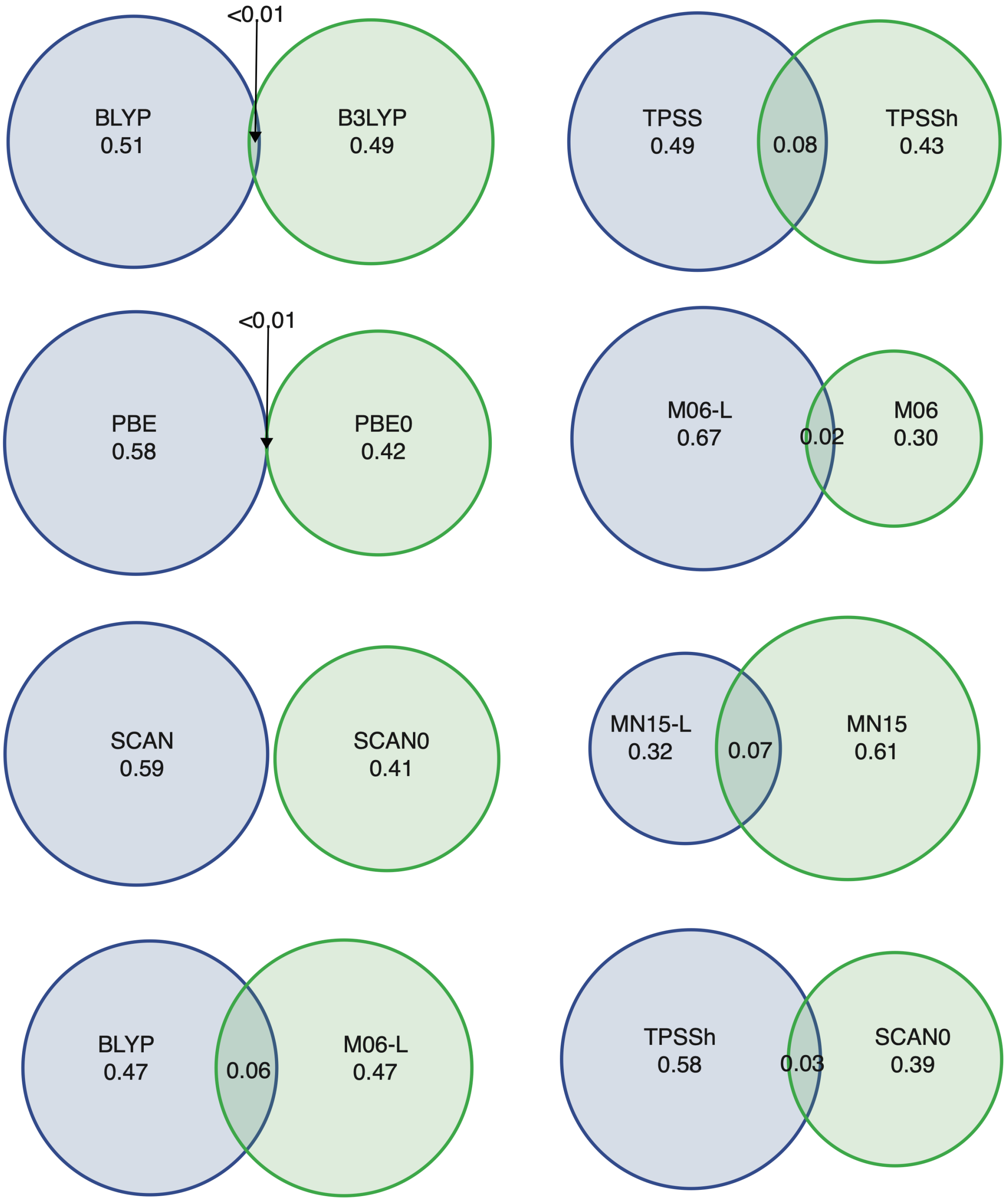

**Figure S26.** Venn diagrams of lead SCO complexes discovered by different pairs of DFAs as indicated.

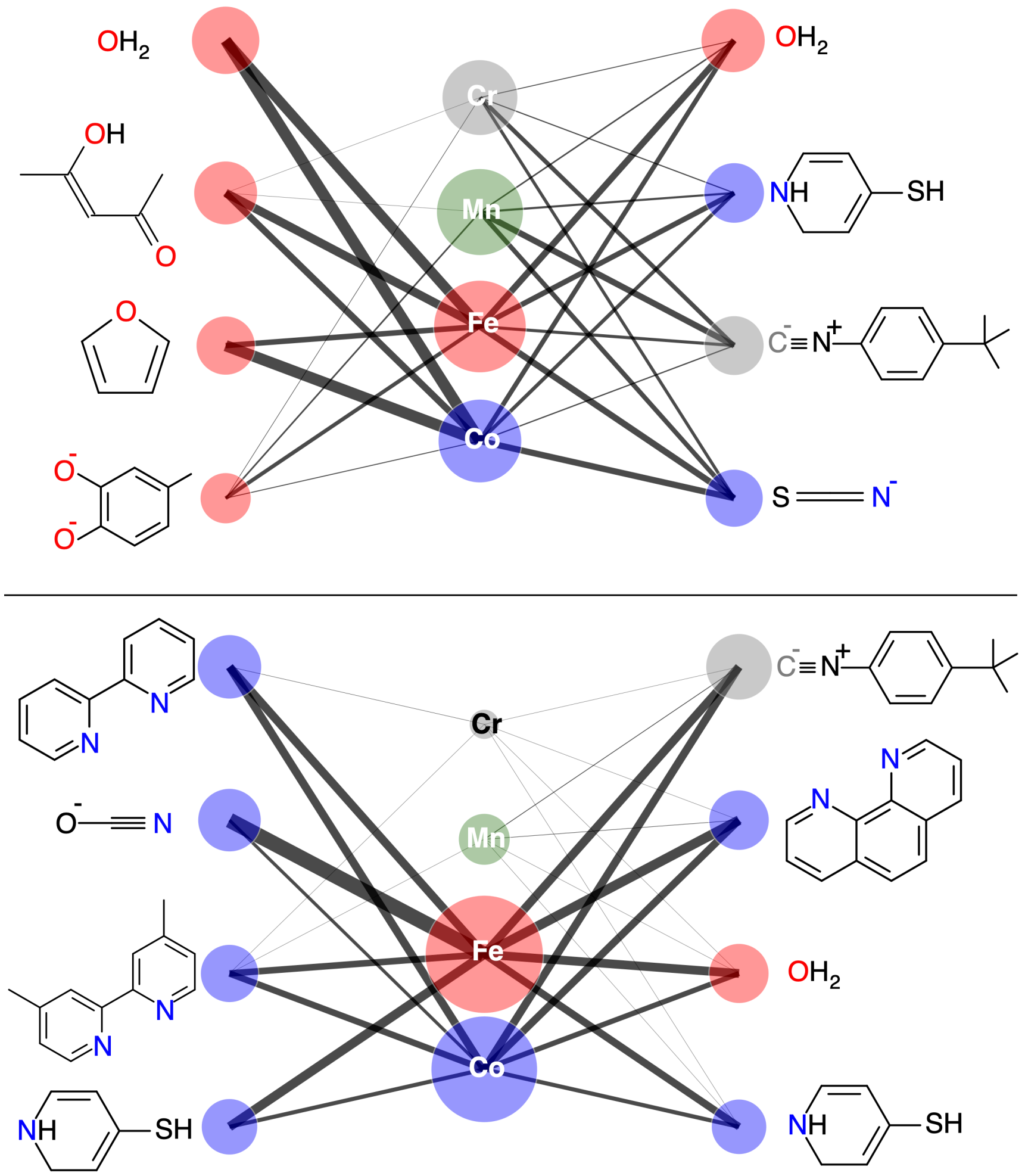

**Figure S27.** Network graph of the statistics of lead SCO complexes from the BLYP FT-ANN (top) and B3LYP ANN (bottom). The scale of the circle indicates the relative abundance of the metal or equatorial/axial ligand appearing in the leads, and the width of a line connecting a metal and a ligand shows the relative abundance of this metal-ligand combination appearing in the leads. Metals are colored as the following: gray for Cr, green for Mn, red for Fe, and blue for Co. Coordinating atom types are colored as the following: gray for C, blue for N, and red for O.

**Text S1.** Extraction of candidate SCO complexes from the literature.
Identification of candidate experimental spin crossover (SCO) complexes was performed largely similarly to automated mining of Fe(II) SCOs previously performed.[5] Briefly, a search through the Cambridge Structural Database (CSD version 5.41(Nov. 2019) + 3 Data updates) was performed for all octahedral mononuclear transition metal (M = Cr, Mn, Fe, Co) complexes (n=29,540).[5-6] From this set of complexes, structures with identical 6-letter refcodes but taken at different temperatures were selected. Multiple temperatures for identical structures are frequently reported in tests for spin crossover behavior (n=5,093 structures from 1,768 6-letter refcodes).[5] For the set of identical structures from each refcode the user-labelled oxidation states were verified to be 2 or 3 and the lowest temperature and highest temperature structures were identified (n=1934 structures from 967 6-letter refcodes). For each of the lowest-temperature structures the abstracts and titles were mined using pybliometrics[7] and SCO keywords and sentiment were analyzed.[5] Structures that contained SCO keywords and positive sentiment were retained resulting in 452 pairs of structures. From this set of likely SCO complexes we removed structures in which at least one of the structures was identified as having user-assigned charges and where the low temperature and high temperature structures were from the same paper resulting in 279 pairs of structures we refer to as candidate experimental SCO complexes.

**Table S16.** Identified experimental SCO complex counts by metal and oxidation state.

| metal | ox | count |
|---|---|---|
| Co | 2 | 30 |
| | 3 | 1 |
| Cr | 3 | 1 |
| Fe | 2 | 153 |
| | 3 | 70 |
| Mn | 3 | 24 |

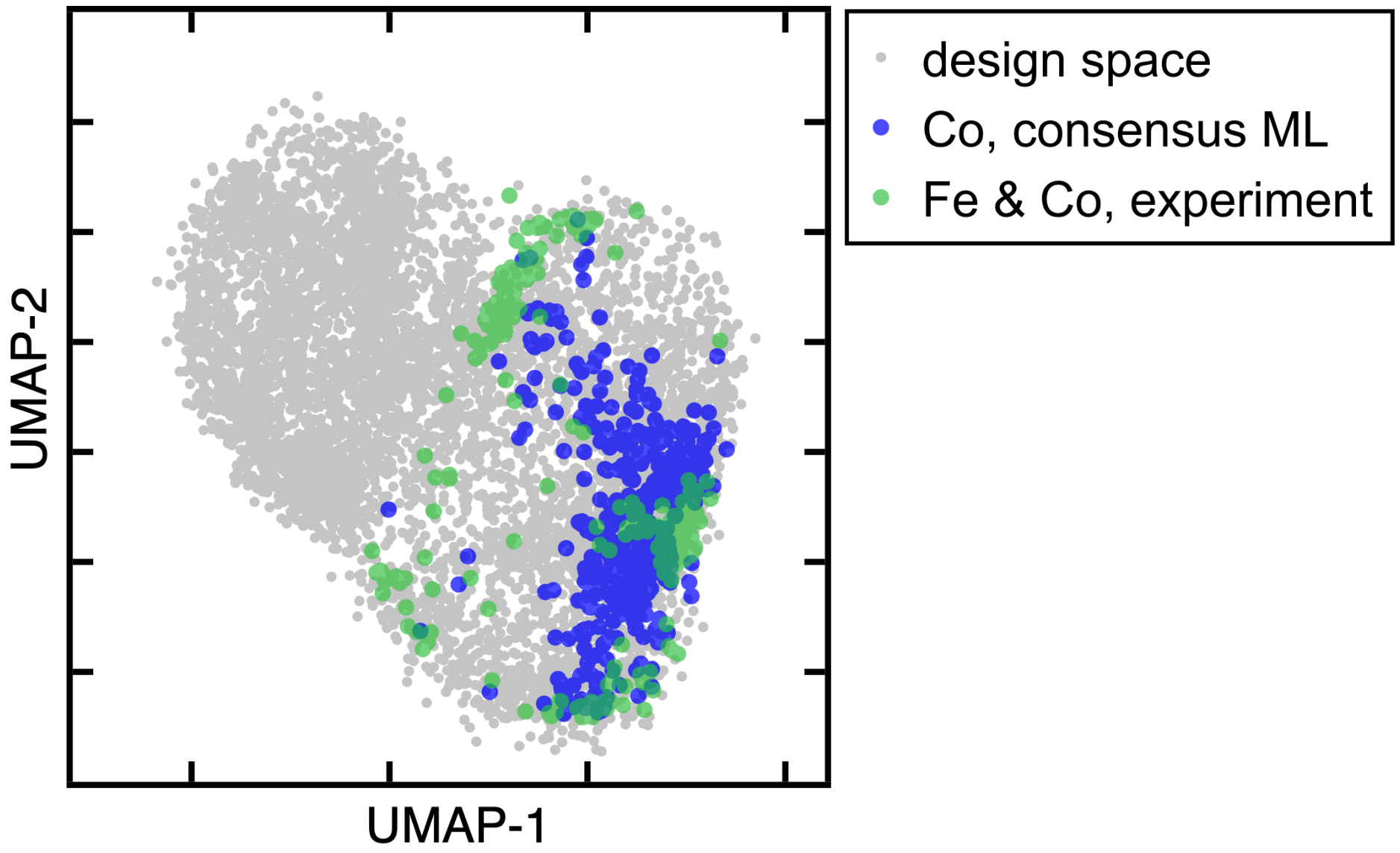

**Figure S28.** UMAP 2D visualization of $\Delta E_{\text{H-L}}$ for the design space of 187,200 complexes (gray), lead Co SCO complexes discovered with the consensus of all 23 DFAs considered in this work (blue), and experimentally observed Fe and Co SCO complexes (green).

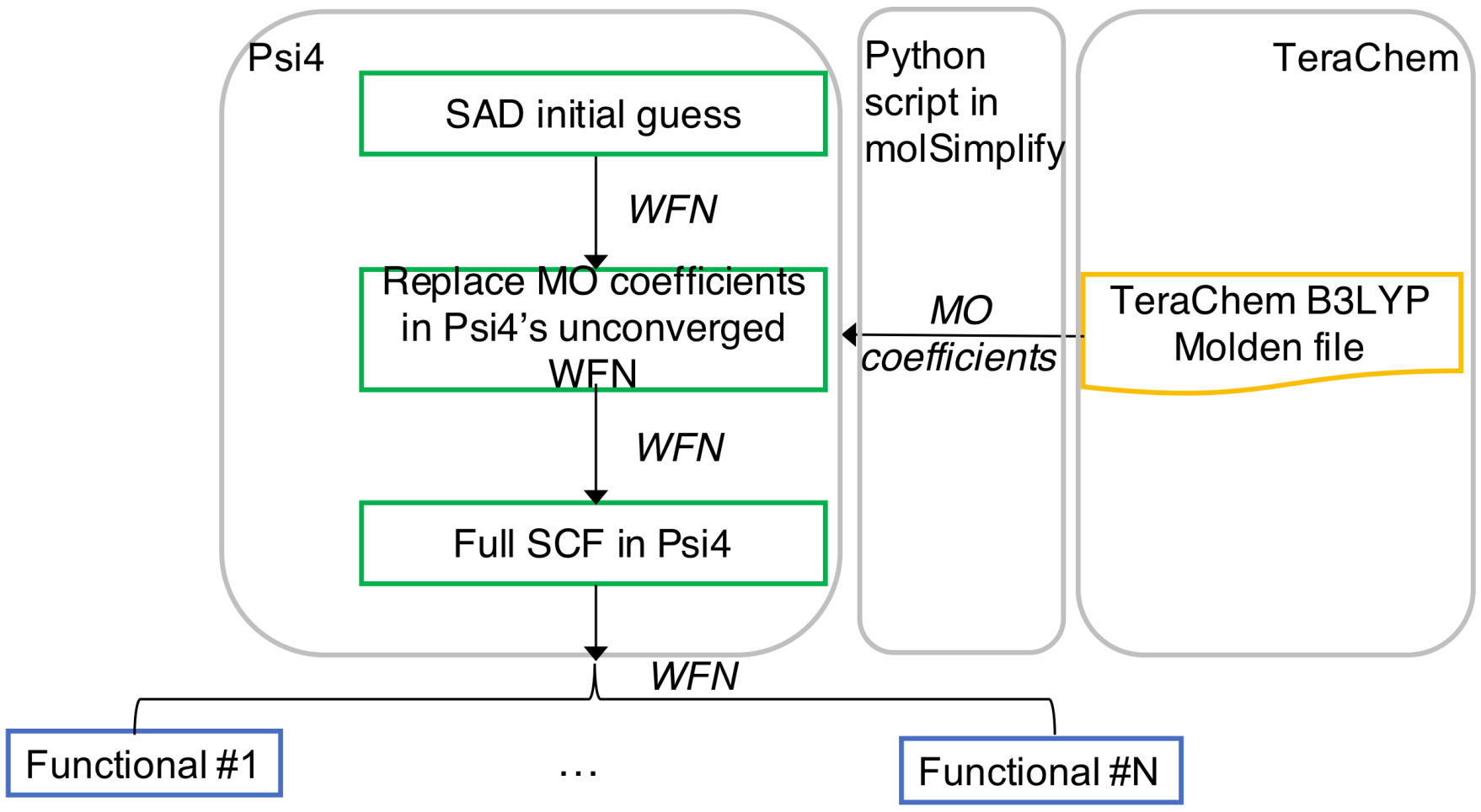

**Figure S29.** Computational workflow for setting up multiple single-point calculations. The molecular orbital (MO) coefficients of the B3LYP converged wavefunction obtained in TeraChem[8-9] are first extracted by a routine in our open-source package molSimplify[10-11]. These MO coefficients are used to replace the MO coefficients that would normally be obtained from an initial guess by the superposition of atomic density (SAD) for the wavefunction in Psi4[12]. A self-consistent field calculation with B3LYP is then performed to obtain the converged B3LYP wavefunction in Psi4. This wavefunction is used as the initial guess for the single-point energy calculations in the 22 functionals other than B3LYP to maximize correspondence of the electronic states converged across different functionals. For larger basis sets, a similar procedure is employed but using basis set projection inside Psi4 starting from the Psi4 B3LYP/LACVP* converged wavefunction.

**Table S17**. Summary of the default DFT calculation parameters and those used in this work in Psi4[13-14]. We used a smaller maximum SCF iteration because we read in the converged B3LYP wavefunction as the initial guess for all functionals. We also chose 3e-5 Ha as the density convergence threshold to be consistent with the TeraChem default setup, which was used to obtain the equilibrium geometries in *MD1* and *OHLDB*.

| | This work | Default |
|---|---|---|
| Number of radial points | 99 | 75 |
| Number of spherical points | 599 | 302 |
| DFT pruning scheme | robust | robust |
| Maximum SCF iterations | 50 | 100 |
| Density convergence threshold | 3e-5 Ha | 1e-6 Ha |

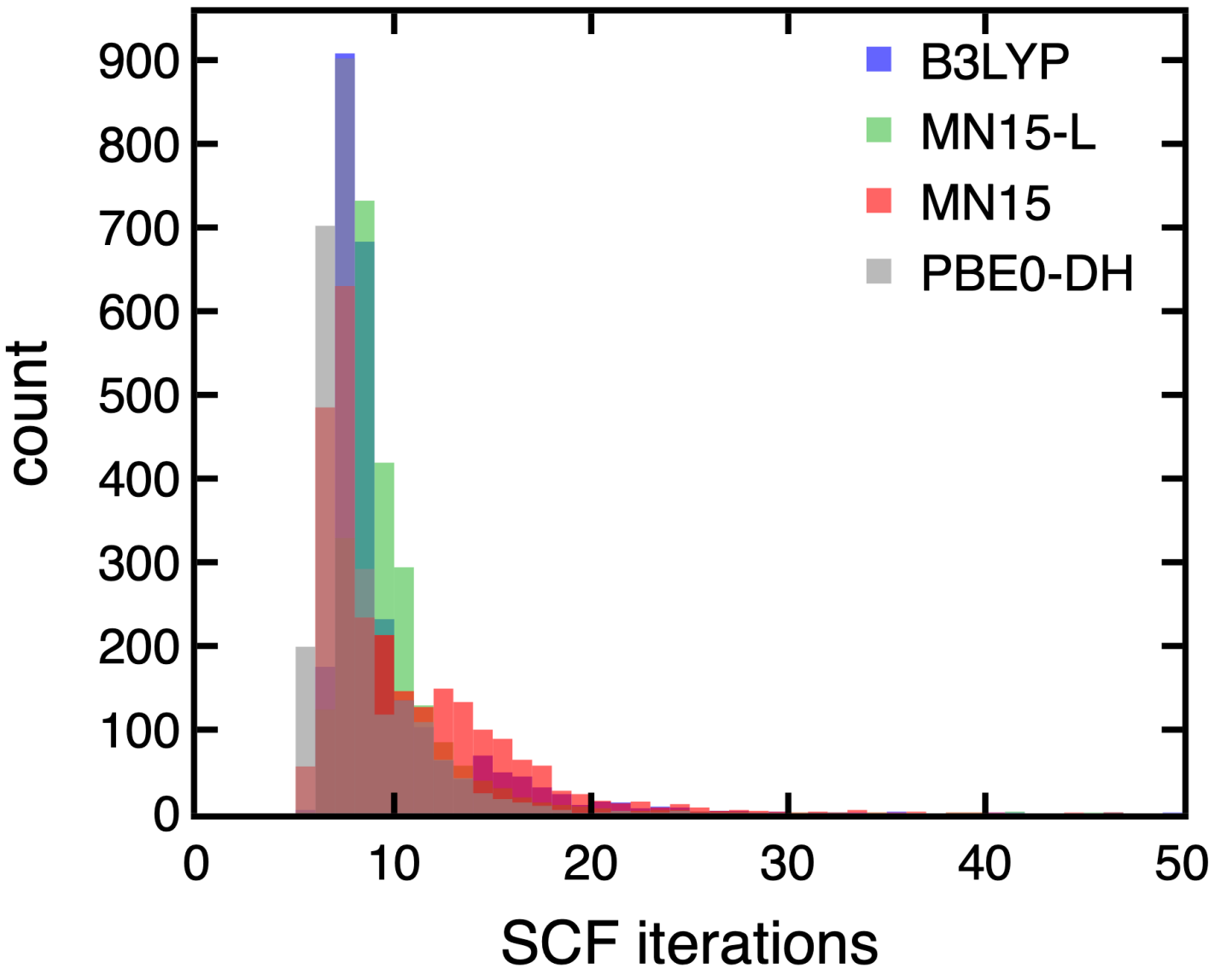

**Figure S30.** Unnormalized histograms of the number of SCF iterations needed for convergence of the N-electron systems calculated with the LACVP* basis set for four representative functionals, B3LYP (blue), MN15-L (green), MN15 (red), and PBE0-DH (gray). Results were obtained on the unique complexes in *MD1+OHLDB*. The bin size is one SCF iteration for all four DFAs.

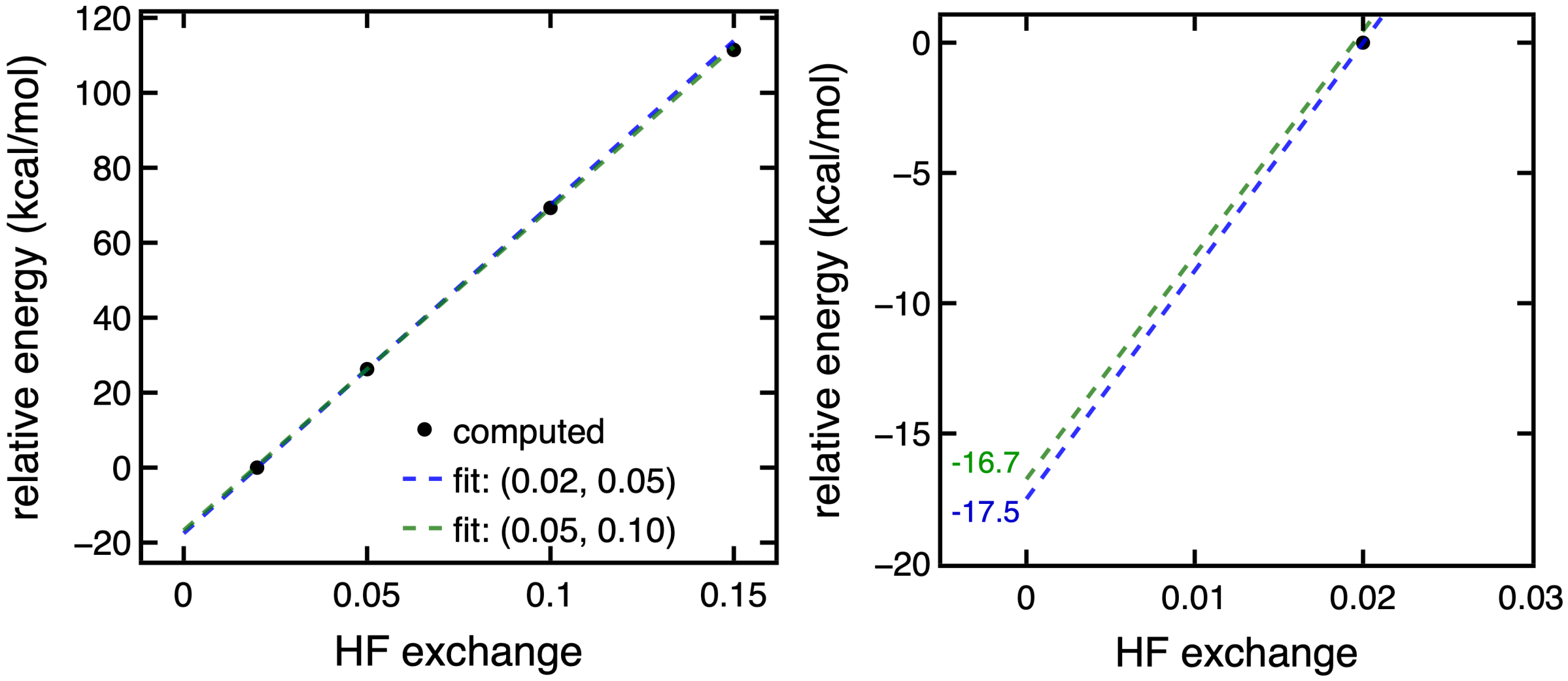

**Figure S31.** Example of the Hartree-Fock (HF) linear extrapolation scheme (left) and expanded approaching the 0.00 HF exchange fraction (right) for a representative quartet Cr(III)$(CO)_4$$(furan)_2$ complex to obtain the TPSS/LACVP* result that did not initially converge. The extrapolated value at 0.00 HF exchange from the linear fit of TPSS with 0.02 HF exchange and TPSS with 0.05 HF exchange fraction (blue dashed line) is used as an approximation for the TPSS energy of this complex. This result would deviate slightly from the green line obtained between the 0.05 and 0.10 HF exchange fraction. The relative energy at 0.02 HF exchange fraction is set to zero for comparison.

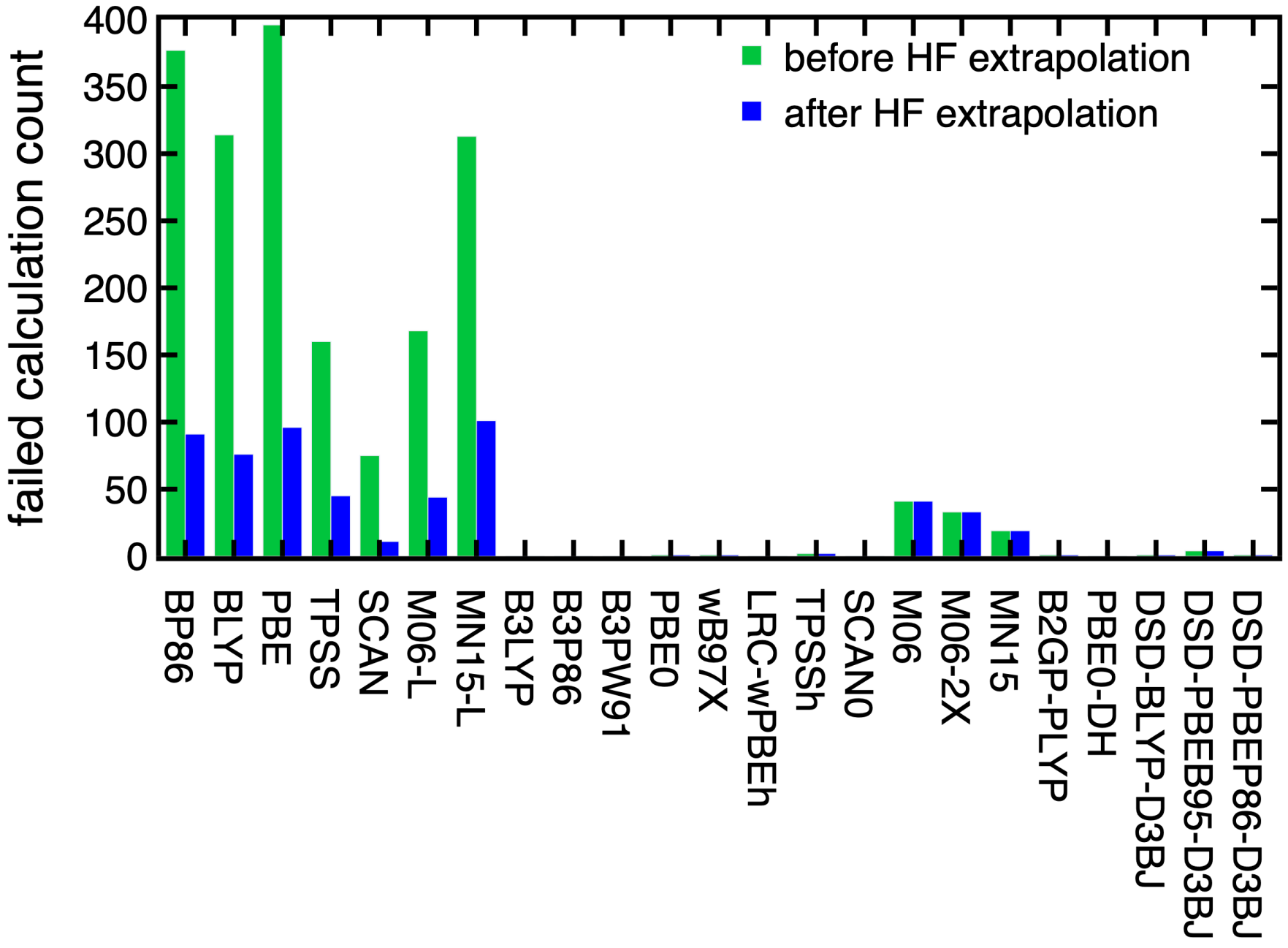


**Figure S32.** Numbers of calculations for which the converged SCF energy cannot be obtained with the LACVP* basis set before (green) and after (blue) HF linear extrapolation with each functional for the *N*-electron *MD1+OHLDB* complexes.

**Text S2.** Description of Hartree-Fock resampling procedure for converging DFT single point energies.

If a single-point calculation for a pure GGA or meta-GGA did not converge, we automatically performed single-point calculations with the hybrid version of that GGA or meta-GGA at a series of Hartree-Fock (HF) exchange percentages, starting at 15, then decreasing to 10, 5, and 2. Each subsequent HF exchange percentage calculation uses the wavefunction from the prior HF exchange percentage and is only performed if the calculation of the prior HF exchange percentage converges in the default maximum number of iterations. We then obtained the line formed by two points of the last three total energies converged (e.g., one with 10 and 5 and one with 5 and 2). We extrapolated the two lines to 0 percent HF exchange. If the two extrapolated total energy values did not deviate by more than 2.5 kcal/mol, the extrapolated value (i.e., to 0) of the last two points (i.e., 5 and 2) was used as an approximate energy for this GGA or meta-GGA. If the 2% calculation was not attempted or not converged or if the two extrapolations disagreed by > 2.5 kcal/mol, we indicated the linear extrapolation failed and removed the complex from the dataset.

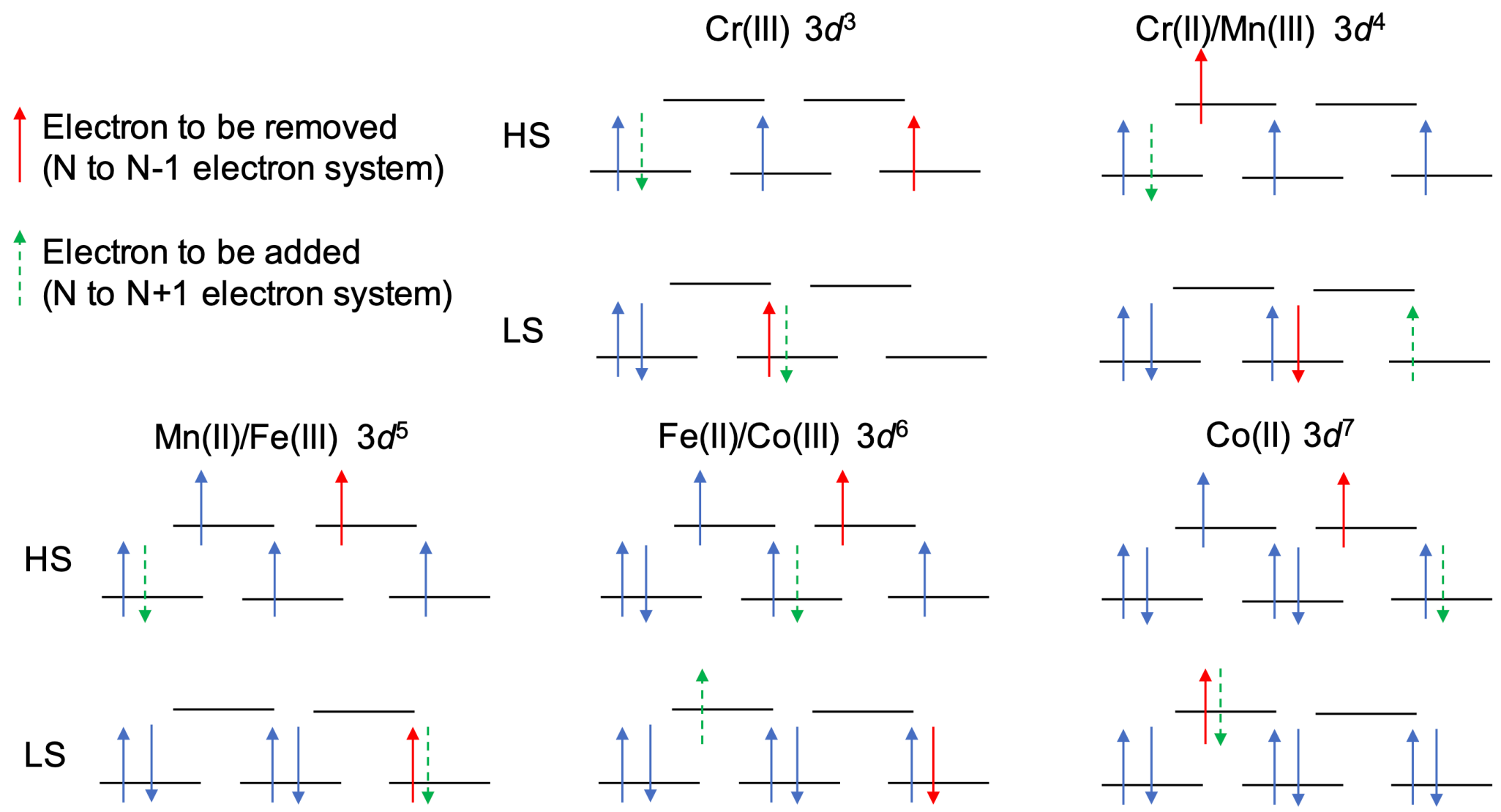


**Figure S33.** Conventions for adding (green dashed arrow) or removing (red solid arrow) electrons to an $N$-electron system to form the $N$+1-electron and $N$-1-electron systems. Both the high spin (HS) and low spin (LS) cases are shown for all $d$ shell configurations ($d^3$ to $d^7$) considered in this work.

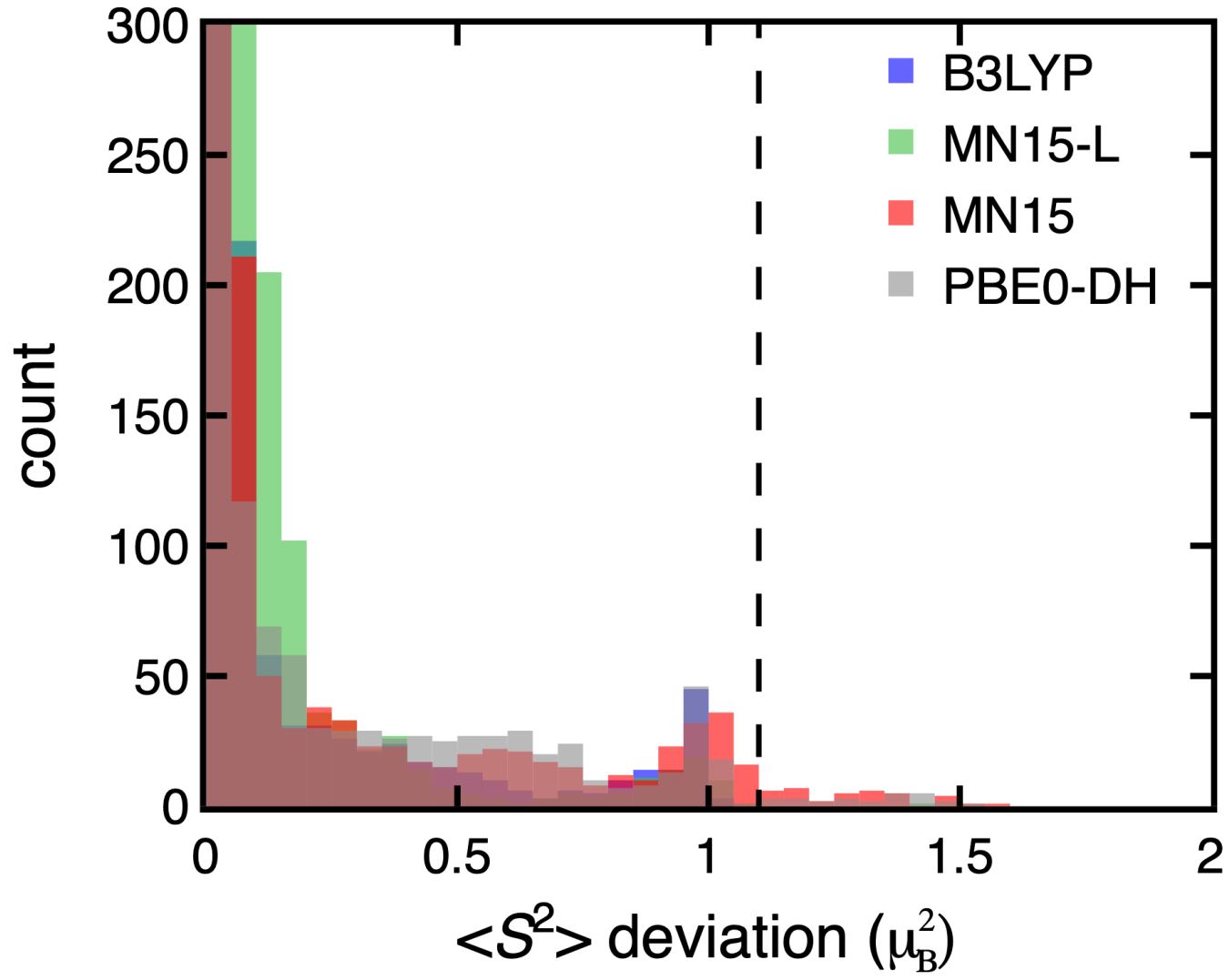


**Figure S34.** Histogram of ⟨$S^2$⟩ deviations from $S(S+1)$ for the N-electron systems calculated with LACVP* basis set for four representative functionals on the unique complexes in *MD1* and *OHLDB*: B3LYP (blue), MN15-L (green), MN15 (red), and PBE0-DH (gray). The bin size is 0.05 for all 4 functionals. The cutoff value of 1.1 $\mu_B^2$ is shown as a dashed black line.

**Table S18.** Number of calculations that were removed either because of the failure to obtain the final energy or an $\langle S^2 \rangle$ deviating from $S(S+1)$ by more than 1.1 $\mu_B^2$ for the original $N$-electron calculation as well as the $N$-1 and $N$+1 calculations. Results are reported for both the LACVP* and def2-TZVP basis sets. In practice, most of the calculations were eliminated because of the failure to obtain the final self-consistent field result. Ideally, we would have 2,639 successful calculations for $N$-1, $N$, $N$+1 electron systems if no calculation failed (i.e., all zeros in this table).

| | | LACVP* | | | def2-TZVP | | |
|---|---|---|---|---|---|---|---|
| | | N-1 | N | N+1 | N-1 | N | N+1 |
| GGA | BP86 | 439 | 377 | 133 | 542 | 403 | 190 |
| | BLYP | 392 | 314 | 128 | 506 | 382 | 179 |
| | PBE | 465 | 397 | 167 | 551 | 394 | 194 |
| meta-GGA | TPSS | 263 | 170 | 101 | 363 | 235 | 125 |
| | SCAN | 161 | 101 | 79 | 145 | 103 | 72 |
| | M06-L | 217 | 169 | 70 | 154 | 232 | 66 |
| | MN15-L | 506 | 314 | 111 | 417 | 107 | 100 |
| hybrid | B3LYP | 98 | 0 | 34 | 68 | 2 | 32 |
| | B3P86 | 103 | 8 | 36 | 72 | 2 | 41 |
| | B3PW91 | 109 | 10 | 40 | 80 | 4 | 37 |
| | PBE0 | 135 | 16 | 47 | 100 | 6 | 41 |
| meta-GGA hybrid | TPSSh | 126 | 15 | 37 | 94 | 7 | 44 |
| | SCAN0 | 199 | 30 | 85 | 169 | 20 | 73 |
| | M06 | 172 | 34 | 56 | 129 | 16 | 48 |
| | M06-2X | 256 | 25 | 57 | 154 | 4 | 57 |
| | MN15 | 215 | 47 | 47 | 116 | 13 | 48 |
| range-separated hybrid | LC-ωPBEh | 143 | 17 | 48 | 101 | 5 | 50 |
| | ωB97X | 131 | 3 | 34 | 95 | 4 | 42 |
| double hybrid | B2GP-PLYP | 284 | 21 | 81 | 267 | 14 | 85 |
| | PBE0-DH | 222 | 27 | 69 | 200 | 15 | 70 |
| | DSD-BLYP-D3BJ | 327 | 27 | 84 | 292 | 20 | 95 |
| | DSD-PBEB95-D3BJ | 314 | 18 | 75 | 248 | 18 | 81 |
| | DSD-PBEP86-D3BJ | 343 | 30 | 86 | 305 | 18 | 103 |

**Table S19.** Number of data points for each property that were obtained for all 23 DFAs with each basis set combination from the unique *MD1+OHLDB* complexes.

| | LACVP* | def2-TZVP | theoretical maximum |
|---|---|---|---|
| $\Delta E_{\text{H-L}}$ | 845 | 862 | 1068 |
| vertical IP | 1406 | 1447 | 2639 |
| Δ-SCF gap | 1214 | 1227 | 2639 |

**Text S3.** We have introduced a systematic approach to featurize molecular inorganic complexes that blends metal-centric and whole-complex topological properties in a feature set referred to as revised autocorrelation functions (RACs).[15] These RACs, variants of graph autocorrelations (ACs),[16-19] are sums of products and differences of atomic properties, i.e., electronegativity ($\chi$), nuclear charge ($Z$), topology ($T$), covalent radius ($S$), and identity ($I$). Standard ACs are defined as

$$P_d = \sum_i \sum_j P_i P_j \delta(d_{ij}, d)$$

where $P_d$ is the AC for property $P$ at depth $d$, $\delta$ is the Dirac delta function, and $d_{ij}$ is the bond-wise path distance between atoms $i$ and $j$.

In our approach, we have five types of RACs:

- ${}^{\mathrm{f}}_{\mathrm{all}}P_d$: standard ACs start on the full molecule (*f*) and have all atoms in the scope (all).
- ${}^{\mathrm{f}}_{\mathrm{ax}}P_d$ and ${}^{\mathrm{f}}_{\mathrm{eq}}P_d$: restricted-*scope* ACs that start on the full molecule (*f*) and separately evaluate axial or equatorial ligand properties

$$ {}^{\mathrm{f}}_{\mathrm{ax/eq}}P_d = \frac{1}{|\text{ax/eq ligands}|} \sum_{i}^{n_{\mathrm{ax/eq}}} \sum_{i}^{n_{\mathrm{ax/eq}}} P_i P_j \delta(d_{ij}, d) $$

where $n_{\mathrm{ax/eq}}$ is the number of atoms in the corresponding axial or equatorial ligand and properties are averaged within the ligand subtype.

- ${}^{\mathrm{mc}}_{\mathrm{all}}P_d$: restricted-scope, metal-centered (mc) descriptors that start on the metal center (mc) and have all atoms in the scope (all), in which one of the atoms, *i,* in the *i,j* pair is a metal center:

$$ {}^{\mathrm{mc}}_{\mathrm{all}}P_d = \sum_{i}^{\mathrm{mc}} \sum_{i}^{\mathrm{all}} P_i P_j \delta(d_{ij}, d) $$

- ${}^{\mathrm{lc}}_{\mathrm{ax}}P_d$ :and ${}^{\mathrm{lc}}_{\mathrm{ax}}P_d$: restricted-scope, metal-proximal ACs that start on a ligand-centered (lc) and separately evaluate axial or equatorial ligand properties, in which one of the atoms, *i*, in the *i,j* pair is the metal-coordinating atom of the ligand:

$$ {}^{\mathrm{lc}}_{\mathrm{ax/eq}}P_d = \frac{1}{|\text{ax/eq ligands}|} \frac{1}{|\mathrm{lc}|} \sum_{i}^{\mathrm{lc}} \sum_{i}^{n_{\mathrm{ax/eq}}} P_i P_j \delta(d_{ij}, d) $$

We also modify the AC definition, *P*′, to property differences rather than products for a minimum depth, d, of 1 (as the depth-0 differences are always zero):

$$ {}^{\mathrm{lc/mc}}_{\mathrm{ax/eq/all}}P'_d = \sum_{i}^{\text{lc or mc}} \sum_{i}^{\text{scope}} P_i P_j \delta(d_{ij}, d) $$

where scope can be axial, equatorial, or all ligands.

We demonstrated these RACs to be predictive for inorganic chemistry properties, such as spin-state splitting and ionization/redox potential. Over all possible origins (i.e. metal-centered, mc, or ligand-centered, lc) there are 42*d*+30 theoretical RAC features, where *d* is the maximum distance in bond paths through which two atoms are correlated in a single descriptor.[15] After eliminating the identity product RACs, There are 30*d*+25 product-based RACs (i.e., 6*d*+6 for each property) that arise from differing starting points (e.g., metal-centered or ligand- centered). After eliminating the identity difference RACs, there are 12*d* additional nontrivial difference RACs. With a bond depth cutoff of 3, this gives 151 RACs in total. Note that a given depth cutoff does not mean that whole-molecule information is excluded since the information can be included through the summation in RACs, but it does allow the user to choose not to directly correlate in a single feature the product of properties of two atoms farther apart than a certain topological distance. In this work, the full definition of the RAC representation also included oxidation state, spin state, and total ligand charge, for a total of 154 features.

**Table S20**. Range of hyperparameters sampled for ANN models trained from scratch with Hyperopt[20]. The lists in the architecture row can refer to one, two, or three hidden layers (i.e., the number of items in the list), and the number of nodes in each layer are denoted as elements of the list. The built-in Tree of Parzen Estimator algorithm in Hyperopt was used for the hyperparameter selection process.

| | |
|---|---|
| Architecture | {[128], [256], [512], [128, 128], [256, 256], [512, 512], [128, 128, 128], [256, 256, 256], [512, 512, 512]} |
| L2 regularization | [1e-6, 1] |
| Dropout rate | [0, 0.5] |
| Learning rate | [1e-6, 1e-3] |
| Beta1 | [0.75, 0.99] |
| Batch size | [16, 32, 64, 128, 256, 512] |

**Table S21**. Range of two hyperparameters for the radial-basis function (RBF) kernel sampled for KRR models with Hyperopt, where the regularization coefficients is the L2 regularization weight and decay width is the standard deviation of the Gaussian distribution in the RBF kernel. The built-in Tree of Parzen Estimator algorithm in Hyperopt was used for the hyperparameter selection process.

| | |
|---|---|
| Regularization coefficient | [1e-8, 100] |
| Decay width | [1e-8, 100] |